\documentclass[
aps,
prx,
showpacs,
preprintnumbers,
twocolumn,
superscriptaddress,
10pt,
]{revtex4-2}

\usepackage[T1]{fontenc}
\usepackage{newtxtext}
\usepackage{newtxmath}

\usepackage{enumitem}

\usepackage{makecell}

\usepackage{physics}
\usepackage{amsmath}
\usepackage{mathtools}
\usepackage{dsfont}

\newcommand{\floquetket}[1]{ \left| \left. #1 \right>\right> }

\newcommand{\timeket}[1]{\left| #1 \right)}
\newcommand{\timebra}[1]{\left( #1 \right|}
\newcommand{\timebraket}[2]{\left(\left. #1 \right| #2 \right)}

\newcommand{\Tramp}{T_\mathrm{ramp}}

\newcommand{\dtilde}[1]{\tilde{\tilde{#1}}}

\newcommand{\operator}[1]{\mathcal{#1}}
\newcommand{\superop}[1]{\hat{\operator{#1}}}

\newcommand{\h}{\operator{H}}
\newcommand{\FF}{F}

\newcommand{\operatorFF}[2]{\operator{#1}_{\FF,#2}}

\newcommand{\hFlambda}{\h_{\FF\!,\lambda}}

\newcommand{\p}{\operator{P}}
\newcommand{\A}{\operator{A}}

\newcommand{\micro}{\p_{\!\lambda}}
\newcommand{\identity}{\mathds{1}}

\newcommand{\ad}{\mathrm{ad}}

\newcommand{\floquethilbert}{\mathscr{F}}
\newcommand{\physhilbert}{\mathscr{H}}
\newcommand{\circhilbert}{\mathscr{L}_\circ}

\usepackage{appendix}

\usepackage{placeins}

\usepackage{color}
\usepackage[dvipsnames]{xcolor}
\usepackage{colortbl}
\usepackage{hyperref}
\hypersetup{colorlinks,citecolor=blue,linkcolor=blue,urlcolor=blue}

\definecolor{darkgreen}{rgb}{0.55, 0.71, 0.00}
\definecolor{Gray}{gray}{0.9}

\usepackage{comment}
\newcommand{\update}[1]{#1}

\newcommand{\supplement}{%
        \setcounter{table}{0}
        \renewcommand{\tablename}{Supplementary Table}
        \renewcommand{\thetable}{\arabic{table}}%
        \setcounter{figure}{0}
        \renewcommand{\thefigure}{S\arabic{figure}} %
        \renewcommand{\theHfigure}{S\arabic{figure}} 
		\setcounter{page}{1}
		\renewcommand{\figurename}{Fig.} 
		\renewcommand{\thesection}{\:S\arabic{section}}
		\setcounter{section}{0}
        \setcounter{equation}{0}
        \renewcommand{\theequation}{S\,\arabic{equation}}
     }

\usepackage{graphicx}
\usepackage[all]{hypcap}
\graphicspath{{supmat_figs/}{main_figs/}}

\begin{document}

\clearpage
\title{Counterdiabatic Driving for Periodically Driven Systems}

\author{Paul M.~Schindler}
\email{psch@pks.mpg.de}
\affiliation{Max Planck Institute for the Physics of Complex Systems, N\"{o}thnitzer Str.~38, 01187 Dresden, Germany}
\author{Marin Bukov}
\email{mgbukov@pks.mpg.de}
\affiliation{Max Planck Institute for the Physics of Complex Systems, N\"{o}thnitzer Str.~38, 01187 Dresden, Germany}

\begin{abstract}
Periodically driven systems have emerged as a useful technique to engineer the properties of quantum systems, and are in the process of being developed into a standard toolbox for quantum simulation. An outstanding challenge that leaves this toolbox incomplete is the manipulation of the states dressed by strong periodic drives. The state-of-the-art in Floquet control is the adiabatic change of parameters. Yet, this requires long protocols conflicting with the limited coherence times in experiments. To achieve fast control of nonequilibrium quantum matter, we generalize the notion of variational counterdiabatic driving away from equilibrium focusing on Floquet systems. We derive a nonperturbative variational principle to find local approximations to the adiabatic gauge potential for the effective Floquet Hamiltonian. It enables transitionless driving of Floquet eigenstates far away from the adiabatic regime. We discuss applications to two-level, Floquet band, and interacting periodically driven models. The developed technique allows us to capture nonperturbative photon resonances and obtain high-fidelity protocols that respect experimental limitations like the locality of the accessible control terms.
\end{abstract}

\maketitle

\begin{figure*}
    \centering
    \includegraphics[width=1\textwidth]{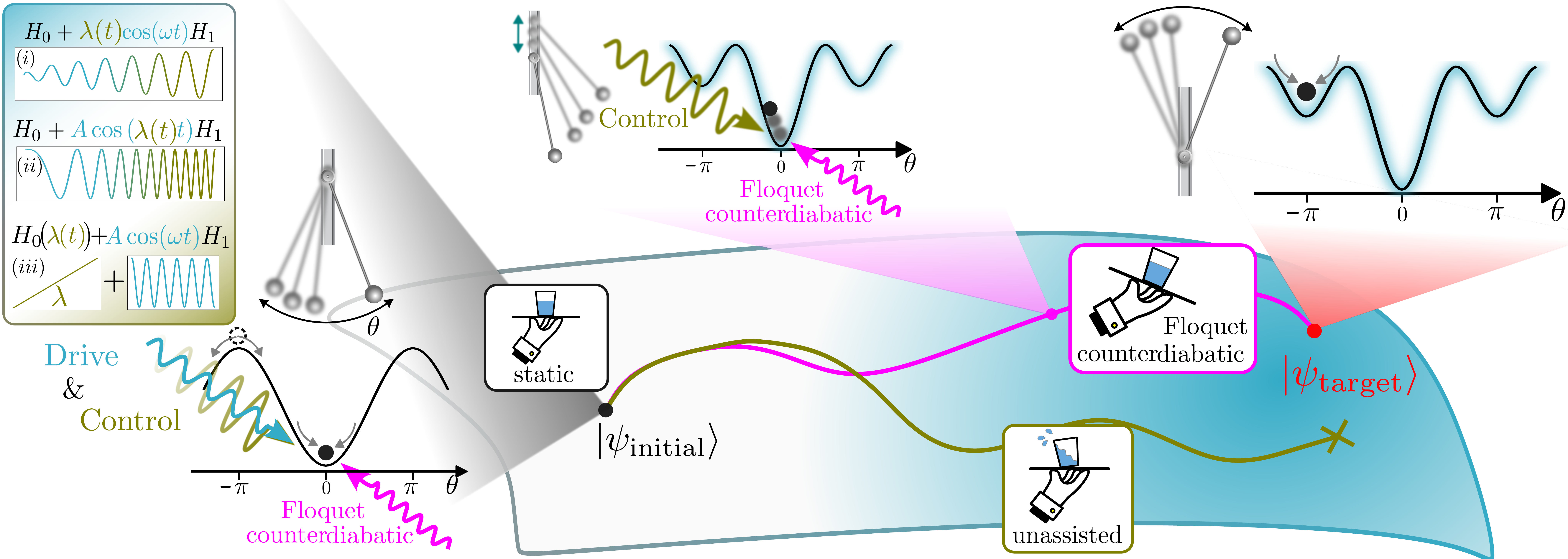}
    \caption{
    \textbf{Sketch: Floquet state manipulation \& Floquet counterdiabatic driving.}
    Periodic drives~(blue) can be used to engineer properties of static systems: e.g., the Kapitza oscillator has unstable fixed points at $\theta{=}{\pm}\pi$~(left black potential curve) stabilized by periodic driving~(other potential curves). 
    Preparing target Floquet states $\ket{\psi_\mathrm{target}}$ from initial states $\ket{\psi_\mathrm{initial}}$ requires additional controls~(yellow) on top of the Floquet drive (left box): for instance, \update{$(i)$ amplitude ramps $\lambda\hat{=}A$, $(ii)$ frequency chirps $\lambda\hat{=}\omega$, and $(iii)$ changes in static Hamiltonian $H_0=H_0(\lambda)$}.
    Direct state manipulation~(unassisted, without additional counterterms) fails away from the adiabatic regime. 
    Floquet counterdiabatic driving~(FCD) gives fast transitionless protocols.
    }  
    \label{fig:intro}
\end{figure*}

The use of periodic drives~\cite{Floquet1883,Shirley1965_FloquetTheory} to design the properties and behavior of quantum matter~\cite{Aidelsburger2011,Struck2012_TunableGaugeField,Struck2013_GaugeFields,Aidelsburger2013_Hofstadter,Aidelsburger2014_MeasuringChernNumber,Miyake2013_Hofstadter,Jotzu2014_Haldane,Wu2016_SpinOrbit,GomezLeon2013_FloquetBlochBandTheory,Goldman2016_Topo,Cooper2019_UltracoldAtomTopoSummary,Rudner2020_FloquetTopo,Leonard2023_FloquetFQH,kaptiza1951_DynamicalStabilization,Landau1976Mechanics,Broer2004_DynamicalStabilization,Apffel2020_FloquetLevitatingLiquid,Grossmann1991_DynamicalLocalization,Lignier2007_DynamicalLocalization,Eckardt2009_DynamicalLocalization,Song2022_FloquetPhaseTransition}, and to create phenomena not found in equilibrium systems~\cite{Khemani2016_DTC,Else2016_DTC,Moessner2017_FloquetThermalization,Khemani2019_DTC,Else2020_DTCReview,Zaletel2023_DTCReview,Rudner2013_AnomalousFloquetEdgeState,Nathan2015_AnomalousFloquetPrediction,Quelle2017_AnomlousFloquetTopo,Liu2019_ExoticTopoPhases,Wintersperger2020_Anomalous,Zhang2023_Anamalous}, has become widely known as Floquet engineering. 
Modern Floquet engineering stands on three pillars:
(i) the capability to engineer an effective Hamiltonian that governs the dynamics of the system~\cite{Goldman2014_FloquetGaugeFields,Bukov_2015_general_HFE,Eckardt2017_FloquetGases_Review,Aidelsburger2018_Review,Oka2019_FloquetMaterials,Weitenberg2021_FloquetGases};
(ii) the feasibility of suppressing unwanted heating out to parametrically long times ( prethermalization)~\cite{Lazarides2014_Heating,abanin2015_heating,abanin2015_heating,Takashi2016_HeatingBounds,Abanin2017_prethermalization,Abanin2017_Rigorous}, and
(iii) the ability to manipulate these metastable steady-states~\cite{Breuer1989_LandauZenerFloquet,Eckardt2008_DressedStatesCrossings,Weinberg_2017_adiabatic_Floquet}.

Designing local effective Hamiltonians, that support long-lived prethermal metastable states, is necessary yet insufficient to render Floquet engineering a self-contained toolbox for quantum simulation. 
An essential prerequisite for investigating the physics of the effective Hamiltonian is the ability to prepare and probe its eigenstates.
However, manipulating states under the presence of strong periodic drives remains an outstanding challenge, for a variety of reasons:
(1) The state-of-the-art approach to control periodically-driven systems is the {adiabatic} change of parameters; yet, the adiabatic limit for the effective Hamiltonian does not exist~\cite{Weinberg_2017_adiabatic_Floquet,Eckardt2008_DressedStatesCrossings}.
Moreover, (2) in practice adiabatic state preparation requires slow protocols to suppress excitations due to diabatic transitions. This stands in contrast to the limited coherence timescales in present-day quantum simulators~\cite{Preskill2018_nisqEra,Gross2017_QuantumSimulation} that require fast processes to avoid decoherence.
On the other hand, (3) fast control terms explicitly break the time-periodicity of the Hamiltonian which precludes a direct application of Floquet's theorem~\cite{Floquet1883} -- the cornerstone of Floquet engineering.
For these reasons, developing a theory for the control of periodically driven systems is a difficult yet important problem, whose solution has the potential to directly advance quantum simulation. 

In this work, we lay the foundations of a quantum control theory away from equilibrium, by focusing on periodically driven systems. We address the problems inherent to adiabatic Floquet control by generalizing the concept of transitionless driving~\cite{Berry2009_CD,Demirplak2003_CD,Demirplak2005_CD,Demirplak2008_CD,Xi2010_FewLevelSTA,Guerin2011_OptimalPulses,delcampo2012assistem,delCampo2013_CD,saberi2014adiabatic,Du2016_STAExperiment,Kolodrubetz2017_GeometryReview,GuryOdelin2019_STA,Santos2018_EnergyCost_STA,Hu2018_ExperimentGeneralizedSTA} to Floquet systems.
This theory of Floquet counterdiabatic driving~(FCD) [Fig.~\ref{fig:intro}] yields transitionless control protocols that transfer population between Floquet states in finite time. In practice, this is achieved by introducing additional terms to counteract diabatic transitions.

To this end, we derive a non-perturbative variational least-action principle for the adiabatic gauge potential associated with the effective Floquet Hamiltonian. Provided with a suitable ansatz, this method suppresses diabatic excitations for all avoided crossings in the quasi-energy spectrum -- including non-perturbative photon resonances.
We illustrate our theory on two paradigmatic and genuinely nonequilibrium control problems without static counterparts.
First, we apply FCD driving to the experimentally relevant case of Floquet state manipulation in a fermionic ultracold atom quantum simulator. Taking experimental constraints into account, we demonstrate that FCD driving leads to a significant reduction in diabatic transitions during a frequency chirp.
Second, we use our theory to enhance the fidelity of many-body state population transfer in a non-integrable Ising model. We showcase how periodic drives open up new ways to aid and improve state preparation even in static interacting systems. 
Ample details are provided in the extensive SM~\cite{SM}.

\textit{Floquet Counterdiabatic Driving---}Consider a periodically-driven Hamiltonian $\h_\lambda(t){=}\h_\lambda(t{+}T)$ with period $T{=}(2\pi)/\omega$, frequency $\omega$, and control parameter $\lambda$, e.g., drive amplitude, frequency, or external field [see Fig.~\ref{fig:intro}].
At fixed $\lambda$, stroboscopic evolution~($t{=}nT$, $n{\in}\mathbb{Z}$) is generated by the time-independent {Floquet Hamiltonian} $\hFlambda$ defined by $e^{-i T \hFlambda} {=} U_{F,\lambda}(T,0)$, where $U_{F,\lambda}(t,0){=}\operator{T} e^{-i \int_0^T \h_\lambda(t) \mathrm{d}t}{=}\micro(t) e^{-i t \hFlambda}$ with a micromotion unitary $\micro(t)$. 
Our goal is to achieve transitionless driving between eigenstates of Floquet Hamiltonians \update{$\h_{\mathrm{F}, \lambda(t)}$} upon varying the control parameter $\lambda\pqty{t}$ from an initial $\lambda_\mathrm{i}$ to a final $\lambda_\mathrm{f}$ value in finite time $\Tramp$. 
\update{
In general, the parameter $\lambda$ and periodic time dependency $t$ can change independently of each other, therefore, we refer use $\lambda(t)$ to indicate whenever they evolve jointly.
}
The difficulty comes from the control protocol which breaks time periodicity $\h_{\lambda(t+T)}(t{+}T){\neq}\h_{\lambda(t)}(t)$ and renders Floquet's theorem inapplicable.

\update{In the adiabatic limit~($T_\mathrm{ramp}{\to}\infty$) the time evolution under $\h_{\lambda(t)}(t)$ follows the instantaneous Floquet eigenstates. For finite ramp times, we suppress the diabatic transitions in the presence of the drive by adding the Floquet adiabatic gauge potential (FAGP)~\cite{Weinberg_2017_adiabatic_Floquet} to the Hamiltonian:}
\begin{equation}
\label{eq:HCD-main}
    \update{
        \h_{\mathrm{CD}\!,\lambda(t)}(t) = \h_{\lambda(t)}(t) +\dot{\lambda}\A_{\lambda(t)}(t), 
    }
\end{equation}
where $\A_\lambda(t) = i \partial_\lambda(\micro(t) W_\lambda)W_\lambda^\dagger \micro^\dagger(t)$, and $W_\lambda$ diagonalizes $\hFlambda$.
\update{Note that, exact FCD, Eq.~\eqref{eq:HCD-main}, leads to transitionless driving independent of the protocol $\lambda(t)$; in fact, any details of the protocol only enter via the velocity $\dot{\lambda}$.
However, the performance of approximations to Eq.~\eqref{eq:HCD-main} depends on the specifics of the protocol $\lambda(t)$.
}
As expected, in the adiabatic limit, $\dot{\lambda}\to 0$, all diabatic transitions vanish since $\dot{\lambda}\A_\lambda \to 0$.
\update{
A straightforward calculation shows that $\A_\lambda(t)$ satisfies the equation~(see SM~\cite{SM})
}
\begin{equation}
\label{eq:GFloquet}
\update{
    i\comm{\h_\lambda\pqty{t}}{\operator{G}\pqty{\A_\lambda\pqty{t}}} + \partial_t \operator{G}\pqty{\A_\lambda\pqty{t}} = 0\, , 
    }
\end{equation}
\update{
where $\operator{G}\pqty{\A_\lambda(t)} = i \comm{\h_\lambda(t)}{\A_\lambda(t)} + \partial_t \A_\lambda(t) - \partial_\lambda \h_\lambda$.
 }
 
The FCD protocol \eqref{eq:HCD-main} guarantees transitionless driving of Floquet eigenstates by construction~\cite{SM}. However, it requires 
(i) solving the Floquet problem to obtain $\micro$, and 
(ii) fully diagonalizing $\hFlambda$ for all parameter values $\lambda(t)$. 
These are hard tasks, as obtaining the full spectrum in many-body systems is exponentially hard in the system size, while finding closed-form expressions for the Floquet Hamiltonian is challenging already in few-level systems due to time-ordering. 
Thus, computing the exact FAGP $\A_\lambda(t)$ is practically out of reach for almost all models.
Therefore, the relevant question for experimental applications concerns finding an optimal local approximation to the exact CD Hamiltonian~\eqref{eq:HCD-main} (l-FCD).

Note that the widely used inverse-frequency expansion (IFE) can be adapted to map the time-dependent problem $\h_\lambda(t)$ to an approximate static CD problem~\cite{SM} for which a variational adiabatic gauge potential can be derived using static theory~\cite{yague2023shortcut,SM}. However, this approach is limited by the validity of the IFE which breaks down whenever photon resonances occur~\cite{Eckardt2015_VanVleck,goldman2015_ResonantFloquet,Bukov2016_resonances}, and is particularly problematic for interacting systems, or when using frequency chirps $\lambda(t){=}\omega(t)$. We thus propose a variational principle for the FAGP unrelated to the IFE. 

\textit{Variational Principle---}The central conceptual result of this work is the formulation of a variational principle to compute an approximate FAGP $\A_\lambda{\approx} \operator{X}_\lambda$. In fact, Eq.~\eqref{eq:GFloquet} is equivalent to extremizing the action $S$, i.e., $\delta S =0$, where
\begin{eqnarray}
\label{eq:varl-main}
        S\bqty{\operator{X}_\lambda} &=& \int_0^{T} \Tr(\operator{G}^2(\operator{X}_\lambda(t)))\, \mathrm{d}t, \\
        \operator{G}\pqty{\operator{X}_\lambda}
        &=& i\comm{\h_\lambda(t)}{ \operator{X}_\lambda(t)}+ \partial_t \operator{X}_\lambda(t) - \partial_\lambda \h_\lambda(t)\, \nonumber ,
\end{eqnarray}
where the integral over time and the partial derivative $\partial_t$ are evaluated at fixed $\lambda$. \update{The special case of a frequency chirp, $\omega=\omega\pqty{\lambda}$, must be treated separately due to the interplay of parametric~($\lambda(t)$) and periodic~($\omega t$) time-dependency. This requires (i) to replace the frequency $\omega$ by the instantaneous frequency $\nu {=} {\dd (\omega t) }/{\dd t}$ throughout, and (ii) to consider the explicit time-dependency in the variational action~\eqref{eq:varl-main}. In summary, one has to replace $\partial_\lambda \h_\lambda(t)$ by $\pqty{ \dot{\lambda} \partial_\lambda - \dot{\nu}/\nu } \h_\lambda(t)$ in Eq.~\eqref{eq:varl-main}, and Eq.~\eqref{eq:HCD-main} is replaced by $\h_{\mathrm{CD}\!,\lambda(t)}(t) = \h_{\lambda(t)}(t) + \A_{\lambda(t)}(t)$. A detailed discussion is found in the SM~\cite{SM}.} 
Crucially, Eq.~\eqref{eq:varl-main} allows us to compute an approximate {local} FAGP without knowing the Floquet Hamiltonian $\hFlambda$. 

The ansatz $\mathcal{X}_\lambda(t){=}\mathcal{X}_\lambda(t{+}T)$ for the variational FAGP carries an explicit periodic time dependence; therefore, in addition to all operators acting on the Hilbert space, a complete basis to expand $\operator{X}_\lambda(t)$ in, also includes a complete set of periodic functions. In general, $\operator{X}_\lambda(t)$ contains infinitely many terms arising from the Fourier harmonics of its periodic time dependence. In practice, it suffices to consider finite numbers $N_h$ of Fourier harmonics, and $N_O$ of local operators:
\begin{equation}
\label{eq:varl_ansatz-main}
    \operator{X}_\lambda(t) =  \sum_{m=1}^{N_O}  \sum_{\ell=-N_h}^{N_h} \chi_{\ell m} e^{i \ell \omega t}  \operator{O}_m \, ,
\end{equation}
where $\chi_{\ell m}$ are the variational parameters determined by minimizing the action $S\bqty{\operator{X}_\lambda}$. The operators $\operator{O}_m$ are chosen to reflect any constraints, e.g., locality or accessibility in the lab~\cite{Bloch2008_QuantumSimulationUltraColdAtoms}.
Since the action is quadratic in $\chi_{\ell m}$, the minimization is convex and thus guaranteed to converge to a global optimum~\cite{Sels2017_LCD}.

Unlike approaches based on the IFE, the variational principle~\eqref{eq:varl-main} is a nonperturbative method that allows for the direct determination of a local approximate FAGP applicable in the lab. In particular, it captures correctly photon resonances, and does not suffer from the asymptotic character of the IFE~\cite{SM}.

\textit{Applications---}We apply the variational principle of l-FCD driving to resonant drives~\cite{SM} and frequency chirps, chosen to emphasize the out-of-equilibrium character of the controlled system. The lack of static counterparts for such control setups brings about genuinely nonequilibrium effects absent in static control problems. 
\update{
Throughout, we quantify the advantage brought in by l-FCD driving, compared to unassisted driving (i.e., without CD terms), by computing the instantaneous fidelity $F(t)=\abs{\braket{\psi_{0, \lambda(t)}}{\psi(t)}}^2$ during the protocol or the final fidelity at the end of the ramp, $F(T_\text{ramp})$, between the time-evolved state $\ket{\psi(t)}$ and the instantaneous Floquet eigenstate $\ket{\psi_{0, \lambda(t)}}$. The instantaneous Floquet eigenstate is defined via $\ket{\psi_{0, \lambda(t)}}=P^\dagger_\lambda(t) \ket{\psi_{0, \lambda(t)}^F}$ where $ \ket{\psi_{0, \lambda(t)}^F}$ is the eigenstate of $\h_{F,\lambda(t)}$ that is adiabatically connected to the initial state of the time-evolution.
}
In general, the closer the fidelity is to unity, the better the FAGP approximation.

\textit{Fermionic Band Models.---}The experimental realization of topological band models~\cite{Cooper2019_UltracoldAtomTopoSummary} is a milestone of ultracold atom quantum simulators.
However, studying topologically non-trivial states in experiments requires swift, high-fidelity state preparation procedures.
FCD driving provides such state preparation schemes at low extra cost. Indeed, the FCD variational principle~\eqref{eq:varl-main} is straightforward to apply to generic Floquet-engineered (topological) band systems. 

We illustrate this using the setup from a recent experiment demonstrating Floquet topological pumping in a system of one-dimensional non-interacting fermions~\cite{Viebahn_etal_FloquetPump_2022}.
The system is described by a free-fermion Hamiltonian in momentum space
$\h_\lambda(t) = \sum_{q} \boldsymbol{\Psi}_q^\dagger \cdot \boldsymbol{h}_{\lambda}(q,t) \cdot \boldsymbol{\Psi}_q$,
where $\boldsymbol{\Psi}_q^\dagger =(c_{p,q}^\dagger,\, c_{s,q}^\dagger )$ with $c_{s(p),q}^\dagger$ the fermionic creation operator at quasimomentum $q$ in the $s$ $(p)$ Bloch band. The fermions are exposed to a periodic force of frequency $\omega$ which results in a time-dependence of the Bloch Hamiltonian $\boldsymbol{h}_\lambda(q,t)$. The corresponding effective Hamiltonian exhibits a Floquet-Bloch band topological pump. The precise details of the model are not important for the present discussion and can be found in the SM~\cite{SM}.

Starting from the filled $s$-band of the system which is topologically trivial in the absence of the drive, a three-stage sequence with amplitude and frequency ramps is applied to prepare the initial state for the Floquet pump. The overall bottleneck for the entire preparation protocol in the experiment~\cite{Viebahn_etal_FloquetPump_2022} is set by the smallest gap encountered during the frequency ramp stage. 
For frequency chirps, the size of the Floquet zone varies with time which can lead to photon absorption resonances. 

\begin{figure}[t]
    \centering
    \includegraphics[width=0.5\textwidth]{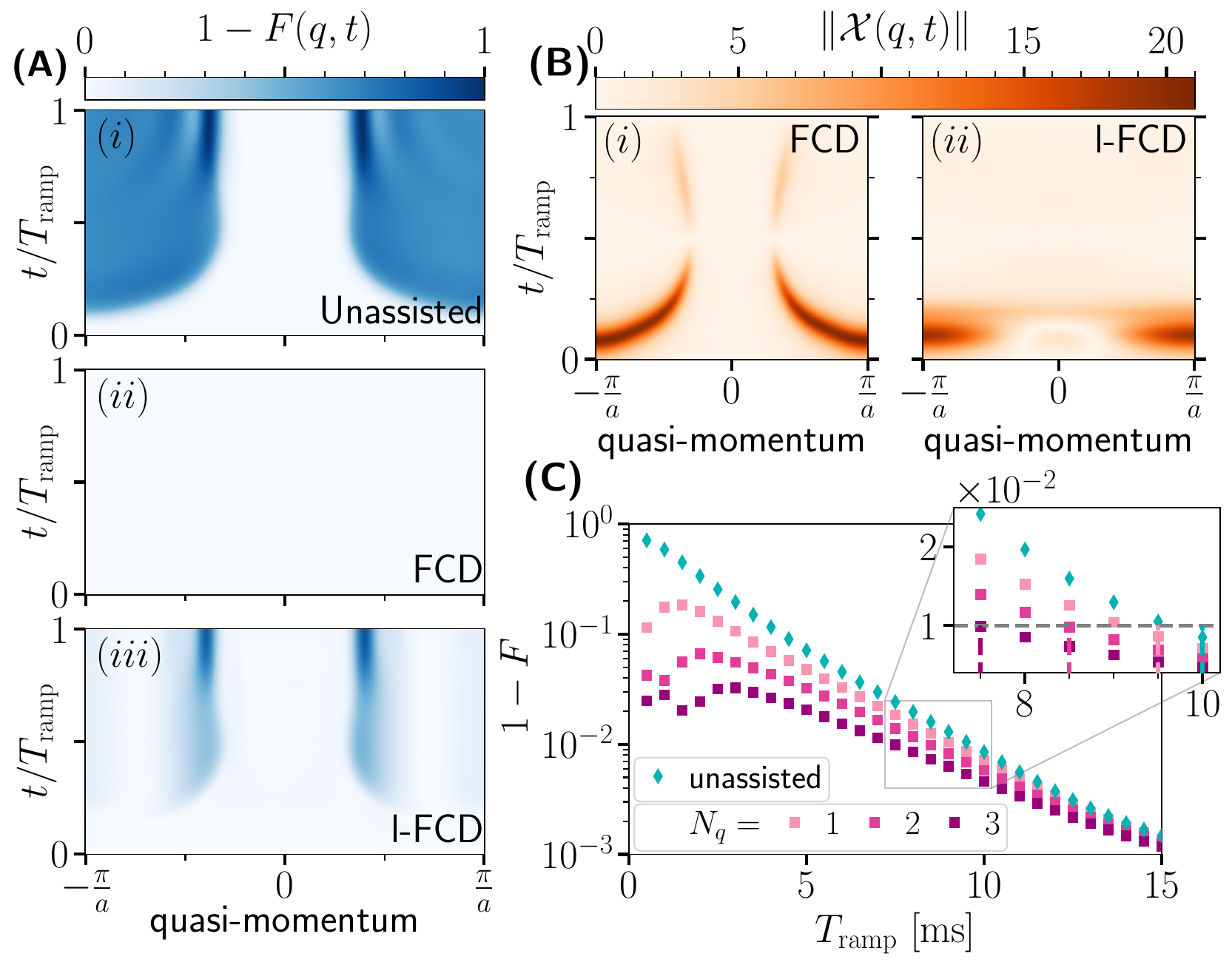}
    \caption{
        \textbf{Fermionic Floquet band model FCD assisted state preparation:} 
        \textbf{(A)} Momentum resolved instantaneous Infidelity $1{-}F(q,t)$ for unassisted~$(i)$, exact FCD~$(ii)$ and l-FCD~$(iii)$ for $N_q{=}2$ protocol for $T_\mathrm{ramp}{=}1\,\mathrm{ms}$.
        \textbf{(B)} Amplitude $\norm{\mathcal{X}(q,t)}$ of corresponding FAGP potential for FCD~$(i)$ and l-FCD~$(ii)$.
        \textbf{(C)} Final infidelity $1{-}F$ as a function of $T_\mathrm{ramp}$ for unassisted~(cyan diamonds), and l-FCD protocols~(purple squares) for $N_q{=}1,2,3$~(light to dark). Inset shows zoom into region around the $1\%$ infidelity threshold~(horizontal dashed line).
        Variational Floquet driving suppresses all diabatic transitions but requires non-local counter-terms [A(ii)]; local approximations yield a significant boost in fidelity [A(iii)].
        \textit{Parameters:}
        cubic chirp with $\omega_\mathrm{i}{=}2\pi {\times} 5\,\mathrm{kHz}$, $\omega_\mathrm{f}{=}2\pi{\times} 7.5 \,\mathrm{kHz}$, $N_f{=}500$ fermions and $N_h{=}32$.
    }
    \label{fig:TopoPump}
\end{figure}

To ensure that the final state corresponds to an eigenstate of the Floquet Hamiltonian at the final drive frequency~\cite{SM}, here we adopt the cubic chirp $\lambda(t){=}\omega(t) {=} \omega_\mathrm{f} {+} \pqty{\omega_\mathrm{i} {-} \omega_\mathrm{f}} \bqty{({t_\mathrm{f} {-} t}) / ({t_\mathrm{f} {-} t_\mathrm{i}})}^3$. Due to the hybridization of energy levels during the chirp, unassisted state preparation in a finite duration $T_\mathrm{ramp}$ leads to a significant loss of fidelity $F(q,t)$, for those quasimomenta where the ramp duration is small compared to the inverse gap: $T_\mathrm{ramp} \not{\gg} 1/\Delta_q$ [Fig.~\ref{fig:TopoPump}A(i)]. This points at the necessity to look into alternative schemes, such as variational FCD driving.

To compute the FAGP, we consider each fixed momentum $q$ individually, and make the ansatz 
\begin{equation}
\label{eq:chi_topo}
    \mathcal{X}_{q;\lambda}(t) = \sum_{\alpha=x,y,z} \sum_{\ell=0}^{N_h} \pqty{\chi^{\alpha,\ell}_{q;\lambda} e^{i\ell \omega t}}\; \sigma^\alpha.
\end{equation}
The variational parameters $\chi^{\alpha,\ell}_{q;\lambda}$ are smooth functions of $(q,\,\lambda)$; we find their values numerically by minimizing the action $S\bqty{\operator{X}_\lambda}$~\cite{SM}. The magnitude, $\norm{\mathcal{X}(q,t)}$, of the resulting variational FAGP is shown in Fig~\ref{fig:TopoPump}B(i).
While ansatz~\eqref{eq:chi_topo} cannot reproduce the Floquet gauge potential exactly, due to the finite number $N_h$ of Fourier harmonics kept, as seen in Fig.~\ref{fig:TopoPump}A(ii), the resulting FCD protocol leads to almost perfect transitionless driving at all times.

The FAGP coefficients $\chi^{\alpha,\ell}_{q;\lambda}$ are localized in momentum space [Fig~\ref{fig:TopoPump}B(i)], involving many {lattice harmonics}, $e^{i j qa}$. 
Since the $j$'th lattice harmonic $e^{i j qa}$ corresponds to a $j$'th-neighbor tunneling in real space, the exact FAGP is delocalized in real space. Therefore, to implement the exact FCD protocol in an experiment, precise control over all individual $j$'th neighbor couplings is required, which may not be feasible~\footnote{Arbitrary-range hopping is feasible using the techniques developed for synthetic dimensions~\cite{Boada2012_SyntheticDimensions,Celi2014_SyntheticDimensions,Mancini2015_SyntheticDimensionsRealization,Dutt2019_SyntheticSpectroscopy}.}. 
To alleviate this locality issue, we make a local approximation~(l-FCD) to the exact FCD protocol restricting the control to a limited number, $N_q$~($q{=}1,2,\dots$), of lattice harmonics. As anticipated, while the l-FCD counterterms take a simpler form in momentum space they cannot suppress all excitations, leading to finite infidelity, cf.~Fig.~\ref{fig:TopoPump}B(ii) and Fig.~\ref{fig:TopoPump}A(iii).
Nevertheless, compared to the unassisted protocol, l-FCD driving gives a noticeable increase in fidelity [Fig.~\ref{fig:TopoPump}C].

The versatile character of the variational FAGP approach allows us to build any experimental restrictions directly into the ansatz, so long as one is willing to tolerate a certain fidelity degradation. 
Frequent experimental constraints reflect the accessibility of control channels or the locality of available control terms.
For example, depending on the experimental platform, imaginary-valued terms $\propto \sigma^y$ required for the gauge potential may be difficult to implement. However, rotating to a moving frame we can replace any $\propto \sigma^y$ term using $\sigma^x$ and $\sigma^z$ terms~\cite{Sels2017_LCD}. Another interesting way to generate the gauge potential terms is using Floquet engineering~\cite{Claeys2019_FloquetCDProtocols}.

\begin{figure}[t]
\centering
\includegraphics[width=0.5\textwidth]{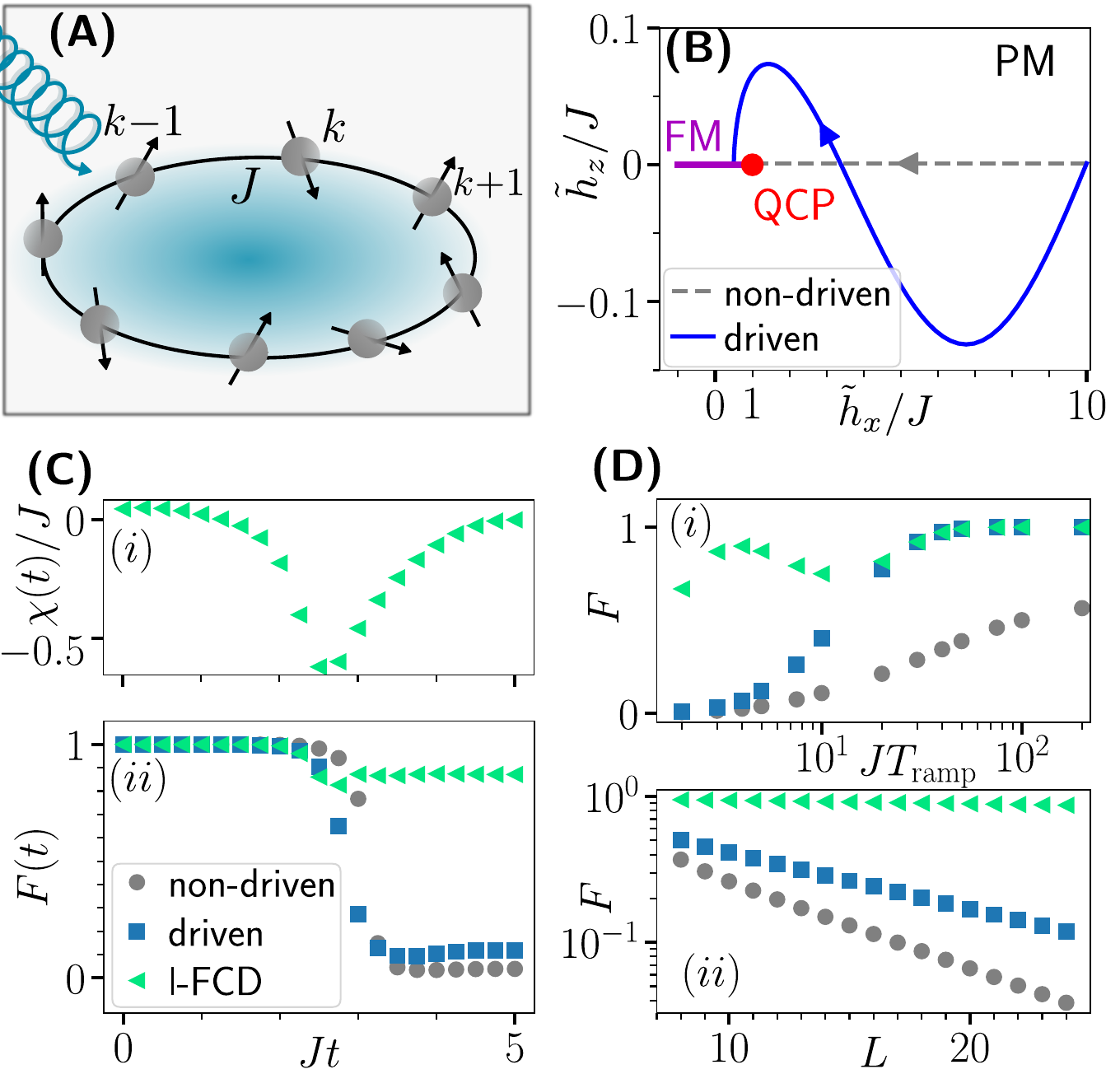}
\caption{
    \textbf{Local Floquet CD driving for circularly driven many-body spin chain}
    \textbf{(A)}.
    \textbf{(B)} Rotating frame fields in $\tilde{h}_x$-$\tilde{h}_z$ plane. The non-driven protocol~(grey dashed) ramps through the quantum critical point~(QCP, red) at $(\tilde{h}_x,\tilde{h}_z){=}(J,0)$. The driven protocol~(blue solid) yields an effective non-zero $\tilde{h}_z$ component allowing to circumvent the QCP.
    \textbf{(C)} Time-dependence of state-preparation schemes.
        (i) Variational parameter $\chi(t)$~\eqref{eq:circTFI_ansatz} computed using the algorithm detailed in~\cite{SM}.
        (ii) Instantaneous fidelity $F\pqty{t}$ for non-driven~(grey circles), driven~(blue squares) and l-FCD assisted driven protocols~(turquoises triangles).
    \textbf{(D)} Final fidelity for protocols from (C) for different ramp durations $T_\mathrm{ramp}$~(i), and different system sizes $L$~(ii).
    Floquet-engineered state preparation leads to significant increase in fidelity. Around $JT_\mathrm{ramp}=5$ the l-FCD protocol enables state-preparation close to unit fidelity almost independent of system size.
    \textit{Parameters:}
    $A(t)/J{=}\bqty{10{-}9.5\times\lambda\pqty{t}}$ and $\omega(t){=}\bqty{\omega_\mathrm{max}\times\sin(\pi \lambda\pqty{t})}$ where $\omega_\mathrm{max}{=}0.2\,J$; $\lambda\pqty{t}$ is the cubic ramp from $\lambda_\mathrm{i}{=}0$ to $\lambda_\mathrm{f}{=}1$; 
    $L{=}24$, $JT_\mathrm{ramp}{=}5$. We used a symmetry-breaking field $h_z/J{=}10^{-3}$ to single out one of the degenerate FM ground states.
}
\label{fig:ManyBody}
\end{figure}

\textit{Evading Many Body Criticality using FCD Drives---}An important problem in quantum many-body control is the preparation of ground states of an ordered phase, starting from a ground state in a disordered phase. This is a hard task since by definition there exists no adiabatic path in the thermodynamic limit connecting the two phases~\cite{Hastings2005_TopoAdiabaticConnection,Chen2010_LocalUnitaryConnection,Bachmann2011_GappedPhaseEquivalence,Bachmann2016_AdiabaticTheoremManyBodySystems} as long as the protecting symmetry is left intact. Therefore, in practice one either makes use of the finite size gap which vanishes in the thermodynamic limit and thus leads to divergent unassisted protocol durations, or one needs to introduce additional control channels to break the protecting symmetry explicitly. Whenever such symmetry-breaking controls cannot be directly implemented, nonequilibrium periodic drives emerge as a useful tool to engineer the implementation of additional terms assisting the state preparation protocol~\cite{Petiziol2018_FloquetFastAdiabatic,Petiziol2019_FloquetFastAdiabatic,Zhou2019_FloquetStateTransfer,Boyers2019_FloquetStateManipulation,Claeys2019_FloquetCDProtocols,morawetz2024efficient}. 
We now use l-FCD driving to demonstrate how periodic drives can be used to improve state preparation in static many-body systems. 

Consider the {transverse field Ising}~(TFI) chain
$\operator{H}_\mathrm{TFI} = \sum_{\ell=1}^L  J \sigma_\ell^z \sigma_{\ell+1}^z + h_x \sigma_\ell^x + h_y \sigma_\ell^y$ 
with ferromagnetic interactions $J{<}0$, and two transverse fields $h_{x,y}$. Independent of the ratio $h_x/h_y$, the system undergoes a quantum phase transition at $h {\equiv} \sqrt{h_x^2+h_y^2} {=} J$ from a disordered paramagnetic~(PM) phase ($h{>}|J|$) to an ordered ferromagnetic~(FM) phase ($h{<}|J|$), where the energy gap closes. Therefore, for the static system in the thermodynamic limit, there exists no adiabatic path connecting the two phases. Since the instantaneous Hamiltonian preserves the underlying $\mathbb{Z}_2$ symmetry for all parameter values, the PM and FM phases are strictly disconnected in equilibrium.

In the following, we assume that we can only access terms already present in $\operator{H}_\mathrm{TFI}$ for control purposes. 
We now demonstrate how one can use nonequilibrium periodic driving to engineer an adiabatic path connecting the two phases using only the $h_x$ and $h_y$ fields.

Consider the {circularly driven} TFI model
$
\label{eq:circTFI}
    \operator{H}(t;A,\omega) 
    = \sum_{j}  J \sigma_j^z \sigma_{j+1}^z + A\pqty{\cos(\omega t) \sigma_j^x + \sin(\omega t)\sigma_j^y}  \, ,
$
where we periodically modulate the $h_x$ and $h_y$ components with amplitude $A{=}A(t)$ and frequency $\omega{=}\omega(t)$ [Fig.~\ref{fig:ManyBody}A]. The time dependence explicitly breaks the integrability of the system, and brings out its intrinsically many-body character. 
Indeed, using the reference frame transformation ${W(t)=\exp(-i\omega(t) t \sum_j \sigma_j^z/2)}$, $\operator{H}(t;A,\omega)$ is mapped to the {non-integrable} {mixed field Ising model}~(MFI)
$
    \operator{H}_\mathrm{MFI}(t;A,\omega) 
    = \sum_{j}  J \sigma_j^z \sigma_{j+1}^z + \tilde{h}_x \sigma_j^x 
    + \tilde{h}_z\sigma_j^z \, ,
$
with transverse field $\tilde{h}_x(t){=}A(t)$ and longitudinal field $\tilde{h}_z(t){=}-\partial_t(\omega(t)t)/2$.
We deliberately choose the circular drive to avoid potential heating processes present in generic periodically-driven many-body systems.

The key insight here is that the circular drive leads to an effective $z$-field which unlocks a new control channel allowing us to circumvent the quantum critical point [Fig.~\ref{fig:ManyBody}B]. 
However, access to a $z$-field control is necessary but not sufficient for high-fidelity state manipulation: a na\"ive application of the circular protocol alone does not lead to a significant increase in fidelity, compared to the unassisted protocol, cf.~Fig.~\ref{fig:ManyBody}C,D.

To improve performance we apply FCD driving. 
The exact AGP in a generic many-body system requires the implementation of highly non-local both long-range and multi-body interactions including Pauli strings of arbitrary length~\cite{Kolodrubetz2017_GeometryReview,Sels2017_LCD}. 
Therefore, obtaining the exact FAGP is computationally expensive. In addition, implementing non-local multi-body terms is experimentally infeasible~\footnote{In fact, the non-local terms become more dominant in the proximity of the QCP. Thus, local CD driving is expected to fail when applied to the static system which inevitably comes close to the QCP.}.
To comply with these constraints, we focus on FCD protocols with only local single-particle counter-drives built from the accessible $h_{x,y}$ fields already present in $\operator{H}_\mathrm{TFI}(t;A,\omega)$. Choosing the ansatz
\begin{equation}
\label{eq:circTFI_ansatz}
    \mathcal{X}(t) = \chi(A,\omega) \sum_{j=1}^L \bqty{ \cos(\omega t)\sigma_j^y - \sin(\omega t) \sigma_j^x  }\,,
\end{equation}
with a single parameter $\chi$ [Fig.~\ref{fig:ManyBody}C(i)], brings a $y$-field $\tilde{h}_y{=}\chi$ in the rotating frame Hamiltonian $\operator{H}_\mathrm{MFI}(t;A,\omega)$~\cite{SM}.

Remarkably, using the numerically computed l-FCD assisted protocol~(cf.~Fig.~\ref{fig:ManyBody}C(i)) leads to an increase of fidelity by almost one order of magnitude compared to the unassisted driven protocol [Fig.~\ref{fig:ManyBody}C]. Notably, for protocols as short as $JT_\mathrm{ramp}{=}5$, the l-FCD assisted protocol leads to almost unit fidelity for a wide range of system sizes $L{=}8{-}24$, see Fig.~\ref{fig:ManyBody}D(ii). 
Already few Floquet cycles are sufficient to yield a nonequilibrium-based enhancement. This emphasizes the advantage of following the Floquet states even in the absence of periodic drives in the original control setup.

\textit{Discussion/Outlook---}Controlling metastable steady-states is an outstanding challenge in nonequilibrium dynamics, of prime importance in Floquet engineering. We propose a novel nonadiabatic control paradigm for Floquet systems based on transitionless driving between Floquet states, generalizing counter-diabatic driving to time-dependent Hamiltonians. 
To compute the Floquet adiabatic gauge potential we derive a nonperturbative variational principle that allows us to incorporate experimental constraints in the control protocol.  
We apply our theory to experimentally motivated control setups without equilibrium counterparts, such as frequency chirps. In a fermionic band model and a nonintegrable many-body spin chain, combining l-FCD driving with ideas from Floquet theory enables the engineering of improved state preparation schemes at no extra cost that are significantly more efficient than unassisted driving. Nonequilibrium drives also allow us to open up new effective control channels via suitably engineered change-of-reference transformations.


Besides having immediate applications in quantum simulation, quantum computing~\cite{yague2023shortcut} and quantum annealing~\cite{Kadowaki1998_QuantumAnnealingTransverseIsing,Das2008_QuantumAnnealingReview}, our work opens up several new directions.
The variational principle can be used as a starting point to generalize other shortcuts-to-adiabaticity techniques~\cite{Torrontegui2013_STA,GuryOdelin2019_STA,Ieva2023_COLD} to periodically driven systems. 
In addition, our ideas can provide new insights into the study of thermalization in nonequilibrium systems, since the AGP is used to detect chaos and integrability breaking~\cite{Mohit_etal2020_AGPChaos,Orlov_etal2023_AGPChaosApplication}.
Last but not least, the relation between the AGP and the quantum geometric tensor~\cite{Kolodrubetz2017_GeometryReview} opens up new avenues to investigate quantum geometry away from equilibrium, with applications in quantum metrology~\cite{yang2022variational}.

Finally, our variational principle applies also to classical periodically driven systems~\cite{Kolodrubetz2017_GeometryReview,higashikawa2018_classicalfloquet}, and can be extended to other nonequilibrium models described by local effective Hamiltonians, such as quasi-periodic~\cite{Dumitrescu2018_QuasiPeriodic} and random multipolar drives~\cite{Hongzheng2021_RMD}. 
\update{
Another promising extension of our variational principle is to the case of periodically driven \textit{open} quantum systems. While static (local) CD can be readily extended to open quantum systems~\cite{Ibanez_STANonHermitian2011,Vacanti_OpenCD2014,Funo_BoundOpenCD2021,Alipour2020_STAOpenSystem,Passarelli_OpenVariationalSTA0222} a similar extension for Floquet systems might be bound to regimes where a Floquet Lindbladian exists~\cite{Schnell_FloquetLindblad2020}.
}
Therefore, our work paves the way for a more general control theory of nonequilibrium systems.

\textit{Acknowledgements.---}We thank P.~W.~Claeys, M.~Kolodrubetz, and A.~Polkovnikov for insightful discussions.
Funded by the European Union (ERC, QuSimCtrl, 101113633). Views and opinions expressed are however those of the authors only and do not necessarily reflect those of the European Union or the European Research Council Executive Agency. Neither the European Union nor the granting authority can be held responsible for them.
The simulations make use of the QuSpin Python package~\cite{quspin2017,quspin2019}. Simulations were performed on the MPI PKS HPC cluster.

\bibliography{bibliography}


\FloatBarrier





\clearpage

\supplement

\onecolumngrid
\begin{center}\large \textbf{\textit{Supplemental Material}\\ Counterdiabatic Driving for Periodically Driven Systems} \end{center}
\vspace{2mm}
\twocolumngrid

\tableofcontents

\FloatBarrier

\vspace{1cm}

\update{
This supplemental material is structured as follows.
In Sec.~\ref{sec:theory}, we give a more detailed introduction to the theory of Floquet counterdiabatic driving. This section serves to introduce the notation and relevant relations that will be needed in the rest of the supplemental material. 
In Sec.~\ref{sec:varl}, we provide more details on the variational principle, including a detailed proof~(\ref{sec:agp_relation}) and the special case of frequency modulation~(\ref{sec:chirp}).
A common approach to Floquet systems is a perturbative treatment in form of an inverse frequency expansion; this can also be used to derive a local FCD protocol as we detail in Sec.~\ref{sec:IFE}.
Complementary to the main text we analyze the paradigmatic example of driven two-level in Sec.~\ref{sec:examples}. Specifically, we provide a class of analytically solvable examples~(\ref{sec:exactlysolvable}) and study the case of photon resonances~(\ref{sec:linear2LS}) where we demonstrate the the non-perturbative nature of the variational approach.
In Sec.~\ref{sec:algo}, we provide a numerical algorithm that we use to compute the variational local Floquet adiabatic gauge potentials reported on in the main text.
Eventually, in Secs.~\ref{sec:floquetpump} and \ref{app:manybody} we give further details on the Floquet topological pump and many body example reported on in the main text, respectively.
}

\section{\label{sec:theory}Theory of Counterdiabatic Driving for Floquet Systems}

\update{
In this section, we extend on the main text discussion of exact Floquet counterdiabatic. In addition, we introduce in great detail all quantities needed for the various detailed proofs in the following sections.
}

Consider a periodically driven Hamiltonian $\h_\lambda(t)=\h_\lambda(t+T)$ with period $T$, frequency $\omega=2\pi/T$ and control parameter $\lambda$. Common examples include switching on a periodic drive by ramping up its amplitude, changing the drive frequency, or another parameter in the static part of the Hamiltonian.
For every fixed value of $\lambda$, the stroboscopic evolution~($t=nT$ $n\in\mathbb{Z}$) is generated by a time-independent \textit{Floquet Hamiltonian} $\hFlambda$,
\begin{equation}
\label{eq:UF}
    e^{-i T \hFlambda} = \operator{T} e^{-i \int_0^T \h_\lambda(t) \mathrm{d}t} \, .
\end{equation}
More generally, there exists a rotating (\textit{Floquet}) frame in which the dynamics (both stroboscopic and non-stroboscopic) are exactly described by the Floquet Hamiltonian, i.e.,
\begin{equation}
\label{eq:FloquetTheorem}
    \hFlambda
    = \micro^\dagger(t) \h_{\lambda}(t) \micro(t) - i \micro^\dagger(t) \frac{\mathrm{d}}{\mathrm{d}t} \micro(t)\, .
\end{equation}
This frame is defined by the micromotion operator $\micro\pqty{t}=\micro\pqty{t+T}\equiv\exp(-i \operator{K}_\lambda(t))$ generated by the kick operator $\operator{K}_\lambda(t)$.

Our goal is to achieve transitionless driving between eigenstates of the Floquet Hamiltonian $\hFlambda$ (a.k.a., Floquet states) upon varying the control parameter $\lambda=\lambda\pqty{t}$ from some initial value $\lambda_\mathrm{i}$ to a final value $\lambda_\mathrm{f}$ in a finite time $\Tramp$. 
The difficulty comes from the observation that the time-dependence of the control parameter need not be periodic in general. Hence, formally it breaks the periodicity of the Hamiltonian, i.e., $\h_{\lambda(t+T)}(t+T)\neq\h_{\lambda(t)}(t)$.

The adiabatic theorem~\cite{Born1928_adiabatic,Kato1950_adiabatic} guarantees transitionless (adiabatic) driving for gapped states only in the limit of infinitely slow variations, i.e., as $\dot{\lambda}\to 0$~($\Tramp\to\infty$). In the absence of quasi-energy level-crossings along the protocol, this also holds for Floquet systems~\cite{Weinberg_2017_adiabatic_Floquet}. However, the strict adiabatic limit is inaccessible in practice. As a result, changing the control parameter faster than the inverse quasi-energy gap to nearby levels unavoidably leads to sizeable diabatic excitations.
In non-driven systems, such diabatic excitations can be exactly suppressed with the help of counterdiabatic (CD) driving.
By adding additional counterterms to the Hamiltonian, exact CD driving allows us to follow the adiabatically-connected instantaneous state of $\h_{\lambda}$ at all times during the application of the protocol.


\begin{figure}[t]
    \centering
    \includegraphics[width=0.48\textwidth]{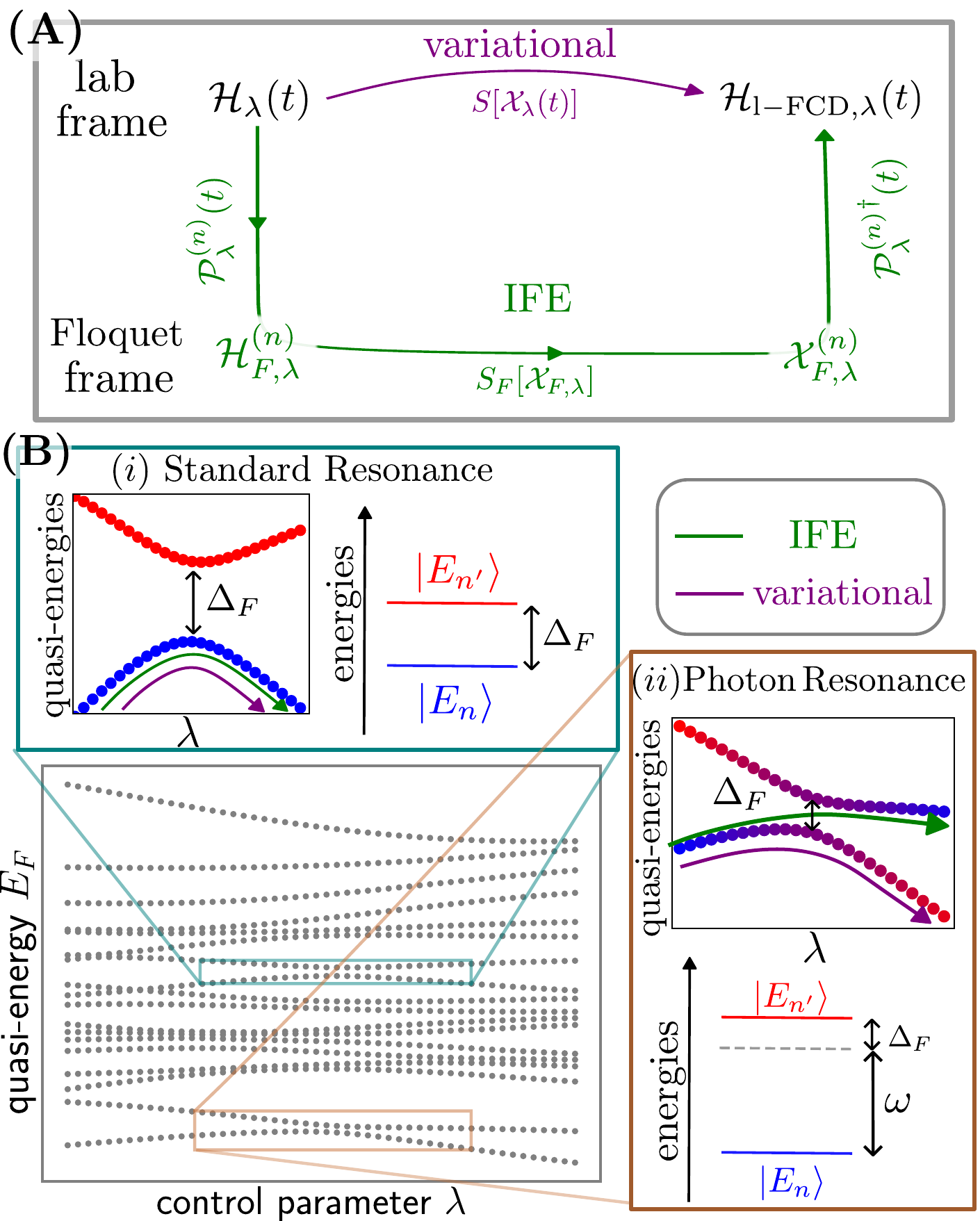}
    \caption{
        \textbf{Sketch: Comparison of local Floquet counterdiabatic driving methods.}
        \textbf{(A)} Comparison of IFE and variational procedure to obtain the (local)-FCD protocol. For IFE a transformation to the perturbative Floquet frame and back is needed, while the variational method operates on the level of the lab-frame.
        \textbf{(B)} Action of the two methods on avoided level crossings in the quasi-energy spectrum. 
        $(i)$ For a standard resonance---where both quasi-energies and non-driven or perturbative energies are close in energy---both l-FCD approaches follow the adiabatic path. 
        $(ii)$ For a photon resonance---where perturbative energy levels which are multiples of the driving frequency~$\omega$ apart hybridize---the variational method still follows the adiabatic path. However, the IFE method leads to a diabatic transition, violating adiabaticity but possibly suppressing heating.
    }
    \label{fig:varl_vs_ife}
\end{figure}

\paragraph*{Floquet Adiabaticity and Floquet Counterdiabatic Driving}

To suppress diabatic excitations, we first identify the cause for diabatic transitions. This requires generalizing the concept of the adiabatic gauge potential~(AGP) $\A_\lambda(t)$~\cite{Berry2009_CD,Demirplak2003_CD,Kolodrubetz2017_GeometryReview} to periodically driven systems and define the Floquet adiabatic gauge potential~(FAGP). To this end, we revise the transformation to the Floquet frame that removes the micromotion dynamics, to include the time-dependence of the protocol $\lambda(t)$~\cite{Weinberg_2017_adiabatic_Floquet} 
\begin{equation}
\label{eq:FloquetFrame}
    \tilde{\h}_\lambda = \hFlambda - \dot{\lambda }\tilde{\A}_{\p\!,\lambda}.
\end{equation}
where we take into account the dependence of $\micro(t)$ on the control parameter
$\lambda$, and $\tilde{\A}_{\p\!,\lambda}(t)=i \micro^\dagger \partial_\lambda \micro$. To make the cause for excitations explicit, we now perform a second transformation to a co-moving frame where the Floquet Hamiltonian $\hFlambda$ is diagonal. Noting the implicit dependence on time of the instantaneous change-of-basis operator $W_\lambda$ via the protocol $\lambda(t)$, we find
\begin{equation}
\label{eq:FloquetEigenframe}
    \dtilde{\h}_\lambda = 
        \tilde{\operator{H}}_{F\!,\lambda} - \dot{\lambda} \pqty{ \dtilde{\A}_{F\!,\lambda} + \dtilde{\A}_{\p\!,\lambda} } \, ,
\end{equation}
where $\tilde{\operator{H}}_{F\!,\lambda}=W_\lambda^\dagger \h_{F\!,\lambda} W_\lambda=\sum_n \ket{n} E_{{F},n}(\lambda) \bra{n}$ is a diagonal matrix containing the instantaneous Floquet (quasi-)energies $E_{{F},n}$, and $\dtilde{\A}_{F\!,\lambda}=i W_\lambda^\dagger\partial_\lambda W_\lambda$.

In this second frame, the Hamiltonian $\tilde{\operator{H}}_{F\!,\lambda}$ is diagonal at all times, and therefore all diabatic transitions are necessarily caused by the off-diagonal \textit{Floquet adiabatic gauge potential}~(FAGP) $\dtilde{\A}=\dtilde{\A}_{F\!,\lambda} + \dtilde{\A}_{\p\!,\lambda}$. 
The two contributions in the FAGP correspond to the change in the eigenbasis of the Floquet Hamiltonian~($\A_F$) and the change of Floquet frame~($\A_\p$).
Therefore, we can suppress all diabatic transitions in the presence of the periodic drive, by adding the gauge potential term. In the original lab-frame, this leads us to the counterdiabatic Hamiltonian
\begin{equation}
\label{eq:HCD}
    \begin{aligned}
        \h_{\mathrm{CD}\!,\lambda}(t)        &= \h_\lambda(t) +\dot{\lambda}\A_\lambda(t), \, \\
        \A_\lambda(t)  &= i \partial_\lambda(\micro(t) W_\lambda)W_\lambda^\dagger \micro^\dagger(t).
    \end{aligned}
\end{equation}
Hence, Eq.~\eqref{eq:HCD} provides the sought-after Floquet counterdiabatic protocol. 
We emphasize that the structure of the operator $\A_\lambda$ is independent of the specific form of the protocol $\lambda(t)$~[We note that the FAGP is independent of the form of the protocol only if the frequency is constant during the protocol, see Sec.~\ref{sec:chirp}]. Therefore, in the adiabatic limit $\dot{\lambda}\to 0$, all diabatic transitions vanish since $\dot{\lambda}\A_\lambda \to 0$.

Notice that the definition of the FCD protocol \eqref{eq:HCD} requires 
(i) solving the Floquet problem, and 
(ii) fully diagonalizing the Floquet Hamiltonian for all parameter values $\lambda(t)$. 
These are notoriously hard tasks, as obtaining the full spectrum in many-body systems is exponentially hard in the system size, while finding closed-form expressions for the Floquet Hamiltonian is challenging even in few-level systems due to time-ordering, let alone in generic many-body systems. 
Thus, computing an exact FAGP using Eq.~\eqref{eq:HCD} is out of reach for almost all models in practice.
Moreover, even if one could obtain the exact FAGP, it often contains long-range or multi-body interactions~\cite{Kolodrubetz2017_GeometryReview} which are challenging to implement in an actual experiment~\cite{Bloch2008_QuantumSimulationUltraColdAtoms}. 
Therefore, the relevant question for experimental applications concerns finding an optimal \textit{local} approximation to the exact CD Hamiltonian~\eqref{eq:HCD}. We refer to this procedure as \textit{local Floquet counterdiabatic driving}~(l-FCD).

\section{Details on Variational Procedure}
\label{sec:varl}

\update{
In this section, we present additional results for the variational principle. 
This section is structured as follows. In Subsec.~\ref{sec:agp_relation}, we give a detailed derivation of the defining relation for the FAGP, Eq.~\eqref{eq:GFloquet}. In Subsec.~\ref{sec:variational} we discuss the application of the variational principle for approximate FCD in more detail.
In Subsec.~\ref{sec:chirp} we discuss frequency chirps in detail.
Finally, in Subsec.~\ref{sec:rigorous_varl} we give an alternative derivation of the variational principles based on the Samb\'e space formulation that connects static CD with FCD.
}

\update{
\subsection{\label{sec:agp_relation} Defining Equation for Floquet Adiabatic Gauge Potential }
}

\update{
In this section, we derive in detail the defining equation of the Floquet Adiabatic Gauge Potential, Eq.~\eqref{eq:GFloquet}. For concreteness, we prove the validity of equation
\begin{equation}
\label{SM:eq:GFloquet}
    i\comm{\h_\lambda\pqty{t}}{\operator{G}\pqty{\A_\lambda\pqty{t}}} + \partial_t \operator{G}\pqty{\A_\lambda\pqty{t}} = 0\, , 
\end{equation}
where 
$$
\operator{G}\pqty{\A_\lambda(t)} = i \comm{\h_\lambda(t)}{\A_\lambda(t)} + \partial_t \A_\lambda(t) - \partial_\lambda \h_\lambda \, ,
$$
and $\A_\lambda(t) = i \partial_\lambda(\micro(t) W_\lambda)W_\lambda^\dagger \micro^\dagger(t)$.
To this end, let us consider matrix elements of $\A_\lambda(t)$ with respect to the Floquet functions $\ket{u_n(t)} = P^\dagger W^\dagger \ket{n}$:
\begin{equation}
\label{SM:eq:identity_states}
    \begin{aligned}
         \left( \A_\lambda(t) \right)_{nm} 
            &= \bra{u_n(t)} \A_\lambda(t) \ket{u_m(t)}
        \\
            &= \bra{u_n(t)} i \partial_\lambda \ket{u_m(t)}
        \\
            &\overset{n \neq m}{=} \frac{\bra{u_n(t)} i \partial_\lambda\pqty{\p_\lambda \h_{F, \lambda}\p^\dagger_\lambda } \ket{u_m(t)}}{E_n^F - E_m^F} \, ,
    \end{aligned}
\end{equation}
where in the last line we have used $\h_{F,\lambda}[t]\equiv \p_\lambda \h_{F, \lambda}\p^\dagger_\lambda$ which is the Floquet Hamiltonian starting the evolution from $t_0=t$ and by definition $\h_{F,\lambda}[t] \ket{u_n(t)} = E_n^F \ket{u_n(t)}$ and orthogonality $\braket{u_n(t)}{u_m(t)}=\delta_{nm}$.
}

\update{
In terms of operators, Eq.~\eqref{SM:eq:identity_states} can be recast as
\begin{equation}
    \label{SM:eq:identity_operators}
    i \comm{\h_{F,\lambda}[t]}{\A_\lambda(t)} = \partial_\lambda\pqty{\h_{F,\lambda}[t]} + \operator{F}_\lambda
\end{equation}
with generalized force operator $\operator{F}_\lambda=-\sum_{n} \pqty{\partial_\lambda E_n^F(\lambda)} \ketbra{u_n(t)}$ which commutes with $\h_{F,\lambda}[t]$ as they share a common eigenbasis. 
We can get rid of the generalized force operator in Eq.~\eqref{SM:eq:identity_operators} by using the commutator with $\h_{F,\lambda}[t]$:
\begin{equation}
\label{SM:eq:GFloquet_pre}
    \comm{\tilde{\operator{G}}\pqty{\mathcal{A}_\lambda} }{ \h_{F,\lambda}[t] } = 0 \, ,
\end{equation}
with $\tilde{\operator{G}}\pqty{\mathcal{A}_\lambda} = i \comm{\h_{F,\lambda}[t]}{\A_\lambda(t)} - \partial_\lambda\pqty{\h_{F,\lambda}[t]}$.
Note the similarity of Eqs.~\eqref{SM:eq:GFloquet} and \eqref{SM:eq:GFloquet_pre}.
}

\update{
In fact, it is straightforward to show that 
$$
-i\comm{\tilde{\operator{G}}\pqty{\mathcal{A}_\lambda} }{ \h_{F,\lambda}[t] }= -i\comm{\tilde{\operator{G}}\pqty{\mathcal{A}_\lambda} }{ \h_{\lambda}(t) }+\partial_t \tilde{\operator{G}}\pqty{\mathcal{A}_\lambda} 
$$
and that $\tilde{\operator{G}}\pqty{\mathcal{A}_\lambda} = \operator{G}\pqty{\mathcal{A}_\lambda}$
using the relations
\begin{equation}
\label{SM:eq:relations}
    \begin{aligned}
        \h_{F,\lambda}[t] &= \h_\lambda(t) - \A_t(t) \\
        i \partial_t \h_{F,\lambda}[t] &= \comm{\A_t(t)}{\h_{F,\lambda}[t]}\\
        i\comm{\A_t}{\A_\lambda} &= \partial_t \A_\lambda - \partial_\lambda \A_t
    \end{aligned}
\end{equation}
with the Floquet adiabatic gauge potential with respect to time $\A_t(t) = \pqty{i\partial_t \p}\p^\dagger$. The first two relations in Eq.~\eqref{SM:eq:relations} are a direct consequence of the definition of $\h_{F,\lambda}[t]$ and $\A_t(t)$, and for the last relation it is convenient to use $\A_\mu = \sum_n \ketbra{i \partial_\mu u_n(t)}{u_n(t)} = \sum_n \ketbra{u_n(t)}{- i \partial_\mu u_n(t)}$.
}

\update{
This proves the validity of Eq.~\eqref{SM:eq:GFloquet}.
}

\subsection{Discussion of Variational Principle}
\label{sec:variational}
Let us recal the variational principle
\begin{equation}
\label{eq:varl}
    \begin{aligned}
        S\bqty{\operator{X}_\lambda} &= \int_0^{T} \Tr(\operator{G}^2(\operator{X}_\lambda(t)))\, \mathrm{d}t,\\
        \operator{G}\pqty{\operator{X}_\lambda}
        &= i\comm{\h_\lambda(t)}{ \operator{X}_\lambda(t)}+ \partial_t \operator{X}_\lambda(t) - \partial_\lambda \h_\lambda(t) \, .
    \end{aligned}
\end{equation}

Note that, Equation~\eqref{eq:varl} allows for the determination of an approximate FAGP using only lab-frame quantities. 
Notably, the ansatz $\mathcal{X}_\lambda(t)$ for the variational FAGP carries an explicit periodic time dependence.
Therefore, a complete basis to expand $\operator{X}_\lambda(t)$ in must, in addition to all operators acting on the Hilbert space, also include a complete set of periodic functions; we will in the following focus on the natural choice of Fourier harmonics $\Bqty{e^{-i\omega \ell t}}_{\ell \in \mathbb{Z}}$.
Hence, in general, $\operator{X}_\lambda(t)$ contains infinitely many terms arising from the Fourier harmonics of the periodic time dependence.

However, in practical applications, it is often sufficient to truncate the infinite sum to a finite-dimensional subset including only a finite number $N_h$ of Fourier harmonics. 
Then, we can choose an ansatz with $N_h$ harmonics and $N_O$ local operators
\begin{equation}
\label{eq:varl_ansatz}
    \operator{X}_\lambda(t) =  \sum_{m=1}^{N_O}  \sum_{\ell=-N_h}^{N_h} \chi_{\ell m} e^{i \ell \omega t}  \operator{O}_m \, ,
\end{equation}
where $\chi_{\ell m}$ are the variational parameters. The operators $\operator{O}_m$ are chosen to reflect any external constraints, e.g., locality or accessibility in the lab.
In order to close the truncated algebra under multiplication we ignore any Fourier harmonics $e^{i\ell \omega t}$ with $\abs{\ell}>N_h$ resulting from a product of two time-dependent functions. 

For the variational principle, we have to truncate $\operator{G}$ to the given number of harmonics $N_h$ before computing the action $S$. This leads to the truncated variational principle
\begin{eqnarray}
    \label{eq:varl_projection}
    S^{(N_h)}\bqty{\operator{X}_\lambda} &=& \int_0^{T} \Tr( \operator{G}^{(N_h)}\pqty{\operator{X}_\lambda(t)}^2 ) \mathrm{d}t\\
    \operator{G}^{(N_h)}\pqty{\operator{X}_\lambda}
    &=&  i\comm{\h_\lambda(t)}{ \operator{X}(t)}+ \partial_t \operator{X}_\lambda(t) - \partial_\lambda \h_\lambda(t) \big|_{N_h} \nonumber \, ,
\end{eqnarray}
which is equivalent to Eq.~\eqref{eq:varl} if a complete ansatz~($N_h\to\infty$) is considered.
A detailed algorithm to compute an approximate variational FAGP numerically is found in Sec.~\ref{sec:algo}.

Note that the action~\eqref{eq:varl} is quadratic in the gauge potential such that the minimization is convex and guaranteed to converge to a global optimum.
Let us re-iterate that the variational principle, Eq.~\eqref{eq:varl}, is a fully non-perturbative method that allows for the direct determination of a local approximate FAGP in the lab-frame. 
As such, it overcomes all of the shortcomings of the IFE approach, at the expense of dealing with an infinite-dimensional variational space. 
However, as we will see below, in practical applications, a truncation to a finite number of harmonics proves sufficient, see Eqs.~\eqref{eq:varl_ansatz} and \eqref{eq:varl_projection}.
Recently, alternative truncation schemes have been introduced for many-body systems, e.g., in Krylov space~\cite{takahashi2023shortcuts}, which can be extended to Floquet systems.

\subsection{\label{sec:chirp}Counter-Diabatic Frequency Modulation}

The frequency modulation~($\lambda=\omega$), often referred to as \textit{chirping}, is commonly considered for state-manipulation in experimental setups~\cite{Viebahn_etal_FloquetPump_2022,Wintersperger2020_Anomalous,Song2022_FloquetPhaseTransition} as it can lead to drastic changes in the system's response. 
However, we stress that the Floquet adiabaticity of chirps requires extra caution as it leads to a change in the driving period: $T=T(t)$. 

Before going into a detailed rigorous derivation of the adjusted action, see Sec.~\ref{sec:chirp_rigorous}, we give some intuitive arguments for the adjustments needed when considering frequency modulations.

To this end, let us consider the micromotion contribution to the adiabatic gauge potential, $\A_\mathcal{P}=-i\p_{\!\omega} \partial_{\omega}\p_{\!\omega}^\dagger$. 
In general, the micromotion operator $\p_{\!\omega}(t)$ need not reduce to the identity $\p_{\!\omega}(t)\not\equiv\identity$ at any time $t\neq nT$($n\in\mathbb{Z}$), and we can write
\begin{equation*}
    \p_{\!\omega}(t) = \sum_\ell \p_{\ell,\omega} e^{i \ell \omega t} \, .
\end{equation*}
Hence, $\norm{\A_\p} \sim t$ as $t{\to}\infty$, since $\partial_\omega e^{i \ell \omega t} = t \times i\ell e^{i \ell \omega t}$.
This leads to a non-zero contribution $\dot{\lambda}\norm{\A_\p}\propto \dot{\lambda} t {\not\to} 0$ in the FAGP even in the adiabatic limit, where $\dot{\lambda}{\to} 0$ and $T_\mathrm{ramp}{\to} \infty$.
Therefore, even in the adiabatic limit, the evolution does not follow the instantaneous eigenstates of the Floquet Hamiltonian~\eqref{eq:FloquetFrame}. This suggests, that the infinite slow evolution follows the eigenstates of a different effective Hamiltonian. In fact, one can convince oneself that including the explicitly time-dependent term -- $\p^\dagger_{\!\omega}\pqty{\sum_\ell \p_{\ell,\omega}\partial_\omega e^{i \ell \omega t}} {\subset} \p^\dagger_{\!\omega}\partial_\omega \p_{\!\omega}$ -- in the Floquet Hamiltonian rather than the FAGP, adiabaticity is restored, i.e., $\dot{\lambda}\norm{\A_\p}{\to} 0$ as $\dot{\lambda}{\to}0$ and $T_\mathrm{ramp}{\to} \infty$.

In practice, this can be achieved by replacing the time-modulated frequency $\omega$ by the \textit{instantaneous} frequency $\nu$:
\begin{equation}
\label{eq:frequency_correction}
    \omega(t) \to \nu(t) =  \derivative{t} \pqty{\omega(t) t}=\omega(t) + t \dot{\omega}(t)\,
\end{equation}
in the Floquet frame Eq.~\eqref{eq:FloquetFrame}.

The instantaneous frequency $\nu$ being the relevant frequency for the Floquet problem instead of the time-modulated frequency $\omega$ can have unexpected and counterintuitive consequences. To illustrate this, consider a linear ramp $\omega(t) = \omega_0 + (\omega_1-\omega_0)t/T_\text{ramp}$: the adiabatic protocol follows the instantaneous frequency $\nu$, i.e., the system will evolve following the instantaneous Floquet eigenstate corresponding to $\nu(t) = \omega_0 +2\times(\omega_1-\omega_0)t/T_\text{ramp}$, cf.~Eq.~\eqref{eq:frequency_correction}. This comes in stark contrast to the na\"ive expectation that the system will trace the Floquet eigenstates corresponding to $\omega(t)=\omega_0+(\omega_1-\omega_0)t/T_\text{ramp}$ [note the missing factor of $2$ compared to $\nu(t)$].

\paragraph*{Variational Principle.---}
As a consequence, the variational principle~\eqref{eq:varl} also needs to be adjusted accordingly if the (instantaneous) frequency depends on the control parameter $\nu=\nu(\lambda)$.
\update{
Note that, the instantaneous frequency $\nu$ does not explicitly appear in the lab frame Hamiltonian $\h_\lambda(t)$ that only depends on the frequency $\omega$. 
}

\update{
We can make the instantaneous frequency explicit by using the dimensionless time $\phi=\omega t$. Then, in the time-dependent Schr\"odinger equation,
\begin{equation}
\label{eq:schroedinger_equation}
    i \frac{\dd }{\dd t} \ket{\psi(t)} = H_\lambda(t) \ket{\psi(t)}\, ,
\end{equation}
we have to replace the time-derivative $\frac{\dd }{\dd t}$ by $\frac{\dd }{\dd t}=\frac{\dd \phi}{\dd t} \frac{\dd }{\dd \phi} = \nu \frac{\dd }{\dd \phi}$.
Therefore, the dimensionless time-dependent Schr\"odinger equation reads
\begin{equation}
\label{eq:dimless_TDSE}
    i \frac{\dd }{\dd \phi} \ket{\psi(\phi)} =  \frac{H_\lambda(\phi)}{\nu} \ket{\psi(\phi)}\, .
\end{equation}
Note that, now the Floquet theorem reads 
\begin{equation}
\label{eq:dimless_HF}
    \operatorFF{H}{\lambda} = \micro^\dagger(\phi) \nu^{-1}\h_\lambda(\phi) \micro(\phi) - i \micro^\dagger(\phi) \derivative{\phi} \micro(\phi) \, ,
\end{equation}
where the instantaneous frequency $\nu$ now appears directly and the operator $\micro^\dagger(\phi) \derivative{\phi} \micro(\phi)$ naturally includes all contributions of the derivative of the harmonic function, i.e., $e^{i\ell \omega t}= e^{i \ell \phi}$.
Note that, we can identify $\nu^{-1}H_\lambda(\phi)$ as the relevant lab-frame Hamiltonian, where again $\nu$ appears directly.
}

\update{
Note that, if $\omega=\omega(\lambda)$ the instantaneous frequency $\nu(\lambda) = \frac{\dd}{\dd t}\pqty{\omega(\lambda) t} = \omega(\lambda) + \dot{\lambda} \partial_\lambda \omega t$ directly depends on the specific protocol $\lambda(t)$ and time $t$. 
Therefore, we can not treat the change in parameter $\lambda(t)$ and periodic time dependency $\phi(t)$ individually. Hence, we may consider time $t$ as the actual parameter changing throughout, i.e., the FCD equation reads
\begin{equation}
\label{eq:CD_chirp}
    \h_{\mathrm{CD}\!,\lambda(t)}(t) = \h_{\lambda(t)}(t) + \A_{\lambda(t)}(t) \, .
\end{equation}
This explicit protocol dependency in the case of a frequency chirp is the reason why we present two distinct variational principles. Remarkably, as presented in the main text, if the frequency is not varied the Floquet adiabatic gauge potential does not depend on the details of the protocol but only on the value $\lambda$.
}

\update{
Repeating a computation analogous to the case without a frequency chirp, Sec.~\ref{sec:agp_relation}, where we have to consider the Hamiltonian $\nu^{-1}H_\lambda(\phi)$ and $\phi$ as the time-dependency, we find that the FAGP for frequency chirps satisfies a similar relation
\begin{equation}
\label{eq:Gchirp}
        i\comm{\nu^{-1} \h_\lambda\pqty{t}}{\operator{G}_\nu\pqty{\A_\lambda\pqty{\phi}}} + \partial_\phi \operator{G}_\nu\pqty{\A_\lambda\pqty{\phi}} = 0\, , 
\end{equation}
where 
\begin{equation*}
    \operator{G}_\nu\pqty{\A_\lambda\pqty{\phi}} = i \comm{\nu^{-1} \h_\lambda(t)}{\A_\lambda(t)} + \partial_\phi \A_\lambda(\phi) - \frac{1}{\nu}\pqty{\dot{\lambda} \partial_\lambda -  \frac{\dot{\nu}}{\nu}} \h_\lambda \, .
\end{equation*}
Using $\partial_\phi = \nu^{-1} \partial_t$, we can remove an overall factor of $\nu^{-1}$ from the above equations, i.e., Eq.~\eqref{eq:Gchirp} may be equivalently written as
\begin{equation*}
        i\comm{\h_\lambda\pqty{t}}{\bar{\operator{G}}_\nu\pqty{\A_\lambda\pqty{t}}} + \partial_t \operator{G}_\nu\pqty{\A_\lambda\pqty{t}} = 0\, , 
\end{equation*}
with
\begin{equation*}
    \bar{\operator{G}}_\nu\pqty{\A_\lambda\pqty{t}} = i \comm{\h_\lambda(t)}{\A_\lambda(t)} + \partial_t \A_\lambda(t) - \pqty{\dot{\lambda} \partial_\lambda -  \frac{\dot{\nu}}{\nu}} \h_\lambda \, .
\end{equation*}
Therefore, the variational principle for the case of frequency chirps reads
\begin{eqnarray}
\label{eq:varl-chirp}
        S\bqty{\operator{X}_\lambda} &=& \int_0^{T} \Tr(\operator{G}^2(\operator{X}_\lambda(t)))\, \mathrm{d}t, \\
        \operator{G}\pqty{\operator{X}_\lambda}
        &=& i\comm{\h_\lambda(t)}{ \operator{X}_\lambda(t)}+ \partial_t \operator{X}_\lambda(t) - \pqty{ \dot{\lambda }\partial_\lambda - \frac{\dot{\nu}}{\nu}} \h_\lambda(t)\, \nonumber ,
\end{eqnarray}
}


\subsection{\label{sec:rigorous_varl}Derivation of Variational Principle in Samb\'e space}

In this section we provide a rigorous derivation of the variational method using the so-called Samb\'e space or extended Floquet Hilbert space approach also referred to as two-time treatment~\cite{Novicenko2017_FloquetModulatedDriving}. 
\update{
Using the Samb\'e space we can remove the periodic time-dependency entirely from the Floquet problem by promoting Fourier modes to states on a Hilbert space. While this renders the problem time-independent it comes at the expense of introducing an infinite dimensional Hilbert space.
However, as there is no periodic time-dependency we can apply the standard toolbox of static CD driving to the Floquet problem in Samb\'e space, allowing us to derive the FCD variational principle, Eq.~\eqref{eq:varl-main}.
In addition, the Samb\'e space description will allow us in Sec.~\eqref{sec:exactlysolvable} to identify an entire class of exactly solvable periodic drives and identify suitable ans\"atze for FAGPs.
}

\subsubsection{Samb\'e space}
In the following, we denote the periodic time-dependence by $\phi=\omega t$ in order to avoid confusion between periodic time-dependence and the time-dependence associated with the change of parameter $\lambda=\lambda(t)$. 
Using this convention Eq.~\eqref{eq:FloquetFrame} reads
\begin{equation}
\label{eq:HF}
    \operatorFF{H}{\lambda} = \micro^\dagger(\phi) \h_\lambda(\phi) \micro(\phi) - i \dot{\phi} \micro(\phi)^\dagger \derivative{\phi} \micro(\phi) \, .
\end{equation}

Using Eq.~\eqref{eq:HF} the general solution to the time-dependent Schr\"odinger equation~(TDSE)
\begin{equation}
\label{app:eq:TDSE}
    \pqty{i \derivative{t} - \h_{\lambda}(\phi)} \ket{\psi(t)} = 0 \, ,
\end{equation}
reads $\ket{\psi(t)} = \sum_n c_n e^{-i \epsilon_n^F\pqty{\lambda} t } \ket{u_{n,\lambda}(\phi)}$, with the so-called Floquet functions $\ket{u_{n,\lambda}(\phi)}=\p_\lambda(\phi) \ket{\epsilon_n^F\pqty{\lambda}}$ and eigenstates $\ket{\epsilon_n^F\pqty{\lambda}}$ and eigenenergies $\epsilon_n^F\pqty{\lambda}$ of the Floquet Hamiltonian $\operatorFF{H}{\lambda}$, i.e, $\operatorFF{H}{\lambda} \ket{\epsilon_n^F\pqty{\lambda}} = \epsilon_n^F\ket{\epsilon_n^F\pqty{\lambda}}$.

Note that there are two distinct time-dependences in $\ket{\psi(t)}$, a periodic time-dependence inherited from the micromotion and the phase accumulation $e^{-i\epsilon_n^F\pqty{\lambda} t}$ as known from usual static systems. 
Therefore, it is natural to split the time-derivative $\derivative{t}$ in the TDSE into two contributions $\derivative{t} = \partial_t + \derivative{\phi}{t}\partial_\phi$ such that
\begin{equation}
\label{eq:TDSE2}
    i \partial_t \ket{\psi(t)} = \pqty{i\nu\partial_\phi + \h_\lambda(\phi)} \ket{\psi(t)} \, ,
\end{equation}
where we used the \textit{instantaneous} frequency $\nu = \dot{\phi} = \omega + \dot{\omega}t$.
\update{
Note that, the operator on the right-hand-side of Eq.~\eqref{eq:TDSE2} only acts on the periodic time-dependence. In addition, notice that the solutions to Eq.~\eqref{eq:TDSE2} are of the form of a separation ansatz $e^{-i \epsilon_n^F\pqty{\lambda} t } \ket{u_{n,\lambda}(\phi)}$. Hence, we can transform the TDSE~\eqref{eq:TDSE2} into a time-''independent" Schr\"odinger equation
\begin{equation}
\label{eq:TISE}
    \epsilon_n^F\pqty{\lambda} \ket{u_{n,\lambda}\pqty{\phi}} = \pqty{i\nu\partial_\phi + \h_\lambda(\phi)} \ket{u_{n,\lambda}\pqty{\phi}}\, ,
\end{equation}
where we only the periodic time dependency enters.
Equation~\eqref{eq:TISE} can be made formally time-independent by promoting the time-dependency to its own degree of freedom. This can be done, for example, by expanding the periodic function $\ket{u_{n,\lambda}\pqty{\phi}} = \sum_{n,m} c_{nm}\pqty{\lambda} e^{-in\phi} \ket{m}$ in Fourier harmonics and some physical basis $\Bqty{\ket{m}}$ with time-independent coefficients $c_{nm}$, and promoting the Fourier harmonics $e^{in\phi}$ to states $\timeket{n}$.
}
Then, the periodic functions $\Bqty{\timeket{n} \hat{=} e^{in\phi}}_{n\in\mathbb{z}}$ form an orthogonal basis of the Hilbert space of periodic square-integrable functions $\circhilbert$ with inner product
\begin{equation}
\label{app:eq:circ_inner}
    \timebraket{n}{m} = \int_{0}^{2\pi} e^{-in\phi} e^{im\phi} \frac{\dd{\phi}}{2\pi}\, .
\end{equation}

Likewise, expanding the Hamiltonian in Fourier modes $\h_\lambda(\phi)=\sum_n \h_{n,\lambda} e^{-in\phi}$ we can promote the product $e^{-i\phi n}(\cdot)$ to an operator $\operator{E}_n\hat{=}e^{in\phi}(\cdot)$ with $\operator{E}_n\timeket{m}=\timeket{n+m}$ acting as a raising operator. In addition, one can verify that the derivative acts as a number operator on the Fourier harmonics $\operator{N}\timeket{n}\hat{=} -i \partial_\phi e^{in\phi} = n e^{in\phi}\hat{=}n\timeket{n}$.

The combination of physical Hilbert space $\physhilbert$ and Hilbert space of periodic functions $\circhilbert$ is called Samb\'e space or extended Floquet Hilbert space $\floquethilbert$, with states denoted by $\floquetket{\psi}\in\floquethilbert$.

Using the Samb\'e space, we can rewrite Eqs.~\eqref{eq:TDSE2} and \eqref{eq:TISE} as 
\begin{equation}
\label{eq:TDSE_floquet}
    i \partial_t \floquetket{\psi} = \operator{Q}_\lambda \floquetket{\psi} \, ,
\end{equation}
and 
\begin{equation}
\label{eq:TISE_floquet}
    \operator{Q}_\lambda \floquetket{\psi} = \epsilon\pqty{\lambda} \floquetket{\psi}
\end{equation}
with the time-independent quasi-energy operator 
\begin{equation}
\label{eq:Q}
    \operator{Q}_\lambda 
    = \nu(\lambda) \operator{N} + \sum_{n} \operator{H}_n(\lambda) \operator{E}_n \, .
\end{equation}

Hence, diagonalizing the time-independent quasi-energy operator $\operator{Q}_\lambda$ gives the Floquet functions $\ket{u_{n,\lambda}}$ and quasi-energies $\epsilon_n^F\pqty{\lambda}$. 
However, note that, if $\floquetket{q_n}$ is a solution to Eq.~\ref{eq:TDSE_floquet} with quasi-energy $q_n$ so is $\operator{E}_n\floquetket{q_n}$ with quasi-energy $(q_n-\omega)$; however, projecting $\floquetket{\psi(t)}=e^{-i q_n t} \floquetket{q_n}$ and $\floquetket{\psi^\prime(t)}=e^{-i (q-n\omega) t} \operator{E}_n \floquetket{q_n}$ back to the physical Hilbert space the two states are in fact the same $\ket{\psi(t)}{=}\ket{\psi^\prime(t)}$.
Hence, $\floquethilbert$ contains infinitely many copies of the physical problem each shifted by $n\omega$.

\subsubsection{Floquet Counterdiabatic Driving in Samb\'e Space}

Let us come back to the Floquet control of transitionless driving along the protocol $\lambda=\lambda(t)$.
The advantage of the Samb\'e space is that the quasi-energy operator carries no periodic time-dependence. Therefore, we can directly apply static counterdiabatic driving methods using the quasi-energy operator $\operator{Q}_\lambda$ as the Hamiltonian.
Specifically, we can again identify the diabatic transition by transforming to a co-moving frame with respect to the periodic diagonalizing unitary transformation $\operator{W}_{\lambda}$, where $\operator{W}_{\lambda}^\dagger \operator{Q}_\lambda\operator{W}_{\lambda}=\nu\operator{N}+\tilde{\operator{H}}$ is diagonal: 
\begin{equation}
\label{eq:FloquetEigenframe_sambe}
    \Tilde{\operator{Q}}_{\lambda} = \nu\operator{N}+\tilde{\operator{H}} + \dot{\lambda} \tilde{\operator{A}}_\lambda 
\end{equation}
with the \textit{Floquet adiabatic gauge potential} (FAGP) $\tilde{\operator{A}}_\lambda=i\operator{W}_{\lambda}^\dagger \partial_{\lambda} \operator{W}_{\lambda}$.
As before all transitions are driven by the FAGP $\tilde{\operator{A}}_\lambda$, such that the Floquet counterdiabatic quasi-energy operator reads
\begin{equation}
\label{eq:FCD}
    \operator{Q}_\mathrm{CD} = \operator{Q}_\lambda+ \dot{\lambda} \operator{A}_\lambda \, ,
\end{equation}
with $\operator{A}_\lambda= -i \operator{W} \partial_\lambda \operator{W}^\dagger$.
The Floquet theorem in the extended Hilbert space guarantees that, for fixed $\lambda$, $\operator{Q}_\lambda$ can be diagonalized by a periodic unitary $\operator{W}_\lambda$~\cite{Eckardt2015_VanVleck}.
Therefore, also $\operator{A}_\lambda(\phi)$ is a time-periodic operator, for fixed $\lambda$.

\paragraph*{Variational Principle.---}

The variational principle for the FAGP $\operator{A}_\lambda$ can now be derived absolutely analogous to the static case, see for example Ref.~\cite{Sels2017_LCD}, i.e., $\operator{A}_\lambda$ exremizes the action
\begin{equation}
\label{eq:action}
    S[\operator{X}] 
            = \Tr_{\floquethilbert}(\operator{G}^2_\lambda(\operator{X})) \, ,
\end{equation}
with
\begin{equation}
\label{eq:G}
    \operator{G}_\lambda(\operator{X}) = i \comm{\operator{Q}_\lambda}{\operator{X}_\lambda} - \partial_\lambda \operator{Q}_\lambda
\end{equation}
where $\operator{X}_\lambda$ is a periodic and hermitian test operator. Note that for periodic operators the trace over the extended Hilbert space reduces to $\Tr_{\floquethilbert}\pqty{\cdot}=\Tr_{\physhilbert}\timebra{0} \cdot \timeket{0}$. Therefore, we can project Eq.~\eqref{eq:action} back to the physical Hilbert space by using $\Tr_{\floquethilbert}(\cdot) \to \int_0^{2\pi} \Tr_{\physhilbert}(\cdot) \dd{\phi}/(2\pi)$, $\operator{E}_n \to e^{-in\phi}$,and from the product rule $i\comm{\operator{N}}{\operator{X}_\lambda} \to \partial_\phi \operator{X}_\lambda$ to yield
\begin{equation}
\label{eq:action_projected}
    \begin{aligned}
         S[\operator{X}_\lambda] 
            &= \int_0^{2\pi} \Tr_{\physhilbert}(\operator{G}^2_\lambda(\operator{X}_\lambda)) \frac{\dd{\phi}}{2\pi} \, , \\
        \operator{G}_\lambda(\operator{X}_\lambda) &= i \comm{\operator{H}}{\operator{X}_\lambda} + \nu \partial_\phi \operator{X}_\lambda - \partial_\lambda \operator{H} \, ,
    \end{aligned}
\end{equation}
where $\operator{X}_\lambda=\operator{X}_\lambda(\phi)$ is a periodic function. Note that Eq.~\eqref{eq:action_projected} is the same as Eq.~\eqref{eq:varl}, up to the transformation of variable $\phi=2\pi t/T$.

\subsubsection{\label{sec:chirp_rigorous}Frequency Modulations}

Let us now come to the case of a frequency modulation. To this end, let us recall the quasi-energy operator~\eqref{eq:Q},
\begin{equation*}
    \operator{Q}_\lambda 
    = \nu_\lambda\operator{N} + \sum_{n} \operator{H}_{n, \lambda} \operator{E}_n \, .
\end{equation*}
Note that, using the quasi-energy operator $\operator{Q}_\lambda$~\eqref{eq:Q} we would find for a chirp, $\lambda=\nu$, that $\partial_\nu\operator{Q}_\lambda= \operator{N}$ appears in the variational principle~\eqref{eq:action}. The appearance of the number operator in the variational principle is not a fundamental problem. However, the practical computation of the variational FAGP becomes more difficult due to the non-periodicity and unboundedness of $\operator{N}$.

An important observation is that changing the quasi-energy operator by a multiplicative (possibly time-dependent) factor $\operator{Q}\to \operator{Q}/f(\lambda)$ leaves the eigenstates unchanged, hence, does not change the FAGP. Therefore, we can instead consider the re-scaled (dimensionless) quasi-energy operator 
\begin{equation*}
    q_\lambda 
        = \frac{\operator{Q}_\lambda }{\nu_\lambda} 
            = \operator{N} + \nu_\lambda^{-1} \sum_{n} \operator{H}_{n, \lambda} \operator{E}_n \,.
\end{equation*}
Then, the number operator no longer appears in the derivative $\partial_\nu q_\lambda = -\nu_\lambda^{-2} \sum_{n} \operator{H}_{n, \lambda} \operator{E}_n$, making the variational principle easier to solve.

However, note that since the instantaneous frequency $\nu$ explicitly depends on $\dot{\omega}$ and time $t$, the FAGP attains an explicit time dependence. This is in contrast to the mere implicit time dependency it carries via $\lambda=\lambda(t)$ if the frequency is fixed.
Combining everything we arrive at the action of form Eq.~\eqref{eq:action} with
\begin{equation}
\label{eq:varl_chirp}
    \begin{aligned}
        S[\operator{X}_\lambda] 
                &= \int_0^{2\pi} \Tr_{\physhilbert}(
                    \operator{G}^2_\lambda(\operator{X})
                ) \dd{\phi}/(2\pi) \, ,\\
        \operator{G}_\omega(\operator{X}_\lambda) 
            &= i \comm{\operator{H}_\lambda}{\operator{X}_\lambda} 
                + \nu^{-1} \partial_\phi \operator{X}_\lambda 
                + \frac{\dot{\nu}}{\nu} \operator{H}_\lambda
                - \dot{\lambda} \partial_\lambda \operator{H}_\lambda\,,
    \end{aligned}
\end{equation}
where $\partial_\lambda$ does not act on the harmonics. Note that, the last term in Eq.~\ref{eq:varl_chirp} vanishes if $\lambda=\omega$.

\paragraph*{Adjusted Action for Frequency Modulation}

The action~\eqref{eq:varl_chirp} describes the FAGP which leads to transitionless driving along the adiabatic path. Let us recall that for a chirp this adiabatic path is described by the instantaneous frequency $\nu(t)$ instead of the time-dependent frequency $\omega(t)$, see Sec.~\ref{sec:chirp}.

Here, we demonstrate that using an adjusted action we can obtain transitionless driving along the instantaneous Floquet states described by the change in the time-dependent frequency $\omega(t)$; not the instantaneous frequency $\nu(t)$. 

To this end, let us recall that in Sec.~\ref{sec:chirp}, we concluded that parts of the `naive' FAGPT contribution $\mathcal{A}_\p=-i\p_{\omega} \partial_\omega \p_\omega^\dagger$ must be included into the Floquet Hamiltonian. This procedure restores adiabaticity, i.e., $\dot{\lambda}\norm{\mathcal{A}_\p} \to 0$ in the adiabatic limit, but requires the change $\nu(t) \to \omega(t)$.
However, if we insist on using the entire expression $\mathcal{A}_\p=-i\p_\omega \partial_\omega \p_\omega^\dagger$ as part of the FAGP, the state protocol will follow the time-dependent frequency $\omega(t)$. This comes at the expense that the counter term will not vanish in the adiabatic limit, see also example in Sec~\ref{sec:chirp_example}.

In order to derive the variational principle for this case one simply retraces the steps from Eq.~\eqref{app:eq:TDSE} up to Eq.~\eqref{eq:action} setting explicitly $\nu=\omega$, i.e., assuming $\derivative{t}=\omega \derivative{\phi}$.
However, in order to correct for the fact that the phase $\phi=\omega t$ depends on $\omega$, one eventually has to use $\derivative{\omega}=\partial_\omega + t \partial_\phi$ such that the variational principle reads
\begin{equation}
\label{app:eq:varl_chirp_variant}
    \begin{aligned}
        S^\prime[\operator{X}_\omega] 
                &= \int_0^{2\pi} \Tr_{\physhilbert}(
                    \operator{G}^2_\omega(\operator{X}_\omega)
                ) \dd{\phi} \, ,\\
        \operator{G}^\prime_\omega(\operator{X}_\omega) 
            &= i \comm{\operator{H}}{\operator{X}_\omega} 
                + \omega \partial_\phi \operator{X}_\omega
                - \partial_{\omega} \operator{H} 
                - t \partial_\phi \operator{H} \, .
    \end{aligned}
\end{equation}

\section{\label{sec:IFE} Perturbative Floquet Counterdiabatic Driving }

In the main text we presented a non-perturbative variational principle, Eq.~\eqref{eq:varl-main}, to compute a local FCD protocol. Here, we present another possibility to compute a local FCD protocol that relies on a perturbative inverse frequency expansion method~(IFE) to compute the Floquet frame; computing $\A_\p$ and $\A_F$ separately. 

A key observation for deriving the IFE approach is the absence of the periodic time-dependence in the Floquet frame, cf.~Eq.~\eqref{eq:FloquetFrame}.
Therefore, provided we have access to (some approximation to) the Floquet Hamiltonian $\hFlambda$ and its micromotion operator $\micro(t)$, one can apply techniques from static CD driving to find a local approximation $\operator{X}_{F\!,\lambda}$ to the exact gauge potential $\A_{F, \lambda}$.

In particular, we can use local counterdiabatic driving~\cite{Sels2017_LCD}. That is, using a parametrized ansatz, $\operator{X}_{F\!,\lambda}$, for the FAGP in the Floquet frame, the optimal parameters are obtained from the CD variational principle $\delta_{\operator{X}_{F\!,\lambda}}S_F=0$ with action
\begin{equation}
\label{eq:varl_IFE}
    \begin{aligned}
        S_F[\operator{X}_{F\!,\lambda}]
        &= \Tr\left(\operator{G}_F^2\pqty{\operator{X}_{F\!,\lambda}}\right), \\
        \operator{G}_F\pqty{\operator{X}_{F\!,\lambda}}
        &= i\comm{\hFlambda}{ \operator{X}_{F\!,\lambda}} - \partial_\lambda \hFlambda \, .
    \end{aligned}
\end{equation}

In general, we do not have access to the exact Floquet Hamiltonian $\hFlambda$ and micromotion operator $\micro$. However, in Floquet engineering one is mostly interested in large driving frequencies compared to local energy scales. In this regime, a perturbative solution in terms of an inverse-frequency expansion can be obtained to a desired order $O\pqty{\omega^{-n}}$, where $\omega=2\pi/T$: 
\begin{equation}
\label{eq:IFseries}
    \begin{aligned}
        \hFlambda 
        &= \hFlambda^{(n)} + O(\omega^{-n-1})
        = \sum_{j=0}^{n} \omega^{-j} \hFlambda^{[j]} + O(\omega^{-n-1}) \, , \\
        \operator{K}_{F\!,\lambda}
        &= \operator{K}_{F\!,\lambda}^{(n)} + O(\omega^{-n-1})
        = \sum_{j=0}^{n} \omega^{-j} \operator{K}_{F\!,\lambda}^{[j]} + O(\omega^{-n-1}) \,,
    \end{aligned}
\end{equation}
e.g., using a van Vleck~\cite{Eckardt2015_VanVleck,Bukov_2015_general_HFE,Novicenko2017_FloquetModulatedDriving}, Floquet-Magnus~\cite{Feldman1984_FloquetMagnusConvergence,Bandyopadhyay2008_FloquetMagnus,Blanes2009_FloquetMagnus,Goldman2014_FloquetGaugeFields} or Brillouin-Wigner~\cite{Takahiro2016_BrillouinWigner,Hubac2010_BrillouinWigner} expansion.
The final approximate FAGP $\operator{X}$ that should be implemented in the lab-frame, is then obtained by transforming $\operator{X}_F$ back to the lab-frame and adding the additional Floquet-frame contribution $\A_{\p\!,\lambda}$:
\begin{equation}
\label{eq:IFE}
    \operator{X}^{(n)}_\lambda(t) = \p^{(n)}_{\!\lambda} \mathcal{X}^{(n)}_{F\!,\lambda} \p^{(n)\dagger}_{\!\lambda} + \A^{(n)}_{\p\!,\lambda} \, .
\end{equation}
Here, $\operator{X}^{(n)}_{F,\lambda}$ denotes the solution to Eq.~\eqref{eq:varl_IFE} computed with respect to the $n$'th order Floquet Hamiltonian $\h^{(n)}_F$.
Using an inverse frequency expansion for Eq.~\eqref{eq:IFE} and recalling that $\operator{K}_{F\!,\lambda}^{[0]}=0$~[for strongly coupled drives, this can always be achieved by an additional transformation to a rotating frame], we find for the lowest few orders in the inverse frequency:
\begin{subequations}
\label{eq:pert_expansion}
\begin{equation}
    \operator{X}_\lambda^{(0)}(t) = \operator{X}_{\lambda}^{(0)}\, ,
\end{equation}
\begin{equation}
    \operator{X}_\lambda^{(1)}(t) = \operator{X}_{\lambda}^{(1)} 
        + \pqty{
            i\comm{\operator{K}_{\lambda}^{(1)}(t)}{\operator{X}_{\lambda}^{(1)}} 
            + \partial_\lambda \operator{K}_{\lambda}^{(1)}(t)
        }\, ,
\end{equation}
\begin{equation}
    \begin{aligned}
        \operator{X}_\lambda^{(2)}(t) 
        &= \operator{X}_{\lambda}^{(2)} + \partial_\lambda \operator{K}_{\lambda}^{(2)}(t) 
            + \ad_{\operator{K}_{\lambda}^{(2)}(t)}\pqty{\operatorFF{X}{\lambda}^{(2)} - \frac{1}{2} \partial_\lambda \operator{K}_{\lambda}^{(2)}(t) }\\
             &\hspace{0.2cm} + \ad^2_{\operator{K}_{\lambda}^{(2)}(t)}\pqty{\operatorFF{X}{\lambda}^{(2)}} \, ,
    \end{aligned}
\end{equation}
with commutator and nested commutators written as $\ad_X\pqty{Y}{=}i\comm{X}{Y}$ and $\ad^{(n+1)}_X\pqty{Y}{=}i\comm{X}{\ad^n_X(Y)}$, respectively.
The general $n$'th order expression is given by
\begin{equation}
    \begin{aligned}
        \operator{X}_\lambda^{(n)}\pqty{t} 
        &= \sum_{k=0}^n \frac{1}{k!} \ad^k_{\operator{K}_{\lambda}^{(n)}(t)}\pqty{\operatorFF{X}{\lambda}^{(n)}} \\
        &\hspace{0.2cm} + \sum_{k=0}^{n-1} \frac{(-1)^k}{(k+1)!} \ad^k_{\operator{K}_{\lambda}^{(n)}(t)}\pqty{\partial_\lambda \operator{K}_{\lambda}^{(n)}}
    \end{aligned}
\end{equation}
\label{eq:IFE_loworder}
\end{subequations}
where $\operator{X}^{(n)}_{F\!,\lambda}$are obtained from minimizing the action~\eqref{eq:varl_IFE} with respect to $\h_F^{(n)}$.
Note that, to simplify the resulting equations we have chosen the perturbative expansion Eqs.~\eqref{eq:pert_expansion} such that the $n$'th order includes also some higher order terms. However, this is not a problem since the perturbative expansion is only valid up to $O(\omega^{n+1})$ corrections.

Let us emphasize that a similar inverse frequency expansion $\operatorFF{X}{\lambda}^{(n)}$ does not exist in general, i.e., $\operatorFF{X}{\lambda}^{(n)}$ and $\operatorFF{X}{\lambda}^{(n-1)}$ might differ by a contribution that is not $O(\omega^{-n})$. Such a situation occurs if the $n$'th order correction to effective Hamiltonian opens a gap: then the FAGP may scale as $\omega^n$ close to the avoided gap closing. A common scenario are quasi-conservation laws, i.e, where the low order Floquet Hamiltonians preserves a symmetry that is explicitly broken by higher order corrections.

Note that, the IFE approach to the Floquet adiabatic gauge potential, is bound to the validity of the inverse frequency expansion which breaks down whenever photon resonances occur~\cite{Eckardt2015_VanVleck,goldman2015_ResonantFloquet,Bukov2016_resonances}. This is particularly problematic for interacting systems.

Moreover, the IFE may also be problematic from a practical perspective. To see why, recall that the main purpose for introducing the variational ansatz Eq.~\eqref{eq:varl_IFE} is the ability to select the operator structure of $\mathcal{X}_{F,\lambda}$, and hence to incorporate experimental constraints such as locality.
However, within the IFE approach, the variational principle is applied in the Floquet frame. Thus, the extra additive contribution from the transformation $\mathcal{P}_{\!\lambda}^{(n)}$ back to the lab-frame -- $\mathcal{A}_{\mathcal{P},\lambda}^{(n)}$ in Eq.~\eqref{eq:IFE} -- is fixed and cannot be shaped.
Therefore, there exists no obvious way to obtain experimentally implementable local approximations to this additional contribution within the IFE approach.

Nonetheless, we emphasize that the IFE method can still be useful for some Floquet engineering studies that are designed to work in the high-frequency regime, where the $\hFlambda^{(n)}$ is the object of interest. Here, both $\hFlambda^{(n)}$ and $\operator{K}^{(n)}_{F\!,\lambda}$ are already known, making the application of IFE straightforward and hence allowing for an improved state manipulation at a low computational cost.

Let us close by noting that the incapability of the IFE to capture photon resonances can be used as an advantage. 
Specifically, in Floquet engineering applications governed by a perturbative Floquet Hamiltonian $\h_F^{(n)}$, state preparation protocols should ideally follow the eigenstates of the approximate Floquet Hamiltonian $\h_F^{(n)}$ rather than the exact $\h_F$. 
Since photon resonances are not captured by the perturbative Floquet Hamiltonian, the resulting IFE CD protocol will ignore any photon-induced hybridization gaps and traverse them diabatically, as if they are not present. This can help avoid undesired hybridization of quasi-energy eigenstates and some forms of heating in few-particle and weakly-interacting or integrable systems.
Indeed, as shown in Sec.~\ref{sec:linear2LS}, using the IFE protocol and a suitably short ramp duration allows us to pass through the spectral gaps of the approximate $\h_F^{(n)}$ adiabatically, while traversing photon resonances diabatically.

\section{\label{sec:examples} Two Level System Examples}

\subsection{\label{sec:exactlysolvable}Analytically solvable example}

In this section, we report on the analytically solvable circularly driven two-level system
\begin{equation}
\label{eq:circ2LS}
    \h_\circ(t) = \Delta S^z + A \bqty{\cos(\omega t)S^x + \sin(\omega t) S^y } \, ,
\end{equation}
with level splitting $\Delta$, driving amplitude $A$ and driving frequency $\omega$. 
This model is the non-interacting limiting case of the circularly driven Ising model $\operator{H}_\mathrm{TFI}(t;A,\omega)$ considered in the main text. Likewise, it can be mapped to a static problem using the frame transformation $W(t)=\exp(-i\phi(t) S^z)$ with $\phi(t)=\omega t$. This leads to
\begin{equation}
\label{eq:circ2LS_rot}
    \widetilde{\h} = (\Delta - \omega) S^z + A S^x \, ,
\end{equation}
where we used that $\dot{\phi}=\omega$ for fixed frequency. Note that Eq.~\eqref{eq:circ2LS_rot} is not yet the Floquet Hamiltonian, as its spectrum is not folded into a window $\omega$ of energies~\cite{Bukov_2015_general_HFE}. However, since this folding is trivial, i.e., no photon resonances occur, we can skip this step for further consideration.
So far, the analysis is general and applies to any parameter in Eq.~\eqref{eq:circ2LS}. 

\subsubsection{Frequency independent parameters}
In the following, we focus on a constant in time frequency $\derivative{t}\omega=0$. The case of a time-dependent frequency, so-called chirp, is studied in detail in Sec.~\ref{sec:chirp_example}.

In the rotating frame, the Hamiltonian~\eqref{eq:circ2LS_rot} can be diagonalized via $U(\alpha)=\exp(-i\alpha S^y)$, where $\tan(\alpha)=A/(\Delta - \omega)$. Therefore, the FAGPs $\A_\Delta$ and $\A_{A}$ for $\tilde{\h}_\circ$ upon changing $\Delta$ and $A$, respectively, read
\begin{equation}
\label{eq:circ_AGPs}
    \begin{aligned}
        \A_{\Delta}(t)   &= \frac{-A}{(\Delta - \omega)^2 + A^2} \pqty{ \cos(\omega t) S^y - \sin(\omega t) S^x } \, , \\
        \A_{A}(t)        &= \frac{\Delta - \omega}{(\Delta - \omega)^2 + A^2} \pqty{ \cos(\omega t) S^y - \sin(\omega t) S^x } \, .
    \end{aligned}
\end{equation}
Notice that there is no additional contribution due to $W^\dagger \partial_\lambda W$ as $W$ is independent of both $A$ and $\Delta$. 

We can also derive the FAGP in the Samb\'e space. This allows us to gain more insights into the system and the mechanism allowing for an exact solution.
To this end, let us consider the quasi-energy operator
\begin{equation}
\label{eq:circ_Q}
    \operator{Q}_\circ = \omega \operator{N} + \Delta S^z + \frac{A}{2} S^+ \operator{E}^+ + \frac{A}{2} S^- \operator{E}^- \, ,
\end{equation}
where $\operator{E}^\pm\hat{=}e^{\pm i\omega t}$. Additionally, we introduced the spin-$\frac{1}{2}$ raising and lowering operators $S^\pm=(S^x \mp i S^y)$ which fulfill $\comm{S^z}{S^\pm}=\pm i S^\pm$.

Note that, we can define generalized raising and lowering operators $\Sigma^\pm=S^\pm \operator{E}^\pm$ with associated pseudo-spin operators $\Sigma^x=(\Sigma^++\Sigma^-)/2$, $\Sigma^y=(\Sigma^+ - \Sigma^-)/2i$ and $\Sigma^z=S^z$. These operators also fulfill the usual spin-$\frac{1}{2}$ algebra
\begin{equation}
\label{eq:pauli_algebra}
    \begin{aligned}
        \comm{\Sigma^z}{\Sigma^\pm}         &= \pm i\Sigma^\pm \, ,\\
        \comm{\Sigma^+}{\Sigma^-}           &= 2 \Sigma^z \, , \\
        \comm{\Sigma^\alpha}{\Sigma^\beta}  &= i \epsilon_{\alpha\beta\gamma} \Sigma^\gamma \, ,
    \end{aligned}
\end{equation}
with $\alpha,\beta,\gamma \in \Bqty{x,y,z}$ and the fully anti-symmetric Levi-Civita symbol $\epsilon$. Additionally, we find
\begin{equation}
\label{eq:circ_derivative}
    \begin{aligned}
        i\comm{\operator{N}}{\Sigma^\pm}    &= \pm \Sigma^\pm \, ,\\
        i\comm{\operator{N}}{\Sigma^\alpha} &= -i \epsilon_{z\alpha\gamma} \Sigma^\gamma \, .
    \end{aligned}
\end{equation}
Therefore, the set of operators $\Bqty{\operator{N},\, \Sigma^{x},\, \Sigma^{y},\, \Sigma^{z}}$ form a closed algebra akin to the standard Pauli-algebra, with the number operator $\operator{N}$ taking a similar role as $\Sigma^{z}$.

Eventually, let us consider the action of the adjoint representation of $\exp(-i \alpha \Sigma^y)$
\begin{equation}
\label{eq:adjoint_pauli_y}
    \begin{aligned}
        \exp(i \alpha \Sigma^y) \Sigma^z \exp(-i \alpha \Sigma^y) 
            &= \cos(\alpha) \Sigma^z - \sin(\alpha) \Sigma^x \, ,\\
        \exp(i \alpha \Sigma^y) \Sigma^x \exp(-i \alpha \Sigma^y) 
            &= \cos(\alpha) \Sigma^x + \sin(\alpha) \Sigma^z \, ,\\
        \exp(i \alpha \Sigma^y) \operator{N} \exp(-i \alpha \Sigma^y) 
            &= \operator{N} + \left(1 - \cos(\alpha)\right) \Sigma^z + \sin(\alpha) \Sigma^x \, ,\\
    \end{aligned}
\end{equation}
which follows immediately from Eqs.~\eqref{eq:pauli_algebra} and \eqref{eq:circ_derivative}.

Using the generalized raising and lowering operators, we can write the quasi-energy operator~\eqref{eq:circ_Q} as
\begin{equation}
\label{eq:circ_Q2}
    \operator{Q}_\circ = \omega \operator{N} + \Delta \Sigma^z + A \Sigma^x \, .
\end{equation}
Then, Eq.~\eqref{eq:adjoint_pauli_y} implies that $\operator{W}=\exp(-i\alpha \Sigma^y)$ diagonalizes the quasi-energy operator for $\tan \alpha=A/(\Delta-\omega)$.
Hence, the adiabatic gauge potentials in the Samb\'e space read
\begin{equation}
\label{eq:circ_AGPs_ext}
    \begin{aligned}
        \A_{\Delta}   &= \frac{-A}{(\Delta - \omega)^2 + A^2} \Sigma^y \, , \\
        \A_{A}        &= \frac{\Delta - \omega}{(\Delta - \omega)^2 + A^2} \Sigma^y \, ,
    \end{aligned}
\end{equation}
which is equivalent to Eqs.~\eqref{eq:circ_AGPs} upon projecting to the physical Hilbert space.

In summary, the circularly driven two-level system is analytically solvable due to the closed algebra generated by the quasi-energy operator in the Samb\'e space. This insight might be generalized to obtain more complex, analytically tractable, Floquet systems, i.e., by ensuring a closure relation among the operators appearing in the quasi-energy operator.

\subsubsection{\label{sec:chirp_example}Frequency modulation}

\begin{figure}[t]
    \centering
    \includegraphics[width=0.5\textwidth]{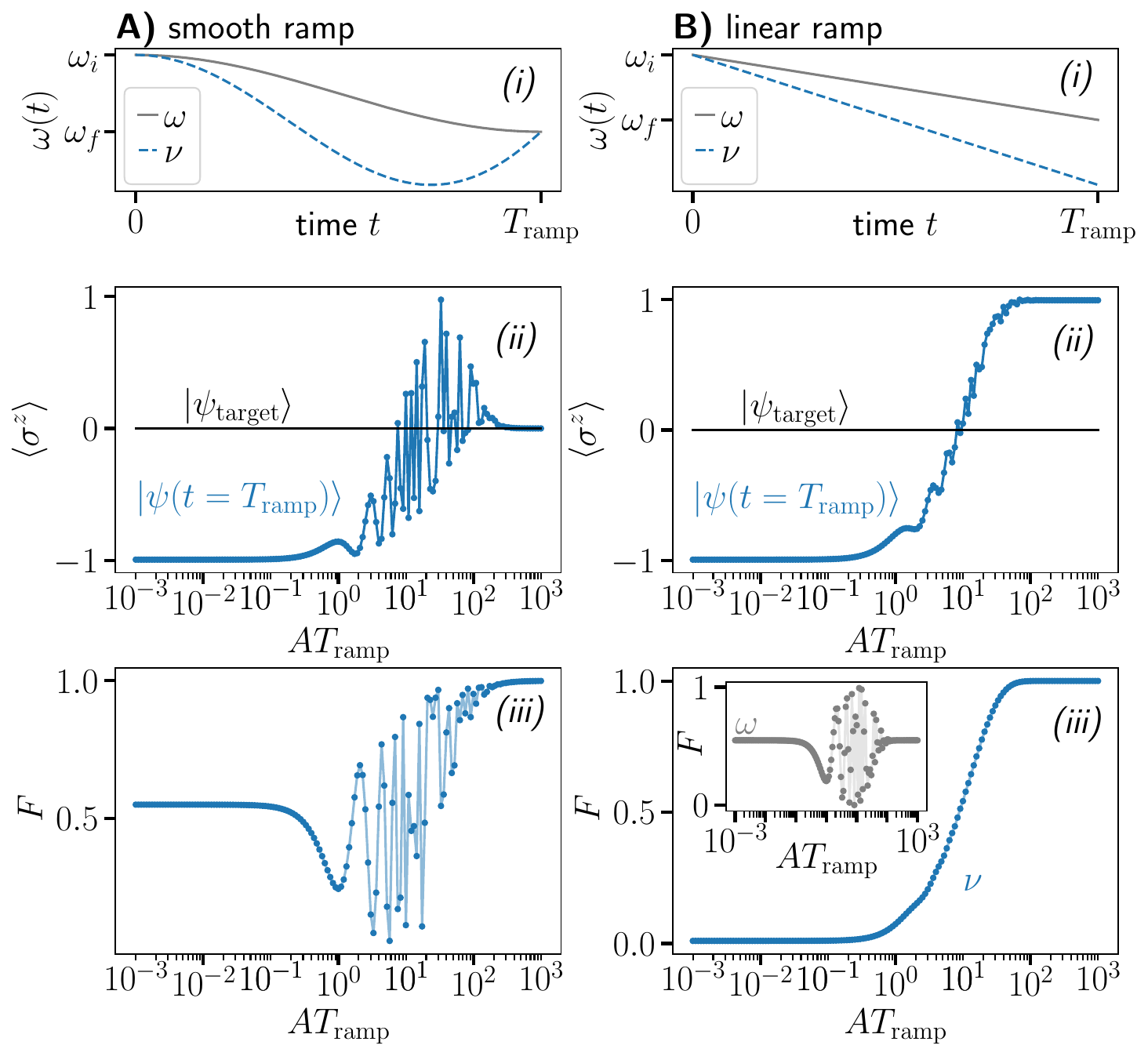}
    \caption{
    \textbf{Protocol Dependence for Chirp.}
    Unassisted state preparation for smooth~\textbf{(A)} and linear~\textbf{(B)} frequency ramp in a circularly driven two-level system~\eqref{eq:circ2LS}.
    \textit{(i)} Driving protocol for time-dependent frequency $\omega(t)$~(gray solid) and instantaneous frequency $\nu(t)$~(blue dashed).
    \textit{(ii)} $z$-polarization of time evolved state~(blue) as a function of the ramp time $T_\mathrm{ramp}$. The constant black line corresponds to $z$-polarization of the target state.
    \textit{(iii)} Fidelity of time evolved state at the end of the protocol with respect to the Floquet eigenstate at instantaneous frequency, $\nu{=}\dot{\omega t}$; for \textbf{B)} also with respect to $\omega$~(inset).
    The adiabatic path does not follow the time-dependent frequency $\omega$-ramp but rather the instantaneous frequency $\nu$, which is different for $t\neq0$ for a linear ramp but coincides with the expected result for the smooth ramp at the beginning and end of the protocol, since $\dot{\omega}(t=T_\mathrm{ramp}{=})0$.
    We used $g=1$ and $\Delta=10$, $\omega_\mathrm{i}=20$ and $\omega_\mathrm{f}=10$; the smooth ramp is given by $\omega(t)=\omega_\mathrm{i} + (\omega_\mathrm{f}-\omega_\mathrm{i})\sin^2(x\pi/2)$ with $x=(t-t_\mathrm{i})/T_\mathrm{ramp}$.
    }
    \label{fig::chirp}
\end{figure}

\begin{figure}[t]
    \centering
    \includegraphics[width=0.5\textwidth]{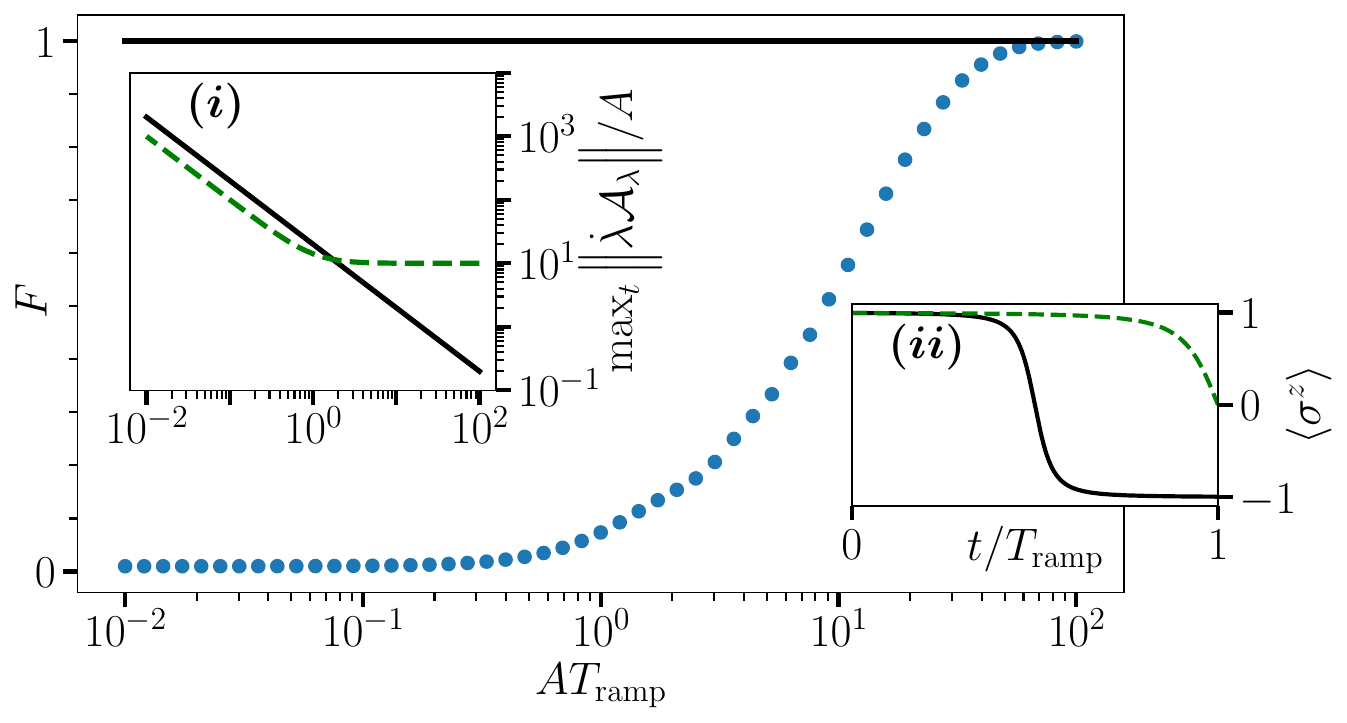}
    \caption{
    \textbf{Floquet Counterdiabatic Driving for Chirp in Circularly Driven Two-level System.}
    \textit{Main Panel.} Fidelity as a function of ramp time $T_\mathrm{ramp}$ for unassisted protocol~(blue dots) and CD drive~(black line), c.f.~Eq.~\eqref{eq:circ2LS_FAGP}. 
    \textit{Inset (i)} Amplitude of counterdiabatic term measured as the norm of the operator \update{$\norm{\dot{\lambda} \operator{A}_\lambda}= \abs{\dot{\lambda}}\sqrt{\Tr(\operator{A}_\lambda^2)} $} where $\dot{\lambda}=\dot{\nu}$ or $\dot{\omega}$ for $\operator{A}$~(black solid) or $\operator{A}^\prime$~(green dashed), respectively.
    \textit{Inset (ii)} $z$-polarization as a function time for $T_\mathrm{ramp}=A$ for CD protocols, colors as in $(i)$.
    Floquet counterdiabatic driving leads to transitionless driving along the adiabatic path. Adjusted FAGP, $\operator{A}^\prime$ allows for following instantaneous states along $\omega$ at the expense of additional finite terms for all ramp times.
    Parameters are as in Fig.~\ref{fig::chirp}.
    }
    \label{fig::chirpCD}
\end{figure}

Let us now consider ramping the frequency from some initial value $\omega(0)=\omega_\mathrm{i}$ to a final value $\omega(T_\mathrm{ramp})=\omega_\mathrm{f}$. 
Then, we would naively expect that the $\dot{\phi}$ contribution which deviates from $\omega$ should be part of the adiabatic gauge potential, i.e., $\dot{\omega } \A \supset t\dot{\omega} S^z$. However, as shown in Sec.~\ref{sec:chirp}, the $t \dot{\omega} S^z$ contribution does not vanish in the adiabatic limit therefore it cannot be part of the adiabatic gauge potential.

\paragraph{Adiabatic Limit}
To restore the adiabatic limit, $t\dot{\omega} S^z$ should be considered a relevant perturbation regardless of the protocol and hence should be included in the Floquet Hamiltonian
\begin{equation}
\label{eq:circ2LS_chirp}
    \widetilde{\h} = (\Delta - \nu(t) ) S^z + A S^x \, .
\end{equation}
Note that, while in the adiabatic limit $\dot{\omega}\to0$ the effective frequency does not approach the actual frequency at all times $\nu \not\to \omega$, however, $\dot{\nu}\to0$. 
Let us further emphasize that the frequency ramp leads to a change in the frequency appearing in the quasi-energy operator~\eqref{eq:Q}, but does not affect the period of the oscillations, i.e., the micromotion operator $W(t)=\exp(i\omega(t)tS^z)$ remains unchanged.

For sufficiently smooth protocols that approach their final value with a vanishing slope $\dot{\omega}(T_\mathrm{ramp})=0$ the two frequencies agree at the final point $\omega(T_\mathrm{ramp})=\nu(T_\mathrm{ramp})$; thus, the final states also agree. However, this may still lead to an increase in the adiabaticity time scale: in general the maximal slope of $\nu$, $\max_t\abs{\dot{\nu}}$, will be larger than the maximal slope of $\omega$, $\max_t\abs{\dot{\omega}}$. This is a direct result of, $\nu$ being a non-convex function in time whenever $\dot{\omega}(T_\mathrm{ramp})=0$.
If, however, $\dot{\omega}(T_\mathrm{ramp})\neq0$ as is the case for a linear frequency ramp, then also $\omega(T_\mathrm{ramp})\neq\nu(T_\mathrm{ramp})$, and the prepared state might differ from the targeted state.

In Fig.~\ref{fig::chirp}, we demonstrate the problem of a linear frequency ramp~(B) in contrast to a smooth ramp~(A). 
Let us consider, that we want to prepare the $x$-polarized state---or more precisely the Floquet state which circles around the $z$-axis in the $xy$-plane---starting from the (almost) $z$-polarized state.
To this end, we choose a frequency chirp with $(\Delta-\omega_\mathrm{i})\gg A$ and $(\Delta - \omega_\mathrm{f})=0$. This leads to $\widetilde{\operator{H}}(t=t_\mathrm{f})=A S^x$ if the state follows the time-dependent frequency $\omega(t)$. 
In Fig.~\ref{fig::chirp}(i), we display the corresponding frequency ramps $\omega=\omega(t)$ and the effective frequency $\nu(t)$ experienced by the system. 

We first estimate the performance of the protocol on a physical observable. Therefore, we consider the $z$-magnetization, $\expval{\sigma^z}$, which should vanish for the target $x$-polarized state. As the $z$-magnetization is unaffected by the micromotion operator $\operator{P}\propto \sigma^z$ a finite $z$-polarization indicates failure of the state preparation procedure. 
With increasing ramp time $T_\mathrm{ramp} \to \infty$ the $z$-magnetization saturates for both ramps indicating that the adiabatic limit has been reached, see Fig.~\ref{fig::chirp}(ii). 
However, for the linear ramp, Fig.~\ref{fig::chirp}B(ii), the $z$-magnetization attains a non-zero value ($\expval{\sigma^z}\to 1$) suggesting that the protocol failed to prepare the target state.

In addition, we consider the fidelity $F$ with respect to the Floquet eigenstate following the instantaneous frequency $\nu$, see Fig~\ref{fig::chirp}(iii). We find for both ramps, linear and smooth, that the fidelity approaches unity for large ramp time $T_\mathrm{ramp}$; confirming that the adiabatic state follows $\nu(t)$ and not $\omega(t)$.
For the smooth ramp, we find that the adiabatic limit is reached only at much larger ramp times compared to the linear ramp. However, the final state in the adiabatic limit agrees with the target state.

\paragraph{Counterdiabatic Driving}

As discussed in Sec.~\ref{sec:chirp_rigorous} there are two ways to perform counterdiabatic driving in the case of chirps. One can either follow the adiabatic path described by instantaneous eigenstates following $\nu(t)$, i.e., eigenstates of \eqref{eq:circ2LS_chirp}. Alternatively, one can follow the naively expected path which corresponds to instantaneous eigenstates with respect to $\omega(t)$, i.e., eigenstates of Eq.~\eqref{eq:circ2LS_rot}.

For both cases, we can readily write down the counterdiabatic term using the results from Sec.~\ref{sec:exactlysolvable}:
\begin{subequations}
\label{eq:circ2LS_FAGP}
    \begin{align}
        \dot{\nu}\operator{A}_\omega(\nu)
            &= \frac{A\dot{\nu}}{(\Delta - \nu)^2 + A^2} 
            \pqty{\cos(\phi) S^y - \sin(\phi) S^x} \, ,\\
        \dot{\omega} \operator{A}^\prime_\omega(\omega) 
            &= \dot{\omega}\operator{A}_\omega(\omega)  +\dot{\omega} t S^z \, ,
    \end{align}
\end{subequations}
where $\operator{A}^\prime_\omega$ denotes the alternative case. From Eq.~\eqref{eq:circ2LS_FAGP}b it is clear that the contribution $\dot{\omega}\operator{A}^\prime_\omega$ will remain finite in the adiabatic limit $\dot{\omega}\to0$.

In Fig.~\ref{fig::chirpCD} the effect of the two FCD protocols is depicted for the linear ramp. 
As expected, the FCD protocols lead to unit fidelity regardless of the ramp time; fidelity is measured with respect to the according targeted `adiabatic' path.
However, in agreement with the quantum speed limit the strength of the counter term strongly depends on the ramp time $T_\mathrm{ramp}$, see Fig.~\ref{fig::chirpCD}(i).
While the usual counter term, $\operator{A}_\omega$, follows the adiabatic curve described in Fig.~\ref{fig::chirp} using the adjusted counter term $\operator{A}_\omega^{\prime}$ allows for state preparation following the naive path $\omega$, see Fig.~\ref{fig::chirpCD}(ii). However, following $\omega(t)$ instead of $\nu(t)$ comes at the expense that the alternative FAGP does not vanish in the adiabatic limit.

\begin{figure}[t]
    \centering
    \includegraphics[width=0.5\textwidth]{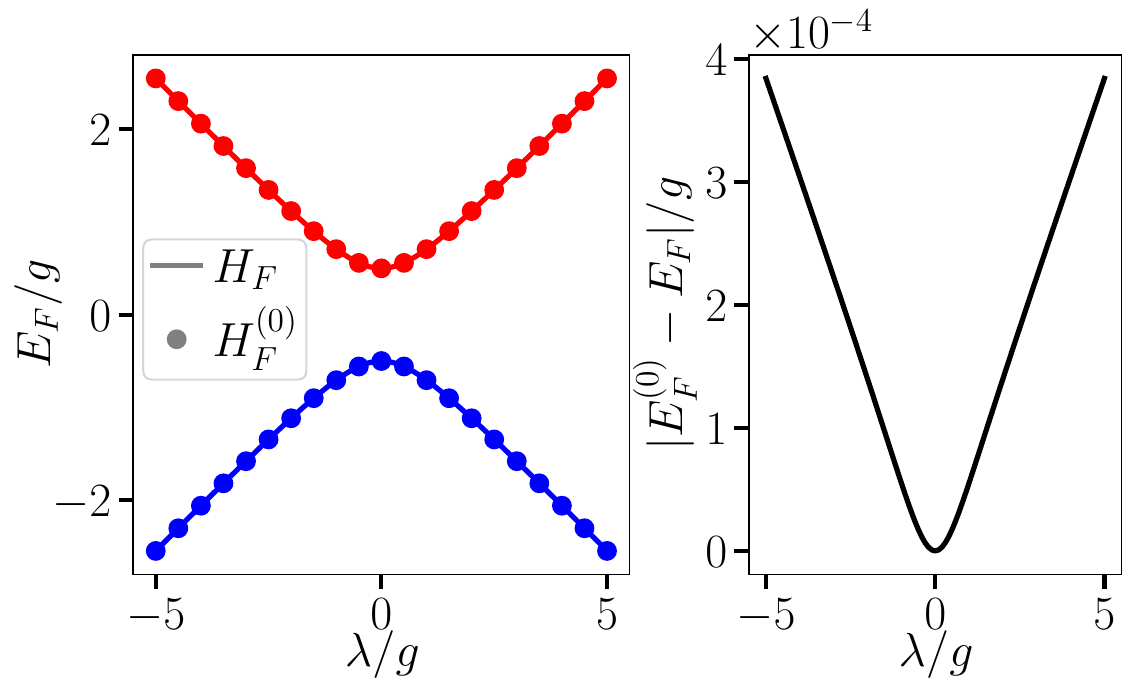}
    \caption{
    \textbf{Performance of high-frequency expansion in linearly driven two-level system} in the large frequency regime.
    \textit{Left Panel:} Exact~(solid line) and lowest order perturbative~(circles) Floquet quasi-energies.
    \textit{Right Panel:} Error in lowest order perturbative Floquet energies compared to exact energies.
    The lowest order perturbative Floquet Hamiltonian accurately describes the exact Floquet quasi-energies.
    Parameters are as in Fig.~\ref{fig:2LS}.
    }
    \label{fig::2LS_HFE}
\end{figure}

\subsection{\label{sec:linear2LS}Example with Photon Resonances}

\begin{figure}[t]
    \includegraphics[width=0.49\textwidth]{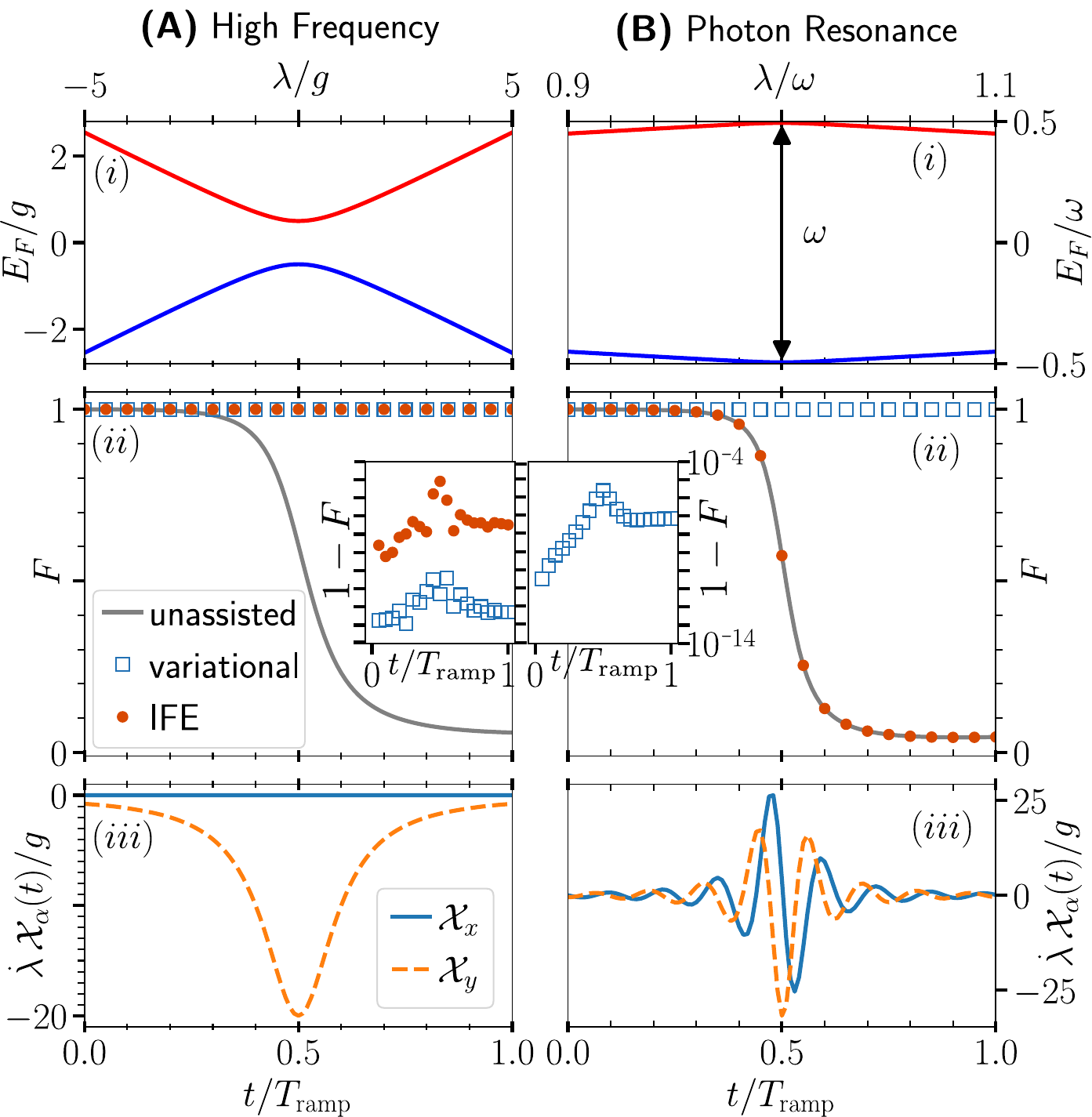}
    \caption{
    \textbf{FCD for linearly driven two-level system}~\eqref{eq:hamiltonian_2LS} in high-frequency~(\textbf{A}) and one-photon resonance~(\textbf{B}) regime. The level-splitting $\lambda$ is varied linearly within $\lambda\in\bqty{-5g,\,5g}$ and $\lambda\in\bqty{0.9\omega,\,1.1\omega}$, respectively.
    $(i)$ Numerically exact instantaneous Floquet energies $E_F$ as a function of time $t$.
    $(ii)$ Instantaneous fidelity $F(t)$ for unassisted~(gray), IFE~(red circles) and variational protocol~(blue boxes). \textit{Inset:} Deviation of fidelity from unity, note the logarithmic scale.
    $(iii)$ $x$~(blue, solid) and $y$~(orange, dashed) component of the variational gauge potential $\mathcal{X}$, see Eqs.~\eqref{eq:2LS_HFE_FAGP} and \eqref{eq:2LS_varl_resonance}, (the $z$ component always vanishes).
    In high-frequency regime~(\textbf{A}) the IFE  method successfully suppresses diabatic transitions but fails in the presence of photon resonances~(\textbf{B}). The variational method suppresses diabatic transitions in all scenarios.
    We use a linear ramp $\lambda(t)=\lambda_\mathrm{i}+(\lambda_\mathrm{f}-\lambda_\mathrm{i}) t /T_\mathrm{ramp}$ for the control parameter $\lambda\in\bqty{\lambda_\mathrm{i}, \lambda_\mathrm{f}}$. Other parameters are $\omega/g=100$, $A/g=2.5$, $T_\mathrm{ramp}g=0.5$.
    }
    \label{fig:2LS}
\end{figure}

A hallmark feature of Floquet systems is the existence of photon resonances in the quasi-energy spectrum. Thus, to build further intuition about the methods introduced in Sec.~\ref{sec:theory}, we investigate a pedagogical example of a linearly driven two-level system. Unlike the analytically solvable circularly driven two-level system studied in the previous section, the linearly driven model possesses photon-resonances.
We are primarily interested in comparing the behavior of the variational approaches with the IFE approach to the FAGP in the regime of photon resonances where the IFE method is expected to break down. 

Consider the Hamiltonian of the linearly polarized two-level system
\begin{equation}
    \operator{H}_\lambda \pqty{t} = \lambda \, S^z + \bqty{g + A \cos(\omega t)}\, S^x \, , \label{eq:hamiltonian_2LS}
\end{equation}
with level-splitting $\lambda$, level-hybridization $g$, drive amplitude $A$(${=}2.5\,g$) and drive frequency $\omega = 2\pi/T$(${=}100\,g$). 
Despite its simple Hilbert-space structure, there exists no known closed-form expression for the Floquet Hamiltonian.

\subsubsection{High-frequency Regime}

Let us first focus on the high-frequency regime. In particular, we consider a linear ramp of the level-splitting $\lambda$ 
\begin{equation}
\label{eq:lin_ramp}
    \lambda(t) = (\lambda_\mathrm{f}-\lambda_\mathrm{i})\frac{t}{T_\text{ramp}} + \lambda_\mathrm{i}
\end{equation}
from $\lambda_\mathrm{i}=-5g$ to $\lambda_\mathrm{f}=5g$ with an duration $T_\mathrm{ramp}=1/(2g)$. 
Such control setups arise naturally when trying to manipulate the behavior of Floquet-engineered systems using external controls, e.g., to demonstrate the presence of a non-zero Berry curvature in a Floquet topological band \cite{Jotzu2014_Haldane,Aidelsburger2014_MeasuringChernNumber}, or to probe the metastability of dynamically stabilized matter~\cite{Zenesini2009_CoherentControlDressedWaves}.
Like in these situations, we assume that the system is initiated in a Floquet eigenstate before the ramp $\lambda(t)$ is turned on.

Note that, due to the invariance of the quasi-energies with respect to a shift by $n\omega$~($n\in\mathbb{Z}$) the Floquet Hamiltonian does \textit{not} have a well-defined ground state. Therefore, we use as an initial state the Floquet eigenstate which has a larger overlap with the ground state of the non-driven Hamiltonian $\lambda_\mathrm{i} \, S^z$. 

Since the frequency is large compared to all other parameters, $\omega \gg g,A,\lambda$, the system is well described by the lowest order Floquet Hamiltonian~(see comment at end of this section)
\begin{equation}
\label{eq:2LS_HFE}
    \operator{H}_F^{(0)} = \lambda S^z + g S^x,\qquad \operator{K}_F^{(0)}=0\, ,
\end{equation}
The Floquet energies exhibit an avoided crossing around $\lambda=0$, see Fig~\ref{fig:2LS}A(i). In fact, in the absence of any counterterms~(referred henceforth as \textit{unassisted} control), the protocol leads to an almost complete loss of fidelity, see grey line in Fig.~\ref{fig:2LS}A(ii). 

Note that, the lowest-order Floquet Hamiltonian~\eqref{eq:2LS_HFE} describes a standard Landau-Zener crossing~\cite{landau1932theorie,landau1965collected,zener1932non,stueckelberg1932theorie,Ivakhnenko2023_LandauZener} as a function of $\lambda$. Hence, the corresponding approximate FAGP for~\eqref{eq:2LS_HFE} takes the well-known form~\cite{Kolodrubetz2017_GeometryReview} 
\begin{equation}
\label{eq:2LS_HFE_FAGP}
    \operator{A}_\lambda^{(0)}(t) = -\frac{g}{\lambda^2 + g^2} S^y \,,
\end{equation}
where, in addition, we also used that to lowest order $\operator{A}_\lambda^{(0)}(t)=\operator{A}_{F,\lambda}^{(0)}$, following Eq.~\eqref{eq:IFE_loworder}.
In contrast to the unassisted protocol, the IFE FCD protocol significantly suppresses diabatic excitations leading to high fidelity throughout the entire protocol duration, see red circles in Fig~\ref{fig:2LS}A(ii).

For comparison, we can also consider a variational FAGP. To this end, we use the ansatz
\begin{equation}
\label{eq:2LS_ansatz}
    \operator{X}(t) = \bqty{y_0 + y_1 \cos(\omega t)}S^y + x_1 \sin(\omega t) S^x + z_1 \sin(\omega t) S^z \, ,
\end{equation}
with variational parameters $\Bqty{y_0, x_1, y_1, z_1}$.
Since real symmetric Hamiltonians are diagonalized by orthogonal transformations, the generating AGP can be chosen imaginary-valued~\cite{Kolodrubetz2017_GeometryReview}. To generalize this observation to Floquet systems for the ansatz above, note that $\cos(\omega t)\propto e^{i\omega t}+e^{-i\omega t}$ can be considered a real-valued, and $\sin(\omega t)\propto i(e^{i\omega t}-e^{-i\omega t})$ an imaginary-valued function [see Sec.~\ref{sec:rigorous_varl} for the details]. The analytic derivation of the variational parameters is given below in Sec.~\ref{sec:linear2LS_analytic}.
Similar to the IFE-FAGP, the variational FAGP fully counteracts the diabatic transitions, see blue squares in Fig~\ref{fig:2LS}A(ii).

To reconcile the IFE and variational approaches, we apply a high-frequency expansion within the variational principle~\eqref{eq:varl_projection}. This allows us to analytically derive the form of the variational FAGP shown in Fig~\ref{fig:2LS}A(iii). As expected, the latter coincides with Eq.~\eqref{eq:2LS_HFE_FAGP} up to $O(\omega^{-1})$ corrections~(see Sec.~\ref{sec:linear2LS_analytic}).

Before we close this section, let us briefly provide evidence for the claim that the lowest order high-frequency expansion~\eqref{eq:2LS_HFE},
\begin{equation*}
    H_F^{(0)} = \lambda S^z + g S^x \,,
\end{equation*}
is sufficient to describe the linearly driven two-level system~\eqref{eq:hamiltonian_2LS} in the high-frequency regime. To this end, we compare the numerically computed spectrum of the exact Floquet Hamiltonian with the spectrum obtained from the lowest order high-frequency expansion~\eqref{eq:2LS_HFE}, see Fig.~\ref{fig::2LS_HFE}.
We find that the exact and high-frequency Floquet quasi-energies agree up to the fourth digit, suggesting that the lowest order Floquet Hamiltonian is sufficient to describe the system at high accuracy.

\subsubsection{Photon Resonance Regime}

Let us now turn to the one-photon resonance regime at $\lambda \approx \omega$. In particular, we consider the linear ramp from Eq.~\eqref{eq:lin_ramp} with $\lambda_\mathrm{i}=0.9\omega$ and $\lambda_\mathrm{f}=1.1\omega$. 

In this regime, the energy levels appear far separated [Fig.~\ref{fig:2LS}B(i)], such that identifying an avoided level crossing is not immediately obvious. However, the energy separation is almost resonant with the external drive $E_{F,1} -E_{F,2} \approx \omega$, leading to a hybridization of the states and an avoided crossing at around $\lambda \approx \omega$, see Fig.~\ref{fig:2LS}B(i). 

Since the ramp duration is small compared to the inverse photon absorption gap, the Landau-Zener condition for adiabatic passage is not satisfied, and the unassisted drive creates excitations.
However, this Floquet resonance is not captured by the high-frequency expansion~[Resonant drives can also be dealt with directly in some cases, using an additional change-of-frame transformation~\cite{goldman2015_ResonantFloquet,Bukov_2015_general_HFE}], and, in addition, the IFE CD protocol fails as well, see Fig.~\ref{fig:2LS}B(ii). 
The variational approach, on the other hand, fully captures this non-perturbative resonance and allows for transitionless driving through the avoided crossing, see Fig.~\ref{fig:2LS}B(ii).
In fact, using the ansatz~\eqref{eq:2LS_ansatz} and taking a large frequency limit in the variational minimization~($\omega {\approx} \lambda {\gg} A,g$), we find~(see Sec.~\ref{sec:linear2LS_analytic})
\begin{equation}
\label{eq:2LS_varl_resonance}
    \operator{X}_\lambda(t) = 
        \frac{2A}{A^2 + 4(\lambda - \omega)^2} \bqty{\cos(\omega t)S^y - \sin(\omega t) S^x} 
        + O(\omega^{-1}) \, ,
\end{equation}
The structure of the variational FAGP is shown in Fig.~\ref{fig:2LS}B(iii).

This simple model already illustrates the potential advantages of the variational over the IFE method. Indeed, using a suitable ansatz, the variational method allows us to capture both perturbative and non-perturbative effects in the quasi-energy spectrum and the Floquet eigenstates.
We expect this technique to be particularly useful in atomic physics where high-precision few-level control can be sped up using FCD drives.

For the purposes of Floquet engineering governed by the \textit{approximate} Floquet Hamiltonian, however, following the adiabatically connected quasi-energy manifold in the presence of photon-resonances can sometimes be undesirable, [see discussion of Fig.~\ref{fig:varl_vs_ife}]. In such cases, one can use the IFE protocol and a short $T_\mathrm{ramp}$ to traverse standard quasi-energy gaps adiabatically, while passing through photon-resonances diabatically. Therefore, depending on the control problem at hand, either the IFE or the variational approach is preferable.

\subsubsection{Analytical Derivation of Variational FAGP}
\label{sec:linear2LS_analytic}

In this subsection, we derive in detail the analytical variational FAGP reported on in the previous Sec.~\ref{sec:linear2LS}.

To this end, let us consider the ansatz Eq.~\eqref{eq:2LS_ansatz} and plug it into the action Eq.~\eqref{eq:varl_projection}, resulting in
\begin{equation}
    \label{eq:action_2LS}
    S[\boldsymbol{X}] = \boldsymbol{X}\cdot \boldsymbol{M} \cdot \boldsymbol{X} - 2 \boldsymbol{X}^T \boldsymbol{b} + \mathrm{const.} \, .
\end{equation}
Hence, minimizing the action $\delta S = 0$ amounts to solving the linear system of equations
\begin{equation}
\label{eq:linearsystem_2LS}
    \boldsymbol{M} \cdot \boldsymbol{X} = \boldsymbol{b}\, ,
\end{equation}
where $\boldsymbol{X}^T=(y_0,\, x_1,\, y_1,\, z_1)$, $\boldsymbol{b}^T = \left( 2 g,\, 0,\, A,\,0 \right)$ and
\begin{equation}
\label{eq:varlmatrix_2LS}
    \boldsymbol{M} = 
        \begin{pmatrix}
            2\lambda^2 + 2g^2+ 4 A^2 
            & 0
            & 3 A g                     
            & -2 A\omega 
        \\
            0 
            &  \omega^2+\lambda^2
            & 2 \lambda \omega           
            & -\lambda g 
        \\
            3 A g     
            & 2 \lambda \omega 
            & g^2+\omega^2+\lambda^2+A^2/2 
            & -2 g \omega  
        \\
            -2 A \omega  
            & -\lambda  g  
            & -2g\omega        
            & \omega^2+\lambda^2  
        \end{pmatrix} \, . 
\end{equation}

While inverting the $4\times4$ matrix may be performed analytically using, e.g., a Gaussian elimination procedure, the general expression is complicated, and not much insight can be gained from it. Also, in general, a larger number of variational parameters may be considered which makes an analytical treatment infeasible.

Therefore, we focus on deriving an expression for the variational parameters for the two regimes considered above, i.e., the high-frequency~(I) and one-photon resonance~(II) regime. 

\paragraph*{1. High-frequency regime}
In the high-frequency regime $\omega$ is the largest energy scale, $\omega \gg \lambda,\,g,\,A$, such that it is favourable to remove the $\propto \omega,\, \omega^2$ contributions in $\boldsymbol{M}$. This is achieved by considering the re-scaling $\tilde{X}^T=(y_0,\, x_1\omega,\, y_1\omega,\, z_1\omega)$ and $\tilde{b}^T=(2g,0,A/\omega,0)$ leading to
\begin{equation*}
    \tilde{\boldsymbol{M}} = 
        \begin{pmatrix}
            2\lambda^2 + 2g^2+ 4 A^2 
            & 0
            & 3 \frac{Ag}{\omega}                    
            & -2 \frac{A}{\omega} 
        \\
            0 
            &  1+ \frac{\lambda^2}{\omega^2}
            & 2 \frac{\lambda}{\omega}        
            & -\frac{\lambda g}{\omega^2} 
        \\
            3 \frac{A g}{\omega^2}     
            & 2 \frac{\lambda}{\omega} 
            & 1 + \frac{g^2+\lambda^2+A^2/2}{\omega^2} 
            & -2 \frac{g}{\omega}  
        \\
            -2 \frac{A}{\omega}  
            & -\frac{\lambda g}{\omega^2}       
            & -2\frac{g}{\omega}       
            & 1+ \frac{\lambda^2}{\omega^2}  
        \end{pmatrix} \, ,
\end{equation*}
taking the limit $\frac{A}{\omega},\, \frac{g}{\omega},\, \frac{\lambda}{\omega} \to 0$ the linear system Eq.~\eqref{eq:linearsystem_2LS} reduces to
\begin{equation*}
    \begin{pmatrix}
        2\lambda^2 + 2g^2+ 4A^2  
        & 0
        & 0                      
        & -2A 
    \\
       0 
        &  1 
        & 0          
        & 0
    \\
        0      
        & 0
        & 1 
        & 0
    \\
        -2A              
        & 0  
        & 0  
        & 1 
    \\
    \end{pmatrix} 
    \cdot
    \tilde{\boldsymbol{X}}
    =
    \begin{pmatrix}
        2g\\
        0\\
        0\\
        0
    \end{pmatrix}\,,
\end{equation*}
which is solved by 
\begin{align*}
    y_0 &= \frac{g}{\lambda^2 + g^2} \\
    x_1 &= 0 \\
    y_1 &= 0 \\
    z_1 &= \frac{A}{\omega} \frac{g}{\lambda^2 + g^2}\, ,
\end{align*}
which is equivalent to the IFE approach Eq.~\eqref{eq:2LS_HFE_FAGP} up to $O(\omega^0)$.

\paragraph*{2. Photon Resonance Regime.}
In the one-photon resonance regime, we have $\lambda\approx \omega$, such that both $\lambda$ and $\omega$ are large scales. Therefore, in order to get rid of the large scales a re-scaling of the parameters $\tilde{X}^T=(y_0\omega,\, x_1\omega,\, y_1\omega,\, z_1\omega)$ and $\tilde{b}^T=(2g/\omega,0,A/\omega,0)$ is needed. Performing the re-scaling and a large frequency expansion $\frac{\lambda}{\omega}\to 1$ and $\frac{A}{\omega},\, \frac{g}{\omega} \to 0$ in Eq.~\eqref{eq:varlmatrix_2LS}, however, leads to the linear system
\begin{equation*}
    \begin{pmatrix}
        1
        & 0
        & 0                      
        & 0
    \\
       0 
        & 1 
        & 1   
        & 0
    \\
        0      
        & 1
        & 1 
        & 0
    \\
        0           
        & 0  
        & 0  
        & 1 
    \\
    \end{pmatrix} 
    \cdot
    \tilde{\boldsymbol{X}}
    =
    \begin{pmatrix}
        0\\
        0\\
        0\\
        0
    \end{pmatrix}\,.
\end{equation*}
Naively one might assume that the ideal parameters are just given by the trivial choice of parameters, i.e., $\tilde{\boldsymbol{X}}_0=0$. However, as the matrix is singular also $\tilde{\boldsymbol{X}}_1=c\pqty{0,1,-1,0}$ with arbitrary $c$ is a solution. Thus, the solution to the variational principle is ill-defined.

To understand whether $\tilde{\boldsymbol{X}}_0$ or $\tilde{\boldsymbol{X}}_1$ is the correct choice of parameters, we consider expanding around these solutions. To this end, we represent the matrix~\eqref{eq:varlmatrix_2LS} in the basis $\Bqty{\hat{\boldsymbol{e}}_1, \pqty{\hat{\boldsymbol{e}}_2+\hat{\boldsymbol{e}}_3}/\sqrt{2}, \pqty{\hat{\boldsymbol{e}}_2-\hat{\boldsymbol{e}}_3}/\sqrt{2}, \hat{\boldsymbol{e}}}_4$, where $\hat{\boldsymbol{e}}_i$ is the $i$'th unit vector.
This leads to the modified linear systems of equations 
\begin{equation}
\label{eq:linearsystem_2LS_basistrafo}
    \boldsymbol{M}^\prime \cdot \boldsymbol{X}^\prime = \boldsymbol{b}^\prime\, ,
\end{equation}
with $\boldsymbol{X}^\prime=(y_0,\chi,\eta,z_0)$, $\boldsymbol{b}^\prime=\pqty{g, -A, A, 0}$, and
\begin{equation}
\label{eq:varlmatrix_2LS_basistrafo}
    \boldsymbol{M}^\prime = 
        \begin{pmatrix}
            2\lambda^2 + 2g^2+ 4 A^2 
            & 3 A g
            & 3 A g                     
            & -2 A\omega 
        \\
            3 A g 
            &  2\Omega_+^2 + G^2
            & G^2    
            & -g \omega - g \Omega_+
        \\
            3 A g     
            & G^2
            & 2\Omega_-^2 + G^2
            & -g \omega - g \Omega_-
        \\
            -2 A \omega  
            & -g \omega - g \Omega_+  
            & -g \omega - g \Omega_-    
            & \omega^2+\lambda^2  
        \end{pmatrix} \, ,
\end{equation}
where we defined $\Omega_{\pm} = \lambda \pm \omega$ and $G^2=g^2+A^2/2$.
Note that, by definition of the one-photon regime $\omega\approx \lambda$, such that $\Omega_-=(\lambda-\omega)$ neither scales with $\omega$ nor $\lambda$. Thus, the $M^\prime_{\eta\eta}$ index in Eq.~\eqref{eq:varlmatrix_2LS_basistrafo} is a finite, non-diverging, contribution even if $\lambda,\omega \to \infty$ is considered.
Therefore, to get rid of the diverging energy scales when taking the infinite frequency limit $\lambda\approx\omega\to\infty$ we should choose the re-scaling $\boldsymbol{X}^\prime\to \tilde{\boldsymbol{X}}^\prime=\pqty{y_0\omega,\chi\omega,\eta,z_0\omega}$. Performing this re-scaling and taking the large frequency limit in Eq.~\eqref{eq:varlmatrix_2LS_basistrafo} we eventually arrive at
\begin{equation*}
    \begin{pmatrix}
        2
        & 0
        & 0                     
        & 0 
    \\
        0
        &  2\frac{\Omega_+^2}{\omega^2} 
        & 0
        & 0
    \\
        0    
        & 0
        & 2\Omega_-^2 + G^2
        & -g 
    \\
        0  
        & 0  
        & -g     
        & 1  
    \end{pmatrix}
    \cdot
    \tilde{\boldsymbol{X}}^\prime
    =
    \begin{pmatrix}
        0\\
        0\\
        A\\
        0
    \end{pmatrix}\,,
\end{equation*}
which is solved by
\begin{align*}
    \boldsymbol{X}^\prime &= \pqty{0,0,\eta,\frac{g}{\omega} \eta}^T\, ,\\
    \eta &= \frac{2A}{4 (\lambda -\omega)^2 + A^2} \, .
\end{align*}

Transforming back to the original basis leads to
\begin{align*}
    y_0 &= 0 \\
    x_1 &= - \frac{2A}{4 (\lambda -\omega)^2 + A^2} \\
    y_1 &= + \frac{2A}{4 (\lambda -\omega)^2 + A^2} \\
    z_1 &= 0 + O(\omega^{-1})\, ,
\end{align*}
which is exactly Eq.~\eqref{eq:2LS_varl_resonance} up to $O(\omega^{-1})$ corrections.


\section{\label{sec:algo}Details of the Numerical Algorithm}

Computing the (approximate) variational adiabatic gauge potential $\operator{X}$, cf. Eq.~\eqref{eq:varl}, analytically quickly becomes infeasible for a large number of variational parameters. Therefore, most variational FAGP's presented in the main text are computed numerically.

In this section, we describe the algorithm that we use to compute the variational Floquet adiabatic gauge potential numerically and comment on some straightforward extensions of our algorithm.

\paragraph{General framework for solving variational principle}

Let us start by recalling the equations that need to be solved. The variational FAGP minimizes the action 
\begin{align*}
    S\bqty{\operator{X}_\lambda} &= \int_{0}^{2\pi} \Tr(\operator{G}^2(\operator{X}_\lambda(\phi))) \mathrm{d}\phi\\
    \operator{G}
    &= i\comm{\h_\lambda(\phi)}{ \operator{X}_\lambda(\phi)} + \partial_\phi \operator{X}_\lambda(\phi)+ \partial_\lambda \h_\lambda(\phi) \, ,
\end{align*}
for each parameter value $\lambda$, see also Eq.~\eqref{eq:varl} in the main text. 

In a numerical approach, we can not compute the variational FAGP for all continuous values of $\lambda$ in the interval $\lambda\in \bqty{\lambda_i,\, \lambda_f}$ but rather need to restrict to a discrete subset. 
Moreover, it would be computationally too demanding to compute the variational FAGP on the fly while computing the time evolution, i.e., while solving the time-dependent Schr\"odinger equation~(TDSE). Therefore, we pre-compute the variational FAGP $\operator{X}(\lambda_n)$ for a given set of discrete parameters $\lambda_n \in \bqty{\lambda_i,\, \lambda_f}$, $n=1,\dots,N$. Then, we interpolate the results, $\operator{X}(\lambda_n)\to\operator{X}_\lambda^{\bqty{N}}$, to obtain a continuous function, which is then used to solve the TDSE. 
To ensure that our results are sufficiently converged we assert that the error between the interpolation $\operator{X}_\lambda^{\bqty{N}}(\lambda_{n^\prime})$ and the exact result, evaluated on a set of unseen points $\lambda_{n^\prime}{\neq}\lambda_n$, is below some threshold value $\epsilon$, $\norm{\pqty{\operator{X}_\lambda^{\bqty{N}} - \operator{X}_\lambda}(\lambda_{n^\prime})} \leq \epsilon$.
If the error exceeds the threshold value we increase the number of points used for interpolation and repeat the procedure until the error is below threshold. For simplicity, we draw points for the evaluation equidistantly in the interval $\bqty{\lambda_i,\, \lambda_f}$. More sophisticated sampling techniques, like importance sampling or Gaussian process regression, may lead to a significant reduction in computational resources.

Let us now consider how to obtain the variational FAGP for a fixed value $\lambda$. 
To this end, we first need to make an ansatz for the variational FAGP. Throughout, we will consider an ansatz given by an operator expansion of the form
\begin{equation}
\label{eq:ansatz}
    \hat{\operator{X}}_\lambda = \sum_n x_{n\!,\lambda} \hat{O}_n\pqty{\phi} \, ,
\end{equation}
with variational parameters $x_{n\!,\lambda}$ and linearly independent operators $\hat{O}_n\pqty{\phi}$ which may carry a periodic time-dependency. Later, we consider more concrete examples of operators but in general, one may consider operators that satisfy experimental constraints. In particular, in most experimental setups there are strict constraints on the physical operators which can be implemented but there are hardly any constraints on their time-dependency. Therefore, a separation into time and physical operators might be useful, i.e., $\hat{O}_n(\phi)\equiv \hat{O}_{n=(kl)}(\phi) = \hat{O}_{k} f_{l}(\phi)$.

Note that, since $\operator{G}$ is linear in $\operator{X}_\lambda$ and $S$ quadratic in $\operator{G}$, the action is quadratic in the variational parameters, leading to a convex optimization problem. Moreover, plugging Eq.~\eqref{eq:ansatz} into $\nabla_{\boldsymbol{x}}S=0$ we find that the optimal parameters are given by
\begin{equation}
\label{eq:linear_system}
    \boldsymbol{M} \boldsymbol{x} = \boldsymbol{b} \, ,
\end{equation}
with
\begin{equation}
\label{eq:linear_operators}
    \begin{aligned}
        M_{jk}(\lambda) &= \int\hspace{-5pt} \mathrm{d}\phi \Tr( g_{j\!,\lambda}(\phi) g_{k\!,\lambda}(\phi) )  \\
        b_{j}(\lambda)  &= \int\hspace{-5pt} \mathrm{d}\phi \Tr( g_{j\!,\lambda}(\phi) \partial_\lambda \h_\lambda(\phi)]) \\
        g_{j\!,\lambda}(\phi)
                        &= i \comm{\h_\lambda\pqty{\phi}}{O_j(\phi)} + \partial_\phi O_j(\phi) \, .
    \end{aligned}
\end{equation}
Notice that, for a given Hamiltonian computing the matrix $\boldsymbol{M}$ and vector $\boldsymbol{b}$ can be done easily using an appropriate ansatz and exploiting the underlying Hilbert space algebra.
Therefore, computing the variational FAGP at a fixed parameter $\lambda$ only amounts to solving the linear system of equations~\eqref{eq:linear_system}; thus only scales in the number of variational parameters but not in the Hilbert space dimension.
However, for a generic system, the exact FAGP is expected to be non-local. Thus, to obtain the exact FAGP variationally the number of variational parameters must be comparable to the Hilbert space dimension.

\paragraph{Basis Representation}
If one is interested in only one particular Hamiltonian and a single variational ansatz with few parameters, we suggest to compute the quantities in Eq.~\eqref{eq:linear_operators} (semi-)analytically for the specific operators and solve the corresponding system of linear equations analytically or numerically.
However, if multiple Hamiltonians or ans\"atze with the same underlying Hilbert space need to be solved a different approach might be beneficial. As we show below, one can avoid computing the linear system~\eqref{eq:linear_system} for each individual one by exploiting the linear structure of Eq.~\eqref{eq:linear_system}.

To this end, let us consider an operator basis $\Bqty{\operator{B}_n}_{n}$ of the Samb\'e space $\floquethilbert^2=\physhilbert^2\otimes\circhilbert^2$, with Hilbert space of physical operators $\physhilbert^2$ and the Hilbert space of operators acting on periodic smooth functions $\circhilbert^2$. 
For example, for a spin system of $L$ spins a complete basis is given by the operators $\sigma^{\boldsymbol{\mu}}\cdot e^{in\omega t}$, where $n\in \mathbb{Z}$ and $\boldsymbol{\mu}=\pqty{\mu_1,\dots \mu_L}$ with $\mu_j=0,x,y,z$ and Pauli-strings $\sigma^{\boldsymbol{\mu}}=\sigma^{\mu_1} \otimes \dots \sigma^{\mu_L}$. 

A key insight which we exploit for our derivation is that the commutator $_n\superop{C}\equiv i \comm{\operator{B}_n}{\cdot}:\floquethilbert^2\to \floquethilbert^2$ and derivative $\superop{N}\equiv\partial_\phi\pqty{\cdot}\equiv \comm{\operator{N}}{\cdot}:\floquethilbert^2\to\floquethilbert^2$ are linear (super-)operators acting on the Hilbert space of operators.
Thus, we can expand them in the operator basis $_n\superop{C}(\operator{B}_k) = i\comm{_n\superop{C}}{\operator{B}_k} = \sum_{k} {}_n \superop{C}_{kl}\, \operator{B}_l$ and $\superop{N}(\operator{B}_k)=\partial_\phi \operator{B}_k = \sum_{k} \superop{N}_{kl} \operator{B}_l$ with matrix entries ${}_n \superop{C}_{kl}$ and $\superop{N}_{nm}$.

Note that, computing the matrix elements ${}_n C_{kl}$ and $N_{nm}$ may seem like a significant overhead. However, depending on the choice of basis they are directly related to the structure factor of the operator Lie-algebra, hence involve low to no computational overhead. 

Using the operator basis, $\Bqty{\operator{B}_n}_n$, we can express any Hamiltonian $\operator{H}$ and its derivative $\partial_\lambda \operator{H}$ as a linear combination of basis elements: $\operator{H}=\sum_{n} h_n \operator{B}_n$ and $\partial_\lambda \h = \sum_{n} \partial_\lambda h_n \operator{B}_n$ with sets of real numbers $h_n$ and $\partial h_n$, respectively.
Likewise, we can also choose a variational ansatz of the form $\operator{X} = \sum_{n \in \mathbb{Y}} x_n \operator{B}_n$, which in general only includes a subset of the operator basis $\Bqty{\operator{B}_n}_{n \in \mathbb{Y}} \subset \Bqty{\operator{B}_n}_{n}$.
Then, we can express the operator $\operator{G}$~\eqref{eq:varl} similarly using the basis:
\begin{align*}
    \operator{G}(\mathcal{X}) &= \sum_{j \in \mathbb{Y}} x_j \pqty{ \superop{N}(\operator{B}_j)) + \sum_{m} h_{m}\, {}_m \superop{C}(\operator{B}_j)} + \sum_{n} \partial_\lambda h_n \operator{B}_n \\
                &= \sum_{n} \bqty{ \sum_{j\in\mathbb{Y}} \sum_{m} \pqty{x_j h_m\, {}_m C_{jn}}  + \partial_\lambda h_n } \operator{B}_n \,.
\end{align*}
Thus, the action, Eq.~\eqref{eq:varl}, can be expressed as
\begin{equation}
\label{eq:varl_basis}
    \begin{aligned}
        S\bqty{\boldsymbol{x}} 
            &= \sum_{j,k \in \mathbb{Y}} x_j x_k \sum_{n,m} S_{jn} \eta_{nm} S_{mk}  \\
            &\vspace{10pt} + 2\sum_{j\in \mathbb{Y}} x_j \sum_{n,m} S_{jn} \eta_{nm} \partial_\lambda h_m  + \mathrm{const.}\\
            &= \boldsymbol{x}^T \boldsymbol{S}^T \boldsymbol{\eta} \boldsymbol{S} \boldsymbol{x} 
                + 2 \boldsymbol{x}^T \boldsymbol{S}^T\boldsymbol{\eta} \partial_\lambda \boldsymbol{h} + \mathrm{const.}\\
    \end{aligned}
\end{equation}
where we defined the metric $\eta_{nm}{=}\int\hspace{-3pt} \mathrm{d}\phi \Tr(\operator{B}_n \operator{B}_m)$ and matrix $\boldsymbol{S}$ with entries $ S_{nm}{\equiv} (\sum_p h_p \, {}_p \superop{C}_{nm}) + \superop{N}_{nm}$. Let us emphasize that, in general, $\boldsymbol{S}$ is a rectangular matrix: considering the matrix entries $S_{nm}$, then, $n$ runs over all indices in the basis, $\Bqty{\operator{B}_n}_n$, but $m\in \mathbb{Y}$ only.
In addition, if the basis is orthonormal the metric is diagonal $\eta_{nm}=\delta_{nm}$.

Using the superoperator expressions has the advantage that all model and ansatz dependence enters the variational action~\eqref{eq:varl_basis} simply by changing the dimension and linear coefficients of the matrix $\boldsymbol{S}$ and vector $\partial_\lambda \boldsymbol{h}$.
Therefore, considering a multitude of ans\"atze or models can be done at low cost, justifying the possible overhead caused by computing the superoperators ${}_n \superop{C}$ and $\superop{N}$.

Note that, if $\boldsymbol{S}$ is invertible Eq.~\eqref{eq:linear_system} simplifies to 
\begin{equation}
\label{eq:simple_linear_system}
    \boldsymbol{S} \boldsymbol{x} = \partial_\lambda \boldsymbol{h} \, .
\end{equation}

In general, however, the variational ansatz $\operator{X}$ does not contain all elements of the basis $\Bqty{B_n}_n$ such that $\boldsymbol{S}$ is rectangular and hence non-invertible. 
Therefore, an obvious approximation scheme is to truncate $\boldsymbol{S}$ to a square matrix, i.e., neglecting all operators $\operator{B}_m$ in $\operator{G}=\sum_m \operator{G}_m \operator{B}_m$ which do not appear in $\operator{X}$. This also naturally leads to a truncation in the number of harmonics as discussed in the main text. 
Such a truncation is, however, not strictly necessary, and in general, the matrix $\boldsymbol{M}$ and vector $\boldsymbol{b}$ in Eq.~\eqref{eq:linear_system} are obtained from 
\begin{equation}
\label{eq:general_linear_system}
    \begin{aligned}
        \boldsymbol{M} &= \boldsymbol{S}^T \boldsymbol{\eta} \boldsymbol{S} \\
        \boldsymbol{b} &= \boldsymbol{S}^T\boldsymbol{\eta} \partial_\lambda \boldsymbol{h} \,.
    \end{aligned}
\end{equation}

\section{\label{sec:floquetpump} Additional Material for Floquet Topological Pump Experiment }

\subsection{\label{sec:floquetpump:derivation} Derivation of the Bloch Hamiltonian}

In this section, we report on the details of the model studied in the main text.
For completeness, let us begin with a formal derivation of the Bloch Hamiltonian from the real space lab-frame Hamiltonian.

\begin{figure}[t]
    \centering
    \includegraphics[width=0.5\textwidth]{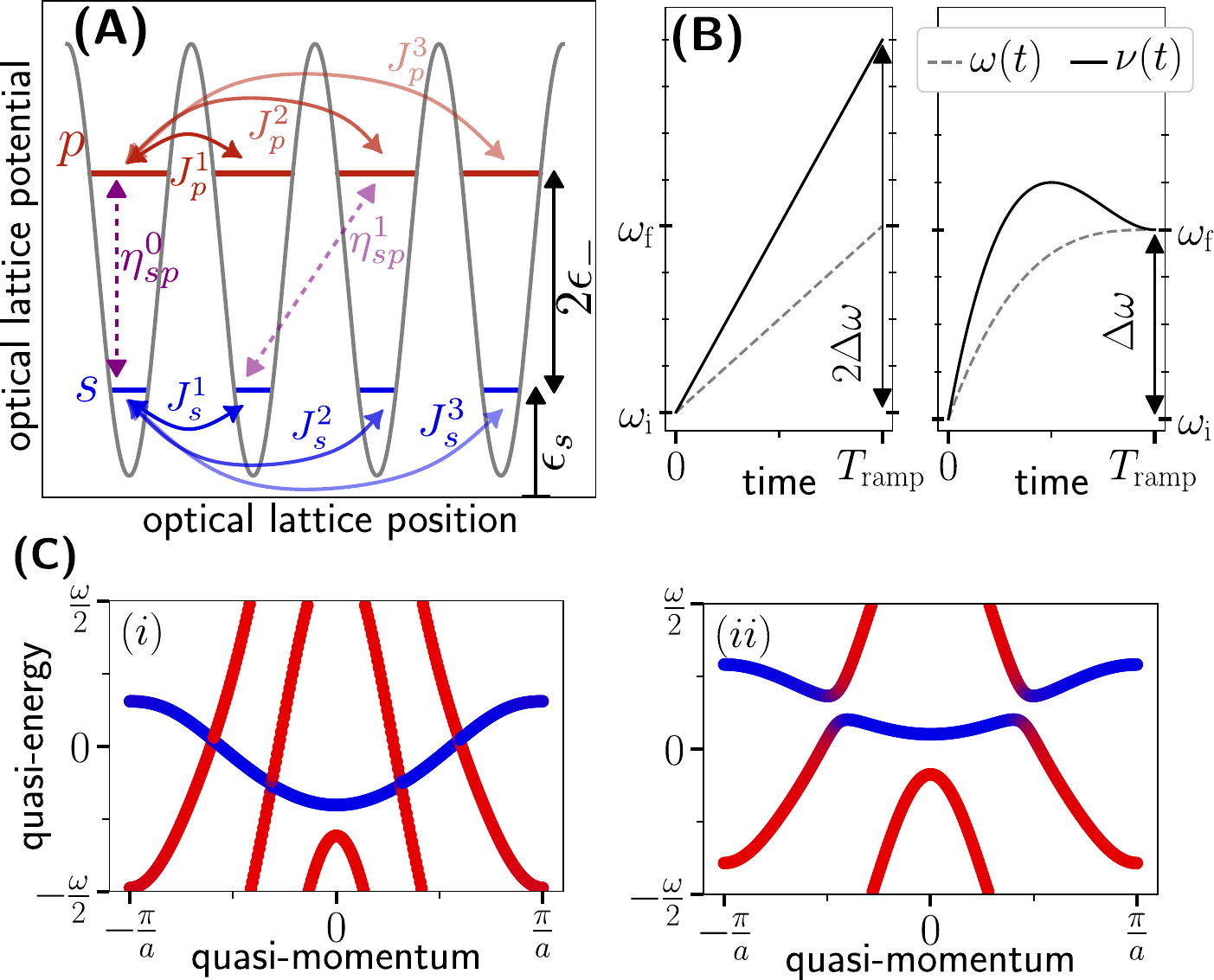}
    \caption{
    \textbf{Fermionic Floquet band model:}
    \textbf{(A)} Schematic of setup with $s$~(blue) and $p$~(red) band energies $\epsilon_{s,p}$, intra-band hoppings $J^{1,2,3}_{s,p}$ and drive-induced inter-band hoppings $\eta^{0,1}_{sp}$.
    \textbf{(B)} In this work we replace the linear drive~(left panel) by a cubic drive~(right panel), see Eq.~\eqref{eq:cubic_drive}, to mitigate the error when considering the time-modulated frequency $\omega(t)$~(gray dashed) instead of the instantaneous frequency $\nu(t)$~(black solid).
    \textbf{(C)} Quasi-energy bands at the beginning~$(i)$ and end~$(ii)$ of the frequency ramp as a function of quasi-momentum, color indicates overlap with $s$~(blue) and $p$~(red) band of the non-driven model. During the state preparation the two bands become $s$ and $p$ band character.
    }
    \label{fig:TopoPump_Intro}
\end{figure}

Following Ref.~\cite{Viebahn_etal_FloquetPump_2022}, the fermionic Hamiltonian in the lab-frame and real space is described by
\begin{equation}
\label{app:eq:fermion_lab}
    H_\mathrm{lab}(t) = \frac{\hat{p}^2}{2m} + V(\hat{x}-x_0(t)) \, ,
\end{equation}
with potential energy $V(x)=V_0 \cos^2(\pi x/a)$ and lattice constant $a$.
The lattice position is driven by the two-tone frequency drive
\begin{equation}
\label{app:eq:topo_drive}
    x_0(t) = \frac{c_\mathrm{exp}}{\omega} \pqty{ K_1 \cos(\phi) + \frac{1}{2} K_2 \cos(2\phi + \varphi) }\, ,
\end{equation}
with phase shift $\varphi$ and time-dependent phase $\phi=\omega t$.
Transforming to the co-moving frame of the shaken lattice, the Hamiltonian reads
\begin{equation}
\label{app:eq:fermion_rot}
    \hat{H}(t) = \frac{\hat{p}^2}{2m} + V(\hat{x}) - F(t)\hat{x} \, ,
\end{equation}
with inertial force $F(t) = - m \ddot{x}_0(t)$.

Using the tight-binding picture, the second quantized operators read
\begin{align*}
    \frac{\hat{p}^2}{2m}+V(\hat{x})               
    \hat{=} \sum_{j,\alpha} \bqty{ 
        \epsilon_\alpha c^\dagger_{j,\alpha}c_{j,\alpha}  
        - \sum_k \pqty{ 
                J_\alpha^k c^\dagger_{j,\alpha}c^\dagger_{j+k,\alpha} 
                + \mathrm{h.c.}
                }    
    }
\end{align*}
and
\begin{align*}
    \frac{\hat{x}}{a} 
    &\hat{=}  \left. \sum_{j\alpha} \right\{
        j c^\dagger_{j,\alpha}c_{j,\alpha} 
    \\
    &\hspace{1.em}\left.
        + \sum_{\beta\neq\alpha}
            \bqty{
                \eta_{sp}^0 c^\dagger_{j,\alpha}c_{j,\beta}
                + \pqty{
                \eta_{sp}^1 c^\dagger_{j,\alpha}c^\dagger_{j+1,\beta}
                + \mathrm{h.c.}
                }
            }
    \right\} \, ,
\end{align*}
with $c^\dagger_{j\alpha}$ ($c_{j\alpha}$) creating (annihilating) a Fermion on site $j$ in the $\alpha(=s,\,p)$ band.

Transforming into a moving frame with respect to the unitary transformation $W(t){=}\exp(i A(t) \sum_{j\alpha} j c^\dagger_{j,\alpha}c_{j,\alpha} )$, with Peierls phase $A(t){=}\frac{-a}{h}\int_0^t F(s) \mathrm{d}s$, the on-site term $\sum_{j\alpha} j c^\dagger_{j,\alpha}c_{j,\alpha} $ is rotated away and the inter-site terms transform as
\begin{align*}
    J_\alpha^k c^\dagger_{j,\alpha}c^\dagger_{j+k,\alpha} 
    &\to 
    e^{-ikA(t))} J_\alpha^k c^\dagger_{j,\alpha}c^\dagger_{j+k,\alpha} 
    \\
    \eta^1_{sp} c^\dagger_{j,\alpha}c^\dagger_{j+k,\beta} 
    &\to 
    e^{-i A(t))} \eta^1_{sp} c^\dagger_{j,\alpha}c^\dagger_{j+k,\beta} \, .
\end{align*}

Using periodic boundary conditions and transforming to quasi-momentum space the Hamiltonian in the rotating frame takes the form:
\begin{equation}
\label{eq:TopoFermion}
    \h_\lambda(t) = \sum_q \boldsymbol{\Psi}_q^\dagger \cdot \boldsymbol{h}_\lambda(q,t) \cdot \boldsymbol{\Psi}_q
\end{equation}
where $\boldsymbol{\Psi}_q^\dagger =(c_{p,q}^\dagger,\, c_{s,q}^\dagger )$ and $\boldsymbol{h}_\lambda(q,t)$ is the Bloch Hamiltonian:
\begin{equation}
\label{eq:TopoBloch}
    \begin{aligned}
         \boldsymbol{h}_\lambda(q,t) 
         &= \pqty{ \epsilon_+ + \sum_{j=1}^3 J_+^k \cos( jqa - jA) } \identity \\
         &- aF(t) \pqty{ \eta^0_{sp} + 2 \eta_{sp}^1 \cos(qa - A) } \sigma^x \\
         &+ \pqty{\epsilon_- - 2 \sum_{j=1}^3 J_-^j \cos(jqa - jA)} \sigma^z
         \, ,
    \end{aligned}
\end{equation}
where $a$ is the lattice spacing constant; the mean and difference between the $s$ an $p$-band in energies is $2\epsilon_\pm = \epsilon_p \pm \epsilon_s$; the intra-band hopping is $2J_\pm^j=J_p^j\pm J_s^j$, and the inter-band hoppings are denoted by $\eta_{sp}^{0,1}$, see sketch in Fig.~\ref{fig:TopoPump_Intro}A.

The external time-periodic driving force $F(t)$ and perierls phase $A(t)$ then read
\begin{equation}
\label{app:eq:force}
    \begin{aligned}
        F(t) &= \omega K_1 \cos(\omega t) {+} 2\omega K_2 \cos(2\omega t + \varphi)\\
        A(t) &= K_1 \sin(\omega t) {+}  2 K_2 \sin(2 \omega t + \varphi)\, .
    \end{aligned}
\end{equation}
respectively.

For the state preparation scheme, we use Eq.~\eqref{app:eq:force} as the definition for the force $F(t)$ and Peierls phase $A(t)$, i.e., also upon introducing additional time-dependencies, e.g., $\omega=\omega(t)$. 
However, in general, they are related to the position displacement $x(t)$, which is the physical quantity that is modulated in the experiment. 
Therefore, in the real experimental setup, the force and Peierls phase may pick up additional terms during the state preparation protocol as they are connected to derivatives of the position displacement, $F\propto \ddot{x}$ and $A\propto \dot{x}$, respectively.
For simplicity, these additional contributions are neglected in our analysis.

\begin{figure}
    \centering
    \includegraphics[width=0.5\textwidth]{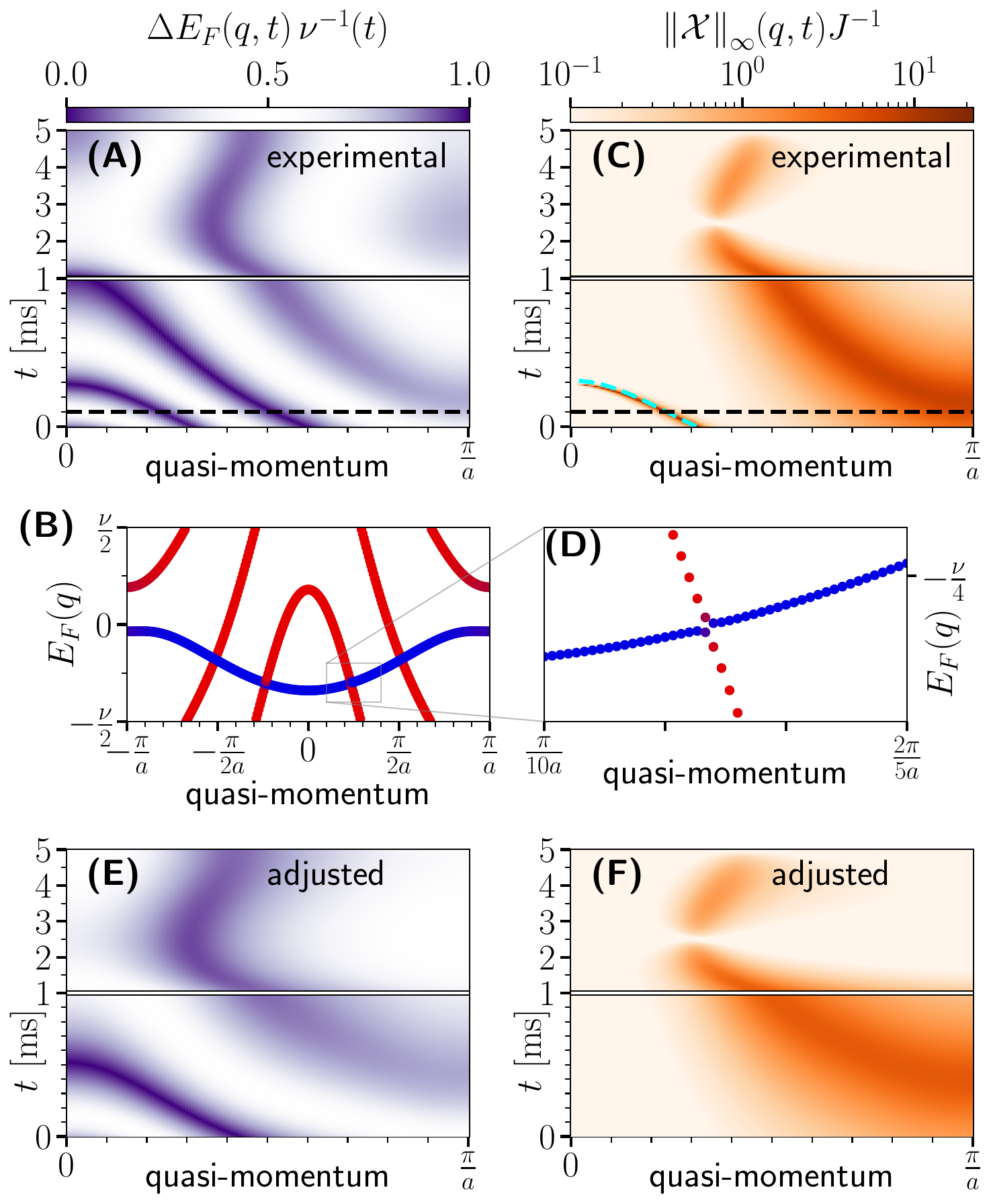}
    \caption{
    \textbf{Additional resonances in state preparation protocol} for topological Floquet pump setup.
    \textbf{(A)} Quasi-momentum resolved quasi-energy gap $\Delta E_F$ normalized by instantaneous frequency $\nu(t)$ for frequency chirp presented in main text. Color map indicates the equivalence of gaps with $\Delta E_F{\equiv} \nu -\Delta E_F $ for quasi-energies.
    \textbf{(B)} Quasi-energy bands at $t=0.1\,\mathrm{ms}$~(see dashed line in A), color indicates overlap with initial $s$~(blue) and $p$~(red) band.
    Due to the folding of the bands level crossings and avoided level crossings occur between $s$ and $p$ band. Avoided crossings with small energy gaps are indistinguishable from level crossings when considering only the quasi-energies.
    \textbf{(C)} $L^\infty$ norm $\norm{\mathcal{X}}_\infty$ of extended Floquet Hilbert space variational FAGP $\operator{X}$. Cyan dashed line indicates the photon resonance computed via Eq.~\eqref{app:eq:topo_photon_resonance}.
    Using the FAGP the level crossings and avoided level crossings are clearly distinguishable.
    \textbf{(D)} Zoom into (B). In agreement with the strong response of FAGP around $t=0$, $q=0.25\frac{\pi}{a}$ we can observe an avoided level crossing with a small gap in the quasi-energies.
    \textbf{(E)} and \textbf{(F)} same as \textbf{(A)} and \textbf{(B)} for adjusted parameters, respectively.
    Using the adjusted parameters we can remove the additional photon resonance.
    Experimental parameters, see table~\ref{tab:TopoPumpParams}, used for A-D, other parameters are as in Fig.~\ref{fig:TopoPump}.
    }
    \label{app:fig:topo_resonance}
\end{figure}

\begin{table}[t]
    \centering
    \caption{
        \textbf{Parameters for Floquet Topological Pump.}
    \update{
        \textit{Left column:} values for the parameters in Eq.~\eqref{eq:TopoBloch} used in the original experiment
        \textit{Right column:}
        adjusted parameters which we use in this work to avoid the photon resonance, i.e., satisfying Eq.~\eqref{app:eq:topo_condition}, see also Sec.~\ref{sec:TopoResonances}.
        The photon resonances can be avoided with minor changes in the experimental parameters.
        }
        The on-site energies $\epsilon_\alpha$ and tunnelings $J_\alpha^j$ are in $\mathrm{kHz}$ and the couplings $\eta_{sp}^{0,1}$ are dimensionless.
    }
    \begin{tabular}{c||c|c}
         Parameter    & Experiment~\cite{Viebahn_etal_FloquetPump_2022} & Adjusted (this work) \\
         \hline
         \hline
        $\epsilon_s$  & $7.523$   & $8.639$\\ 
        $\epsilon_p$  & $20.586$  & $23.639$\\ 
        \hline
        $J_s^0$       & $0.378$               & $0.252$ \\ 
        $J_s^1$       & $-2.620\cdot 10^{-2}$ & $-2.620\cdot 10^{-2}$ \\
        $J_s^1$       & $2.630 \cdot 10^{-3}$ & $2.630 \cdot 10^{-3}$ \\
        \hline
        $J_p^0$       & $-2.037$ & $-1.358$ \\ 
        $J_p^1$       & $-0.357$ & $-0.357$ \\
        $J_p^1$       & $-0.154$ & $-0.154$ \\
        \hline
        $\eta_{sp}^0$ & $0.184$  & $0.184$  \\ 
        $\eta_{sp}^1$ & $-0.059$ & $-0.059$ \\
    \end{tabular}
    \label{tab:TopoPumpParams}
\end{table}

\subsection{
\label{sec:floquetpump:resonance} Detecting photon resonances with the variational Floquet adiabatic gauge potential
}

In this section, we exemplify on the Floquet band model how the variational FAGP can be used to detect unwanted photon resonances. Moreover, exploiting the simplicity of the model we are also able to remove the photon resonance by adjusting the parameters of the Floquet band model.

\subsubsection{General Approach}
Let us first present a general approach that allows for the detection of photon resonances.

The first insight enabling the detection scheme is the sensitivity of the adiabatic gauge potential to the presence of avoided crossings. 
In particular, diabatic transitions only occur near level crossings where the diabatic states strongly hybridize. Thus, the AGP takes its largest values near avoided level crossings and almost vanishes far away from them.
For level crossings, however, the diabatic levels do not hybridize such that the AGP usually vanishes in their vicinity.

Secondly, let us recall that the high-frequency expansion is not susceptible to the presence of photon resonances. In contrast, the variational FAGP is non-perturbative and thus will resolve the photon resonances. Therefore, a photon resonance can be detected by its distinctively different responses for the two FAGPS: a photon resonance is present if the variational FAGP indicates an avoided level crossing, i.e., taking a large value for some parameter regime, and the IFE FAGP takes a small value at the same regime.

This simple approach may break down for many body systems where many avoided level crossings of different levels may appear for the same parameter value. However, if one has access to the perturbative Floquet energies and eigenstates one can always restrict the analysis to a subspace of candidate states where a resonance might occur; that is, those states which are separated by multiples of the driving frequency, such that photon absorption processes become resonant.

\subsubsection{Example: Floquet Topological Pump Setup}
\label{sec:TopoResonances}

\paragraph{Detecting Photon Resonances}
For the example of the Floquet topological pump setup studied in the main text, we use a slightly different approach. In Ref.~\cite{Viebahn_etal_FloquetPump_2022}, the perturbative high-frequency Floquet Bloch bands have already been studied extensively. 
Therefore, we can use those results as a starting point for our analysis.
In particular, the perturbative analysis suggests that during the frequency chirp, the quasi-energy bands hybridize starting at the edge of the Brillouin zone~($q\approx\pm \frac{\pi}{a}$). All other crossings caused by the folding of the energies, see Fig.~\ref{fig:TopoPump_Intro}C, into a quasi-energy Brillouin zone do not lead to hybridization.

Considering the exact instantaneous quasi-energies during the chirp seems to support this picture, see Fig.~\ref{app:fig:topo_resonance}A. Also considering a cut at short times indicates hybridization only at the edge of the Brillouin zone and otherwise no change of the trivial $s$ and $p$ bands, see Fig.~\ref{app:fig:topo_resonance}B.

To distinguish level crossings from avoided level crossings with small gaps, without the need to compute the quasi-energy spectrum with high resolution, we consider the numerically computed non-perturbative variational FAGP~\eqref{eq:chi_topo}, see Fig.~\ref{app:fig:topo_resonance}C. At small times $t\approx 0$ we observe a strong response of the variational FAGP close to the edge of the Brillouin zone, in agreement with the perturbative analysis.
However, we also observe a strong contribution close to the center of the Brillouin zone, see Fig.~\ref{app:fig:topo_resonance}C, suggesting another avoided crossing.
In fact, by drastically increasing the quasi-momentum resolution we can observe the avoided crossing also in the quasi-energy spectrum, see Fig.~\ref{app:fig:topo_resonance}D.
Notice that, this additional photon resonance is not important for the actual experiment as it passed diabatically with almost unit fidelity due to the gap being much smaller compared to the velocity used in the experiment.

\paragraph{Avoiding photon resonances.}

Combining the perturbative Floquet Hamiltonian approach and the insight gained from the variational FAGP we can remove the photon resonance.
Note that, the hybridization of the energy levels at the edge of the Brillouin zone is also engineered using photon resonances. However, this can be accounted for by transforming to a suitable rotating frame, as we describe below.
During the chirp the energy splitting $\epsilon_-$ is comparable to the driving frequency, $\epsilon_- \approx \omega$. By transforming to a rotating frame with respect to $W=\exp(-i\omega \sigma^z t)$ we can remove this resonance: thereby, $\epsilon_-$ is replaced by $\Delta = \epsilon_- - \omega$, and likewise $\sigma^x$ by $\cos(\omega t)\sigma^x + \sin(\omega t)\sigma^y$ in the Bloch Hamiltonian~\eqref{eq:TopoBloch}.

Ignoring any additional resonances, the level splitting in the high-frequency expansion of the Floquet Hamiltonian is in lowest order described by $\Lambda \sigma^z \subset \boldsymbol{h}_F^{\mathrm{rot}, (0)}$ with effective level splitting $\Lambda = \Delta - 2 \sum_k J_-^k \cos(kqa) $.
Due to the rotating frame transformation $\Delta=\Delta(\omega)$ is no longer resonant with $\omega$, such that one might expect no further photon resonances to occur.
However, taking a closer look at the entire expression $\Lambda=\Lambda(\omega,q)$ there exist $\omega^\star$ and $q^\star$ values, such that
\begin{equation}
\label{app:eq:topo_photon_resonance}
    \abs{\Lambda}(\omega^\star,q^\star) = \omega^\star \, ,
\end{equation}
indicating the presence of additional photon resonances.
In fact, the photon resonance location obtained from the matching condition~\eqref{app:eq:topo_photon_resonance} agree well with the avoided gap closings encountered in the FAGP, see Fig.~\ref{app:fig:topo_resonance}C.

To avoid the additional photon resonance we adjust the parameters in the model such that during the ramps Eq.~\eqref{app:eq:topo_photon_resonance} cannot be satisfied. 
In addition, we must preserve the topological properties of the model. To this end, it suffices to ensure that
\begin{equation}
\label{app:eq:topo_condition}
    \begin{aligned}
        \abs{\epsilon_{-} - \omega_\mathrm{i}} &> 2 \sum_k \abs{J_-^k} \\
        \abs{\epsilon_{-} - \omega_\mathrm{f}} &< 2 \sum_k \abs{J_-^k} \\
    \end{aligned}
\end{equation}
where $\omega_\mathrm{i}$ and $\omega_\mathrm{f}$ refer to the frequencies at the beginning and end of the protocol, respectively.
Let us emphasize, that these three conditions are not sufficient to uniquely determine the parameters of the model. In Tab.~\ref{tab:TopoPumpParams}, we present a possible choice of parameters that we used in the main text, \update{which we expect to be within reach for the experimental platform}.

\paragraph*{Summary.}

We demonstrated that the non-perturbative character of the variational FAGP allows us to detect photon resonances. In fact, for the simple example of the Floquet band model, we have even been able to adjust the parameters to avoid the detected photon resonances. This is a key ingredient to yield the high-fidelity FCD state preparation scheme presented in the main text.

In addition, this example demonstrates that the adiabatic gauge potential can be used to distinguish level crossings from avoided level crossings with small gaps, without the need to resolve the small gap -- for both static and Floquet systems.

\begin{figure}[t]
    \centering
    \includegraphics[width=0.5\textwidth]{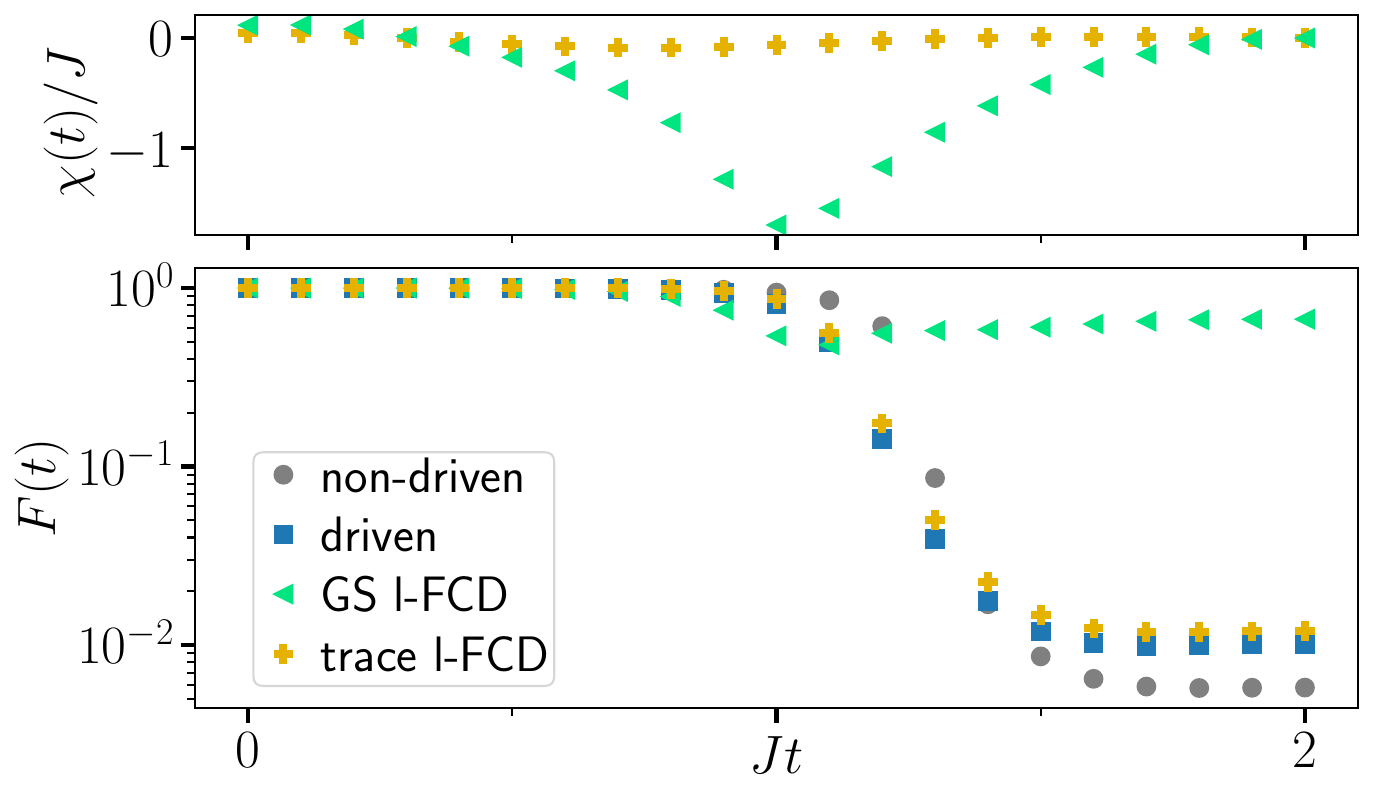}
    \caption{
    \textbf{Groundstate expecation value vs trace l-FCD} for circularly driven transverse field Ising model~$\operator{H}_\mathrm{TFI}(t;A,\omega)$.
    \textit{Upper Panel:} Numerically computed variational parameter $\chi(t)$, see Eq.~$\operator{H}_\mathrm{TFI}(t;A,\omega)$, for l-FCD using the Floquet groundstate~(GS) expectation value~(green triangles) and trace~(yellow pluses).
    The groundstate l-FCD shows a stronger response than the trace l-FCD. 
    \textit{Lower Panel:} Instantaneous fidelities $F(t)$ for the unassisted non-driven~(gray circles) and driven~(blues squares) protocol, and the groundstate l-FCD~(green triangles) and trace l-FCD~(yellow pluses) assisted Floquet protocol. Note the log scale for the $y$-axis.
    The trace l-FCD protocol leads to a small increase in fidelity compared to the unassisted protocol. However, the increase in fidelity is considerably larger for the groundstate l-FCD.
    We use $JT_\mathrm{ramp}=2$, other parameters as in Fig.~\ref{fig:ManyBody}.
    }
    \label{app:fig:manybody_GSvsTrace}
\end{figure}

\begin{figure}[t]
    \centering
    \includegraphics[width=0.5\textwidth]{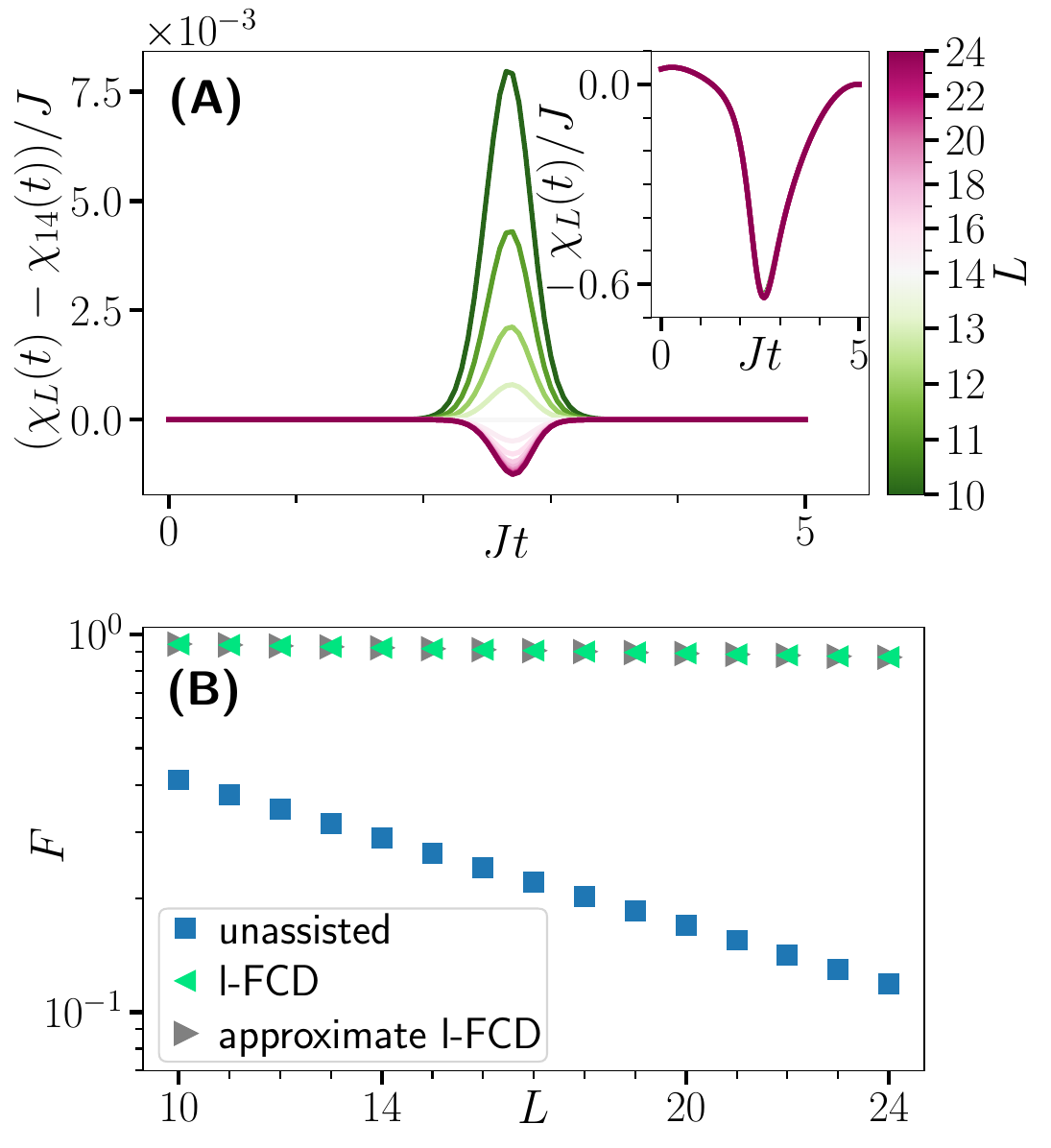}
    \caption{
    \textbf{Approximate groundstate l-FCD} for circularly driven transverse field Ising model~$\operator{H}_\mathrm{TFI}(t;A,\omega)$.
    \textbf{(A)} Deviation, $\pqty{\chi_L(t)-\chi_{14}(t)}$, of variational parameter $\chi_L(t)$ for different system sizes $L$ from variational parameter obtained for $L=14\equiv L_\mathrm{comp}$. 
        \textit{Inset:} Variational parameters for different system sizes $L$.
    The l-FCD protocol hardly changes as a function of system size.
    \textbf{(B)} Final fidelities $F$ for different system sizes for unassisted driven~(blue squares) and l-FCD~(green triangles) protocol, as shown in Fig.~\ref{fig:ManyBody}, and approximate l-FCD for $L_\mathrm{comp}=14$, as described in Sec.~\ref{app:manybody}.
    For this example, computing the l-FCD for a small system size is sufficient to yield a high-fidelity l-FCD protocol for all other considered system sizes.
    Other parameters are as in Fig.~\ref{fig:ManyBody}.
    }
    \label{app:fig:manybody_approxlFCD}
\end{figure}

\begin{figure}[t]
    \centering
    \includegraphics[width=0.5\textwidth]{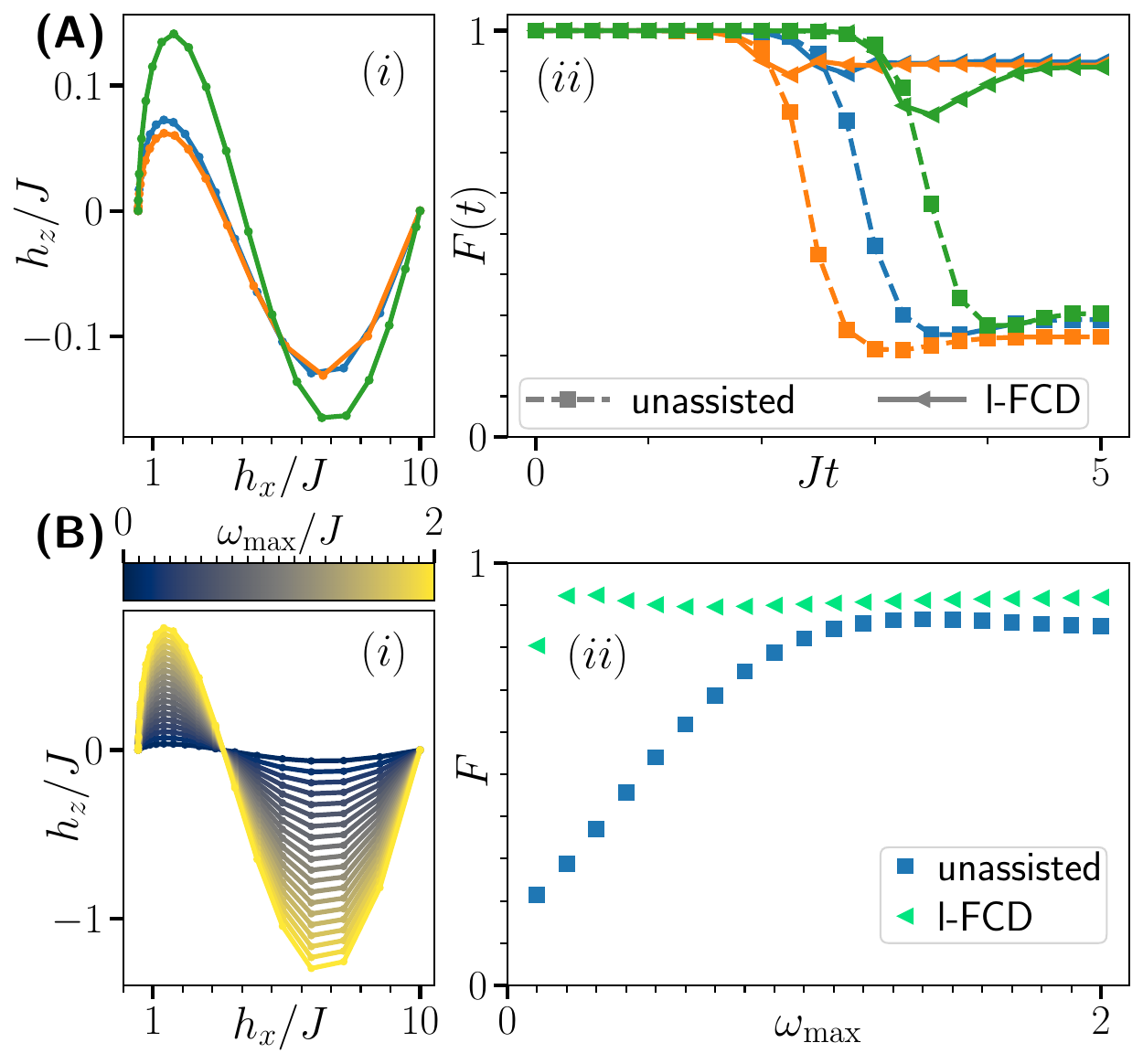}
    \caption{
    \textbf{Protocol Dependence of state preparation} for circularly driven transverse field Ising model $\operator{H}_\mathrm{TFI}(t;A,\omega)$.
    \textbf{(A)} Dependence on the protocol $\lambda(t)$ with $A(t)/J=\bqty{10-9.5\times\lambda\pqty{t}}$ and $\omega(t)=\bqty{\omega_\mathrm{max}\times\sin(\pi \lambda\pqty{t})}$ for cubic~\eqref{eq:cubic_drive}~(blue), quartic~\eqref{app:eq:quartic_drive}~(orange) and quadratic-to-cubic~\eqref{app:eq:quadratic_cubic}~(green) drive.
    (i) Protocols in $h_x$-$h_z$ plane, (ii) fidelity of protocols.
    \textbf{(B)} Dependence on $\omega_\mathrm{max}$ with
    (i) protocols in $h_x$-$h_z$ plane, (ii) corresponding fidelity of protocols.
    Data is obtained for a chain of length $L=14$ and all other parameters are as in Fig.~\ref{fig:ManyBody}, if not stated otherwise.
    }
    \label{fig:app:manybody_protocol}
\end{figure}

\section{\label{app:manybody} Details of Many Body System}

\paragraph{Trace vs groundstate expectation value.}

In the main text, we presented results for l-FCD protocols using the groundstate expectation value instead of the trace in the variational principle~\eqref{eq:varl-main}. Here, we briefly compare the performance of the two approaches.
In Fig.~\ref{app:fig:manybody_GSvsTrace}, we show the performance of both methods compared to the unassisted protocols presented in the main text.
The l-FCD protocol using the trace norm results in an increase in fidelity compared to the unassisted protocol, see lower panel of Fig.~\ref{app:fig:manybody_GSvsTrace}. However, the increase is negligible compared to the increase in fidelity caused by the l-FCD using the groundstate norm, see lower panel of Fig.~\ref{app:fig:manybody_GSvsTrace}. 
This can be understood by taking a closer look at the variational parameters obtained for both methods, upper panel of Fig.~\ref{app:fig:manybody_GSvsTrace}. We find that the groundstate l-FCD has a notably stronger counter term than the trace l-FCD, thus, motivating the reduced performance of the latter.

\paragraph{Approximate groundstate expectation value}

In the previous paragraph, we demonstrated that using the (Floquet) groundstate expectation value in the l-FCD variational principle leads to a notable improvement in fidelity compared to using the trace.
However, in general computing the Floquet groundstate involves computing the exact Floquet Hamiltonian and computing its eigenstate. Therefore, computing the Floquet groundstate and evaluating the action~\eqref{eq:varl-main} is computationally demanding and can in general only be done for small system sizes.

A possible approximation scheme that reduces the computational cost is to compute the l-FCD protocols for some computationally tractable system size $L_\mathrm{comp}$ and then apply it to any target system size $L_\mathrm{target}$.
Using this scheme inevitably leads to degradation in fidelity compared to directly computing the protocol for the target system size, $L_\mathrm{comp}=L_\mathrm{target}$, however, allows us to access arbitrary system sizes.

Let us emphasize, that in order to reach system size of up to $L=24$ spins, as presented in the main text, we need to exploit two properties specific to this model: (i) we can exactly write down the Floquet Hamiltonian saving us to compute the Floquet Hamiltonian numerically and (ii) the parity and translational symmetry of the model reduce the effective Hilbert space dimension notably.
While (ii) may hold in more general scenarios, (i) is a fine-tuned scenario, such that in general the Floquet Hamiltonian must be computed numerically; putting serious restrictions on the achievable system size.

Therefore, we consider the proposed approximation scheme for $L_\mathrm{comp}=14$ -- a system size where computing the Floquet Hamiltonian is numerically tractable. We present results for this approximate groundstate l-FCD in Fig.~\ref{app:fig:manybody_approxlFCD}.
Notably, the l-FCD protocol, characterized by the variational parameter $\chi_L(t)$ in Eq.~\eqref{eq:varl_ansatz-main}, hardly varies for different system sizes $L=10,\dots,24$, see Fig.~\ref{app:fig:manybody_approxlFCD}A. This suggests that using the protocol obtained for some small system size $L_\mathrm{comp}$ also performs well for other system sizes.
In fact, we find no notable difference in the final fidelity between the approximate and exact groundstate l-FCD protocols, see Fig.~\ref{app:fig:manybody_approxlFCD}B.

In summary, using the approximate groundstate l-FCD enables to access high-fidelity counter diabatic protocols at low computational cost. However, the performance of such approximations in general depends strongly on the considered model.

\paragraph{Protocol dependence}

Eventually, let us emphasize, that the choice of protocol for the non-equilibrium drive also can have a significant impact on the performance of the unassisted and l-FCD protocols. 
For the many body model obvious choices of parameters controlling the protocol are the maximum extent in the $z$-direction determined by the range $\omega_\mathrm{max}$ of the frequency $\omega\in\bqty{0,\,\omega_\mathrm{max}}$, and the fundamental protocol $\lambda(t)$.

We demonstrate the dependence of the performance of the state preparation on these parameters of the protocol in Fig.~~\ref{fig:app:manybody_protocol}.
First, let us focus on the dependence on the fundamental drive protocol $\lambda(t)$, see Fig.~\ref{fig:app:manybody_protocol}A. In particular, we compare the cubic ramp~\eqref{eq:cubic_drive}, used in the main text,
\begin{equation}
\label{eq:cubic_drive}
    \omega(t) = \omega_\mathrm{f} + \pqty{\omega_\mathrm{i} - \omega_\mathrm{f}} \pqty{\frac{t_\mathrm{f} - t}{t_\mathrm{f} - t_\mathrm{i}}}^3\, .
\end{equation}
with the quartic ramp
\begin{equation}
\label{app:eq:quartic_drive}
    \lambda(t) = x(t)^4\, .
\end{equation}
and the quadratic-to-cubic ramp
\begin{equation}
\label{app:eq:quadratic_cubic}
    \lambda(t) = \pqty{t_\mathrm{f} - t_\mathrm{i}} \bqty{ \frac{x\pqty{t}^4}{4} - \frac{2x\pqty{t}^3}{3} + \frac{x\pqty{t}^2}{2} }
\end{equation}
where $x{=}\pqty{t - t_\mathrm{i}}/T_\mathrm{ramp}$.
The quadratic-to-cubic drive~\eqref{app:eq:quadratic_cubic} is an adjusted quartic drive with $\dot{\lambda}(t) = x(1-x)^2$ ensuring that $\nu(t)$ is quadratic around $t=t_\mathrm{i}$ and cubic around $t=t_\mathrm{f}$.
The choice of drive can impact the final fidelity for both unassisted and l-FCD protocols, see Fig.~\ref{fig:app:manybody_protocol}A. However, for the chosen protocols the change in final fidelity is negligible for the l-FCD protocol.

Second, we consider different $\omega_\mathrm{max}=0.1J,\,\dots,\,2J$ values, see Fig.~\ref{fig:app:manybody_protocol}B. Again, for l-FCD protocols, the dependence on the details of the protocol is less dominant than for the unassisted protocol. Notably, for l-FCD the optimal protocol is reached around $\omega_\mathrm{max}=0.2J$. In contrast, for the unassisted protocol the optimum is assumed around $\omega_\mathrm{max}=1.4J$, see Fig.~\ref{fig:app:manybody_protocol}B. 
Therefore, the optimization of such hyperparameters may always be performed with respect to the l-FCD protocol in order to yield the highest fidelity protocol.

In summary, if one considers approximate l-FCD state preparation protocols the performance of the protocol may depend on the details of the respective unassisted protocol. Therefore, to find an optimal driving protocol an optimization of the protocol should be performed.

\end{document}